\documentclass[pre,aps]{revtex4}

\usepackage{times}
\usepackage{amsmath}
\usepackage{graphicx}
\usepackage{amssymb}
\usepackage{color}
\usepackage{xspace}

\newcommand{\wb}{\omega_{\mathrm{b}}}
\newcommand{\Tb}{T_{\mathrm{b}}}
\newcommand{\wt}{\wb\, t}

\newcommand{\ii}{\mathrm{i}}

\newcommand{\F}{\mathcal{F}}

\newcommand{\cP}{\ensuremath{\mathcal{P}}}
\newcommand{\cT}{\ensuremath{\mathcal{T}}}
\newcommand{\veps}{\varepsilon}
\newcommand{\vphi}{\varphi}
\newcommand{\Ap}{$\mathrm{A}_\mathrm{\phi}$\xspace}
\newcommand{\Sp}{$\mathrm{S}_\mathrm{\phi}$\xspace}
\newcommand{\Ao}{$\mathrm{A}_\mathrm{0}$\xspace}
\newcommand{\So}{$\mathrm{S}_\mathrm{0}$\xspace}
\newcommand{\At}{$\mathrm{A}_\mathrm{3}$\xspace}
\newcommand{\Atp}{$\mathrm{A}_\mathrm{3}^+$\xspace}
\newcommand{\Atm}{$\mathrm{A}_\mathrm{3}^-$\xspace}
\newcommand{\sn}{{\rm sn}}

\newcommand{\cn}{{\rm cn}}

\newcommand{\wA}{\omega_{\mathrm{A}}}
\newcommand{\wS}{\omega_{\mathrm{S}}}
\newcommand{\wps}{\omega_{\mathrm{\phi +}}}
\newcommand{\wph}{\omega_{\mathrm{\phi -}}}
\newcommand{\wass}{\omega_{\mathrm{AS +}}}
\newcommand{\wash}{\omega_{\mathrm{AS -}}}
\newcommand{\gL}{\gamma_{\mathrm{L}}}
\newcommand{\gPT}{\gamma_{\mathrm{PT}}}
\newcommand{\ga}{\gamma_{\mathrm{a}}}
\newcommand{\gs}{\gamma_{\mathrm{s}}}

\begin{document}

\title{$\cP\cT$-Symmetric dimer in a generalized model of coupled nonlinear oscillators}

\author{Jes\'us Cuevas--Maraver}
\affiliation{Nonlinear Physics Group, Departamento de F\'{\i}sica Aplicada I, Universidad de Sevilla, Escuela Polit{\'e}cnica Superior, C/ Virgen de \'Africa, 7, 41011-Sevilla, Spain \\
Instituto de Matem\'aticas de la Universidad de Sevilla (IMUS).
Edificio Celestino Mutis. Avda. Reina Mercedes s/n, 41012-Sevilla, Spain}

\author{Avinash Khare}
\affiliation{Indian Institute of Science Education and Research (IISER), Pune 411008, India}

\author{Panayotis\ G.\ Kevrekidis}
\affiliation{Department of Mathematics and Statistics, University of
Massachusetts, Amherst, MA 01003-9305, USA}

\author{Haitao Xu}
\affiliation{Department of Mathematics and Statistics, University of
Massachusetts, Amherst, MA 01003-9305, USA}

\author{Avadh Saxena}
\affiliation{Center for Nonlinear Studies and Theoretical Division, Los Alamos National Laboratory, Los Alamos, New Mexico 87545, USA}

\begin{abstract}
In the present work, we explore the case of a general
$\mathcal{PT}$-symmetric dimer in the context of
two both linearly and nonlinearly coupled cubic oscillators. To obtain
an analytical handle on the system, we first explore the rotating
wave approximation converting it into a discrete nonlinear
Schr{\"o}dinger type dimer. In the latter context, the stationary
solutions and their stability are identified numerically but
also wherever possible analytically. Solutions stemming
from both symmetric and anti-symmetric
special limits are identified. A number of special cases are explored regarding the
ratio of coefficients of nonlinearity  between oscillators over
the intrinsic one of each oscillator. Finally, the considerations
are extended to the original oscillator model, where periodic
orbits and their stability are obtained. When the solutions are
found to be unstable their dynamics is monitored by means
of direct numerical simulations.
\end{abstract}

\maketitle

\section{Introduction}

The notion of parity-time ($\cP\cT$) symmetry has recently been
receiving increasing attention over a wide variety of settings; see
e.g. \cite{Bender_review,special-issues,review}. The original proposal
involved a non-Hermitian variant of quantum mechanics which might
still produce real eigenvalues (and hence be associated with measurable
quantities). However, it was instead the analogy of this model with
the paraxial approximation in optics which led to the proposal
that such mathematical constructs can be realized in optical
settings~\cite{Muga,PT_periodic}, and which eventually led
to their experimental realization~\cite{experiment}. This series
of developments, in turn, prompted researchers towards a
more detailed understanding of the stationary states of such
$\cP\cT$-systems (and how they differ from their Hamiltonian analogues),
an effort to appreciate their stability properties and finally
an attempt to quantify their nonlinear dynamics. This effort
emerged both at the level of few-site
configurations~\cite{Dmitriev,Li,Guenter,Ramezani,suchkov,Sukhorukov,ZK,MKK,CW}
(which were chiefly experimentally accessible), as well as at that of
infinite-size lattices~\cite{Pelin1,Sukh,zheng}.

Although the quantum-mechanical and paraxial-optical focal points
of this activity have provided an emphasis on the study of
Schr{\"o}dinger type settings, a number of recent studies,
especially on the experimental side, have led to an increased
interest in oscillator systems (which one can think of as
oligomers -few site settings- of the Klein-Gordon type).
More specifically, a mechanical system realizing $\cP\cT$-symmetry
has been proposed in~\cite{pt_mech}, while a major thrust of
research has focused on the context of electronic circuits; see e.g.
the original realization of~\cite{tsampikos_recent} and
the more recent review of this activity in~\cite{tsampikos_review}.
As an aside, we note that additional intriguing realizations
of $\cP\cT$-symmetry have also emerged e.g. in the realm of
whispering-gallery microcavities~\cite{pt_whisper}. Mostly,
the efforts on this oscillator realm have been limited to the
study of linear systems, yet recently a number of nonlinear
variants have been explored both theoretically/numerically and even
experimentally. As notable such examples, we mention the split-ring
resonator chain in the context of magnetic metamaterials proposed in
the work of~\cite{tsironis}, as well as the experimental realization
of a $\cP\cT$-symmetric dimer of Van-der-Pol oscillators that
arose in the work of~\cite{bend_kott}.

On the theoretical side, some of these studies raised a number of
intriguing theoretical questions. For instance, the theoretical
modeling of the linear $\cP\cT$-symmetric analogue
of the system~\cite{pt_whisper}
led to the realization that such linear oscillator pairs may be
Hamiltonian {\it although} one of them has gain and the other
has loss~\cite{bend_gian}. This, in turn, led the authors of~\cite{bar_gian}
to appreciate that this feature (the Hamiltonian nature of a
$\cP\cT$-symmetric oscillator system) can be extended to the
nonlinear case, if the nonlinearity contains both self- and cross-
interactions between the oscillators and if these interactions
have an appropriate ratio (the ratio utilized between cross- and
self-interactions in that work was $3$). Importantly, the latter
work also extended consideration of that model to the Schr{\"o}dinger
variant thereof (through a multiple scales expansion), finding
that nonlinearity may, in that context, ``soften'' the $\cP\cT$-symmetric
phase transition. That is, it may enable the existence of stable periodic
and quasi-periodic states at any value of the gain-loss parameter
$\gamma$.

Our aim in the present work is to revisit this context of
two coupled nonlinear oscillators, one of which bears gain
and the other loss. We will consider the nonlinear case
(almost exclusively, briefly touching upon the linear case
as a special limit). Importantly, we will also explore
the ratio of cross- to self-interaction of the oscillators
as a free parameter. Interestingly, this will enable us
to identify a series of special cases, {\it including} the integrable one recently explored in~\cite{bar_gian}. For all values of this nonlinear parameter
and as a function of the loss/gain parameter ($\gamma$)
and of the frequency parameter $\wb$, we will study
systematically {\it both} dimer systems. That is, we will first derive
and analyze the discrete nonlinear Schr{\"o}dinger (DNLS)
dimer, to obtain a simplified understanding of the existence,
stability and dynamics properties. Then, in a way
reminiscent of our earlier work (involving no cross
interactions)~\cite{PRA} --and complementing the earlier
work of~\cite{bar_gian} who did not focus on the periodic
orbit solutions of the original nonlinear oscillator dimer--,
we will return to the oscillator system and explore its
own solutions, in terms of their existence, stability
and dynamical properties. When the solutions are identified
as unstable, a brief discussion will also be given of their
dynamical evolution.

This paper is organized as follows.  In the next section (II)
we provide the model
equations, discuss their symmetries and potential
Hamiltonian structure and indicate where
corresponding exact solutions for the generalized coupled
nonlinear oscillators can be obtained.  In Sec. III we invoke the
rotating wave approximation (RWA) and provide the stability equations and
analytical results as well as perform numerical analysis of both the
symmetric and asymmetric solutions for the resulting generalized DNLS dimer.
Section IV contains the corresponding analysis of the Klein-Gordon dimer.
A discussion of the dynamics of unstable solutions is given in Sec. V.
Our main results and conclusions are summarized in Sec. VI, where
a number of directions for future study are also highlighted.
Details of the numerical analysis are relegated to Appendix A.

\section{The model}

As per the above discussion,
we consider the system of coupled oscillators given by:

\begin{eqnarray}\label{eq:dyn}
\ddot{u} &=& - u + k v + \gamma \dot{u} +\epsilon u^3+\delta u v^2, \nonumber \\
\ddot{v} &=& - v + k u - \gamma \dot{v} +\epsilon v^3+\delta v u^2 .
\end{eqnarray}

This model is an extension of that in \cite{PRA}, which can be obtained by taking $\delta=0$. Additionally, it is an extension of the specific case
of $\delta=3 \epsilon$ considered in~\cite{bar_gian}.
In the linear limit, $\delta=\epsilon=0$, there are two branches of solution eigenfrequencies given by:
\begin{equation}\label{eq:linear}
    \omega_\pm= \sqrt{1 - \gamma^2/2 \pm \sqrt{k^2 - \gamma^2 + \gamma^4/4}}
\end{equation}
with $\omega_+$ ($\omega_-$) corresponding to symmetric (anti-symmetric) linear modes at $\gamma=0$.
Here, we proceed with the understanding that $\pm \omega_{\pm}$
are of relevance but we will focus our attention on the positive
frequencies hereafter.
The two  pairs of real (for small $\gamma$) eigenfrequencies will
collide and give rise to a frequency quartet for $\gamma > \gamma_{PT,L}$, where $\gamma_{PT,L}$ satisfies the condition:
\begin{equation}\label{eq:PTlinear}
    \gamma_{PT,L}^4 - 4 \gamma_{PT,L}^2 + 4 k^2=0 .
\end{equation}

\noindent Thus, for fixed $k$, the lowest value of $\gamma_{PT,L}$ corresponds to $\omega=\sqrt[4]{1-k^2}$.

Additionally, to this linear analysis, we observe that the nonlinear
dynamical equations (\ref{eq:dyn}) possess several symmetries that
leave them invariant:

\begin{itemize}

\item $u\rightarrow-u$, $v\rightarrow-v$ ,

\item $u\rightarrow-u$, $k\rightarrow-k$, $v\rightarrow v$ ,

\item $u\rightarrow u$, $k\rightarrow-k$, $v\rightarrow-v$ ,

\item $t\rightarrow-t$, $\gamma\rightarrow-\gamma$ .

\item $u\rightarrow\alpha u$, $v\rightarrow\alpha v$, $\epsilon\rightarrow\epsilon/\alpha^2$, $\delta\rightarrow\delta/\alpha^2$ .

\end{itemize}

In the limit $\gamma=0$, (\ref{eq:dyn}) is a Hamiltonian system, with $H$ given by

\begin{equation}\label{eq:Ham}
    H=\frac{\dot{u}^2+\dot{v}^2+u^2+v^2}{2}-\frac{\epsilon}{4}(u^4+v^4)-kuv-\frac{\delta}{2}u^2v^2,
\end{equation}

\noindent and, for the case $\delta=3\epsilon$, dynamical equations (\ref{eq:dyn}) are Hamiltonian for any value of $\gamma$ \cite{bar_gian}, with
a Hamiltonian of the form:

\begin{equation}\label{eq:Ham2}
H_{2} = p_{u} p_{v} +\frac{\gamma}{2} (u p_{u}-vp_{v})+(1-\frac{\gamma^2}{4})u v
-\frac{k}{2}(u^2+v^2) -\epsilon (u^3 v+ v^3 u)\,,
\end{equation}

In this case, $p_{u} = \dot{v}+\gamma v/2,~p_{v} = \dot{u}-\gamma u/2$.%

The aim of this paper is to identify periodic orbits of frequency $\wb$ of the model (\ref{eq:dyn}) (and to compare them also to the results of the DNLS
approximation). Toward achieving this aim, Fourier space techniques have been utilized in order to expand the solution in time and to obtain its numerically exact form (up to a prescribed numerical tolerance). Finally, Floquet theory has been used to explore the stability of the pertinent configurations. More details about the numerical methods have been given in Appendix \ref{app:numerics}.

An important diagnostic quantity for probing the dependence of the solutions on parameters such as the gain/loss strength $\gamma$, or the oscillation frequency $\wb$, is the energy averaged over a period, defined as:
\begin{equation}
    <H>=\frac{1}{\Tb}\int_0^{\Tb}\ H(t)\,\mathrm{d}t ,
\end{equation}
with the Hamiltonian (of the case without gain/loss) given by (\ref{eq:Ham}) and $\Tb=2\pi/\wb$ being the oscillation period.

In what follows, we will restrict to the values of $\delta/\epsilon=\{1,3/2,3\}$, for which as will be seen below, the solutions and/or dynamical equations possess special properties. In addition, we restrict to $\gamma\geq0$, $0<k<1$ and $|\epsilon|=1$. Unless stated otherwise $k=\sqrt{15}/8\approx0.48$ has been fixed; this value implies $\gamma_{PT,L}=0.5$.

\section{The Rotating Wave Approximation}\label{sec:RWA}

\subsection{The DNLS dimer: the model, stability equations and analytical results}

The RWA provides a means of connecting with the extensively analyzed $\cP\cT$-symmetric Schr{\"o}dinger
dimer~\cite{Li,Ramezani,CW,R44,hadiadd,dwell,barasflach2}. This link
follows a path similar to what has been earlier proposed e.g. in \cite{RWA1,RWA2}. In particular, the following ansatz is used to approximate the solution of the periodic orbit problem as a roughly monochromatic wavepacket of frequency $\wb$ (for $\phi_{1,2}$ in what follows we will seek stationary states).
\begin{equation}\label{eq:RWA0}
    u(t)\approx \phi_1(t)\exp(\ii \wb t)+\phi^*_1(t)\exp(-\ii \wb t),\qquad v(t)\approx \phi_2(t)\exp(\ii \wb t)+\phi^*_2(t)\exp(-\ii \wb t) .
\end{equation}

By supposing that $\dot\phi_n\ll\wb\phi_n$ and $\ddot\phi_n\ll\wb\dot\phi_n$ (i.e., $\phi$ varies slowly on the scale of the oscillation of
the actual exact time periodic state), discarding the terms multiplying $\exp(\pm 3\ii\wb t)$, the dynamical equations (\ref{eq:dyn}) transform into a set of coupled Schr{\"o}dinger type equations:
\begin{eqnarray}\label{eq:DNLS}
2\ii\wb\dot{\phi}_1 &=& [(\wb^2-1)+3\epsilon|\phi_1|^2+2\delta|\phi_2|^2+\ii\wb\gamma]\phi_1 + [k+\delta\phi_1^*\phi_2]\phi_2 , \nonumber \\
2\ii\wb\dot{\phi}_2 &=& [(\wb^2-1)+3\epsilon|\phi_2|^2+2\delta|\phi_1|^2-\ii\wb\gamma]\phi_2 + [k+\delta\phi_2^*\phi_1]\phi_1,
\end{eqnarray}
i.e., forming, under these approximations, a $\cP\cT$-symmetric Schr{\"o}dinger dimer. The stationary solutions of this dimer can then be used in order to reconstruct via Eq.~(\ref{eq:series}) the solutions of the RWA to the original $\cP\cT$-symmetric oscillator dimer. These stationary solutions for $\phi_1(t)\equiv y_1$ and $\phi_2(t)\equiv z_1$ satisfy the algebraic conditions

\begin{eqnarray}\label{eq:RWA1}
    Ey_1&=&(p+qz_1y_1^*) z_1+(|y_1|^2+2q|z_1|^2)y_1+\ii\Gamma y_1 , \nonumber \\
    \label{eq:RWA2}
    Ez_1&=&(p+qy_1z_1^*) y_1+(|z_1|^2+2q|y_1|^2)z_1-\ii\Gamma z_1 ,
\end{eqnarray}

\noindent with

\begin{equation}
    E=\frac{1-\wb^2}{3\epsilon},\quad p=\frac{k}{3\epsilon},\quad q=\frac{\delta}{3 \epsilon},\quad \Gamma=\frac{\gamma\wb}{3\epsilon} .
\end{equation}

\noindent Notice that when $q=1/2$, i.e. $\delta/\epsilon=3/2$, coupling in Eq. (\ref{eq:DNLS}) resembles that in the Manakov model \cite{manakov}.

If we express $y_1$ and $z_1$ in polar form:

\begin{equation}\label{eq:polar}
    y_1=A\exp(\ii \theta_1),\qquad z_1=B\exp(\ii \theta_2),\qquad \varphi=\theta_2-\theta_1 ,
\end{equation}
the stationary equations can be rewritten as

\begin{eqnarray}\label{eq:RWAb1}
    EA&=&pB \cos\ \varphi + qAB^2\cos\ 2\varphi+A(A^2+2qB^2) ,
\\
\label{eq:RWAb2}
    EB&=&pA \cos\ \varphi + qBA^2\cos\ 2\varphi+B(B^2+2qA^2) ,
\\
\label{eq:RWAb3}
- \Gamma B &=& A \sin\ \varphi (p+2qAB\cos\ \varphi),
\\
\label{eq:RWAb4}
- \Gamma A &=& B \sin\ \varphi (p+2qAB\cos\ \varphi) .
\end{eqnarray}

In the case $\gamma=0$, there can be symmetric or anti-symmetric solutions, fulfilling $A^2=B^2$. Contrary to the $\delta=0$ setting,
where $\sin\ \varphi=0$ only, here we have, apart from this case, the possibility of a phase different than 0 or $\pi$, i.e. $\cos\ \varphi=-p/(2qAB)$. Consequently, we have two pairs of symmetric / anti-symmetric solutions with $A=B$ at the Hamiltonian limit:

\begin{eqnarray}\label{eq:RWAgamma0}
    A^2=\frac{E-p}{1+3q}=\frac{1-\wb^2-k}{3(\epsilon+\delta)}, & \varphi=0 & \textrm{\So solution} \\ \nonumber \\
    A^2=\frac{E+p}{1+3q}=\frac{1-\wb^2+k}{3(\epsilon+\delta)}, & \varphi=\pi & \textrm{\Ao solution} \\ \nonumber \\
    A^2=\frac{E}{1+q}=\frac{1-\wb^2}{3\epsilon+\delta},
    & \vphi=\cos^{-1}\left[-\frac{p(1+q)}{2qE}\right]=\cos^{-1}\left[-\frac{k(\delta+3\epsilon)}{2\delta(1-\wb^2)}\right] & \textrm{\Sp solution} \\ \nonumber \\
    A^2=\frac{E}{1+q}=\frac{1-\wb^2}{3\epsilon+\delta},
    & \vphi=\pi+\cos^{-1}\left[\frac{p(1+q)}{2qE}\right]=\pi+\cos^{-1}\left[\frac{k(\delta+3\epsilon)}{2\delta(1-\wb^2)}\right] & \textrm{\Ap solution}
\end{eqnarray}

Recall that $A=B$ in all the previous cases, i.e. the sign of the anti-symmetric solutions has been introduced into the phase. Apart from the previous solutions, there is an asymmetric solution (AS) whose properties strongly depend on $\delta/\epsilon$.
This solution is given by:

\begin{equation}\label{eq:RWAasym}
    A^2=\frac{(1-\wb^2)\pm\sqrt{(1-\wb^2)^2-\frac{4k^2}{(1-\delta/\epsilon)^2}}}{6\epsilon},\ \qquad
    B=\pm \frac{k}{3(\epsilon-\delta)A},\ \qquad \varphi=0 ~(\pi)\ \qquad \textrm{AS solution}.
\end{equation}

Note that the asymmetric solution exists only if $\delta \ne \epsilon$. When
they are equal it is easily checked from RWA equations that there is no
asymmetric solution.


It is easy to show that at $\gamma=0$ and $\epsilon>0$, \So solutions exist for $\wb<\wS=\sqrt{1-k}$, \Ao solutions exist for $\wb<\wA=\sqrt{1+k}$ and both \Sp and \Ap solutions only exist when $\wb\leq\wps=\sqrt{1-k(1+3\epsilon/\delta)/2}$; for $\epsilon<0$, \So solutions exist for $\wb>\wS=\sqrt{1-k}$, \Ao solutions for $\wb>\wA=\sqrt{1+k}$ and both \Sp and \Ap solutions only exist when $\wb\geq\wph=\sqrt{1+k(1+3\epsilon/\delta)/2}$. In addition, asymmetric solutions only exist for $\wb<\wass=\sqrt{1+2k/(1-\delta/\epsilon)}$ if $\wb<1$ and for $\wb>\wash=\sqrt{1-2k/(1-\delta/\epsilon)}$ if $\wb>1$.

{Using the identifications $\phi_1(t)\equiv y_1$ and $\phi_2(t)\equiv z_1$
introduced after (\ref{eq:DNLS}), the averaged energy within the RWA can be written as}:

\begin{equation}
    <H>=(1+\wb^2)(|y_1|^2+|z_1|^2)-2k\mathrm{Re}(y_1z_1^*)-\frac{3\epsilon}{2}(|y_1|^4+|z_1|^4)-\delta[\mathrm{Re}(y_1^2z_1^{*2})+2|y_1|^2|z_1|^2]
\end{equation}

and, by making use of (\ref{eq:polar}), the average energy for each of the previous solutions at $\gamma=0$ is given by the following expressions:

\begin{eqnarray}\label{eq:RWAHgamma0}
    <H> &=& \frac{\wS^4+2\wS^2\wb^2-3\wb^4}{3(\epsilon+\delta)} , \qquad \textrm{\So solution} \nonumber \\
    <H> &=& \frac{\wA^4+2\wA^2\wb^2-3\wb^4}{3(\epsilon+\delta)} , \qquad \textrm{\Ao solution} \nonumber \\
    <H> &=& \frac{1+2\wb^2-3\wb^4}{3\epsilon+\delta}+\frac{k^2}{2\delta} , \qquad \textrm{\Sp and \Ap solutions} \nonumber \\
    <H> &=& \frac{1+2\wb^2-3\wb^4}{6\epsilon}+\frac{k^2}{3(\delta-\epsilon)} . \qquad \textrm{AS solution}
\end{eqnarray}

\noindent Notice that the average
energy of both \Sp and \Ap are the same for every $\delta$ and that also coincide with that of the AS solution for $\delta=3\epsilon$.

When $\gamma\neq0$ only symmetric and anti-symmetric solutions can exist and Eqs. (\ref{eq:RWAb1})-(\ref{eq:RWAb4}) can be simplified as a quartic equation for $A^2$:

\begin{equation}\label{eq:RWAgamma1}
    \sum_{j=0}^4 P_jA^{2j}=0
\end{equation}
with
\begin{eqnarray}
    P_0 &=& (\Gamma^2 + E^2)(\Gamma^2 + E^2 - p^2) , \nonumber \\
    P_1 &=& 2E[(1 + q)p^2 - 2(1 + 2q)(\Gamma^2 + E^2)] , \nonumber \\
    P_2 &=& 4(1 + 2q)^2E^2 + 2(1 + 3q)(1 + q)(\Gamma^2 + E^2)-(1 + q)^2p^2 , \nonumber \\
    P_3 &=& - 4E(1 + q)(1 + 2q)(1 + 3q) , \nonumber \\
    P_4 &=& (1 + 3q)^2(1 + q)^2 , \nonumber
\end{eqnarray}
whereas the phase fulfills the equation:
\begin{equation}\label{eq:RWAgamma2}
    \tan\ \varphi=-\frac{\Gamma}{E-(1+q)A^2} .
\end{equation}

Just as one could give an expression for $A$ without involving $\phi$,
similarly by eliminating $A$, one finds that $\phi$ must satisfy the
constraint
\begin{equation}\label{RWAgamma3}
Eq\sin(2\phi) \pm p(1+q)\sin(\phi)+\Gamma[1+q+2q\cos^{2}(\phi)] =0\,,
~~~~B = \pm A\,.
\end{equation}

\noindent Notice that there is a phase degeneracy that must be removed by applying, e.g., Eq. (\ref{eq:RWAb4}) together with the previous one.

We now turn to the linear stability of different solutions within the RWA. The spectral analysis of the symmetric and anti-symmetric solutions
can be obtained by considering small perturbations [of order ${\rm O}(\veps)$, with $0< \veps \ll 1$] of the stationary solutions. The stability can be determined by substituting the ansatz below into (\ref{eq:DNLS}) and then solving the ensuing [to O$(\veps)$] eigenvalue problem:

\begin{eqnarray}
\phi_1(t) &=& y_1 + \veps (a_1 e^{-\ii\theta t/\Tb} + b^{*}_1 e^{\ii\theta^* t/\Tb}) , \nonumber \\
\phi_2(t) &=& z_1 + \veps (a_2 e^{-\ii\theta t/\Tb} + b^{*}_2 e^{\ii\theta^* t/\Tb}) ,
\end{eqnarray}
with $\Tb=2\pi/\wb$ being the orbit's period and $\theta$ being the Floquet exponent (FE). The FEs can be expressed as:

\begin{equation}
    \theta=\frac{\pi}{\wb^2}\ii\Omega
\end{equation}
with $\Omega$ being the {eigenfrequencies} of the stability matrix $M$, which is defined as $\Omega(a_1,a_2,b_1^*,b_2^*)^T=M(a_1,a_2,b_1^*,b_2^*)^T$. In the case of symmetric and anti-symmetric solutions, the matrix can be written as:
\begin{equation}\label{eq:stabmatrix}
     M=\left(\begin{array}{cccc}
     M_1 & M_2 & M_3 & M_4 \\
     M_2 & M_1^* & M_4 & M_3^* \\
     -M_3^* & -M_4 & -M_1^* & -M_2 \\
     -M_4 & -M_3 & -M_2 & -M_1 \\
     \end{array}\right)
\end{equation}
with the elements being
\begin{eqnarray}
    M_1 &=& (\wb^2-1)+2(3\epsilon+\delta)A^2+\ii\wb\gamma , \\
    M_2 &=& 4\delta A^2\cos\ \vphi+k , \\
    M_3 &=& [3\epsilon\exp(-\ii\vphi)+\delta\exp(\ii\vphi)] A^2 , \\
    M_4 &=& 2\delta A^2 .
\end{eqnarray}

Thus, the non-zero eigenvalues $\lambda$ can be expressed in terms of $A^2$ and $\varphi$, which must be determined by solving Eqs. (\ref{eq:RWAgamma1})-(\ref{eq:RWAgamma2}):
\begin{equation}\label{eq:lambda}
    \Omega^2/2=[\delta^2(1-16\cos^2\vphi)-6\epsilon\delta(5-2\cos^2\vphi)-27\epsilon^2]A^4-
    [8k\delta\cos\ \vphi-4(\wb^2-1)(3\epsilon+\delta)]A^2-
    [(\wb^2-1)^2+k^2-\gamma^2\wb^2] ,
\end{equation}

\subsection{Numerical analysis of symmetric and anti-symmetric solutions}

\begin{figure}
\begin{tabular}{cc}
$\epsilon=1$, $\delta=\epsilon$ &
$\epsilon=-1$, $\delta=\epsilon$ \\
\includegraphics[width=6cm]{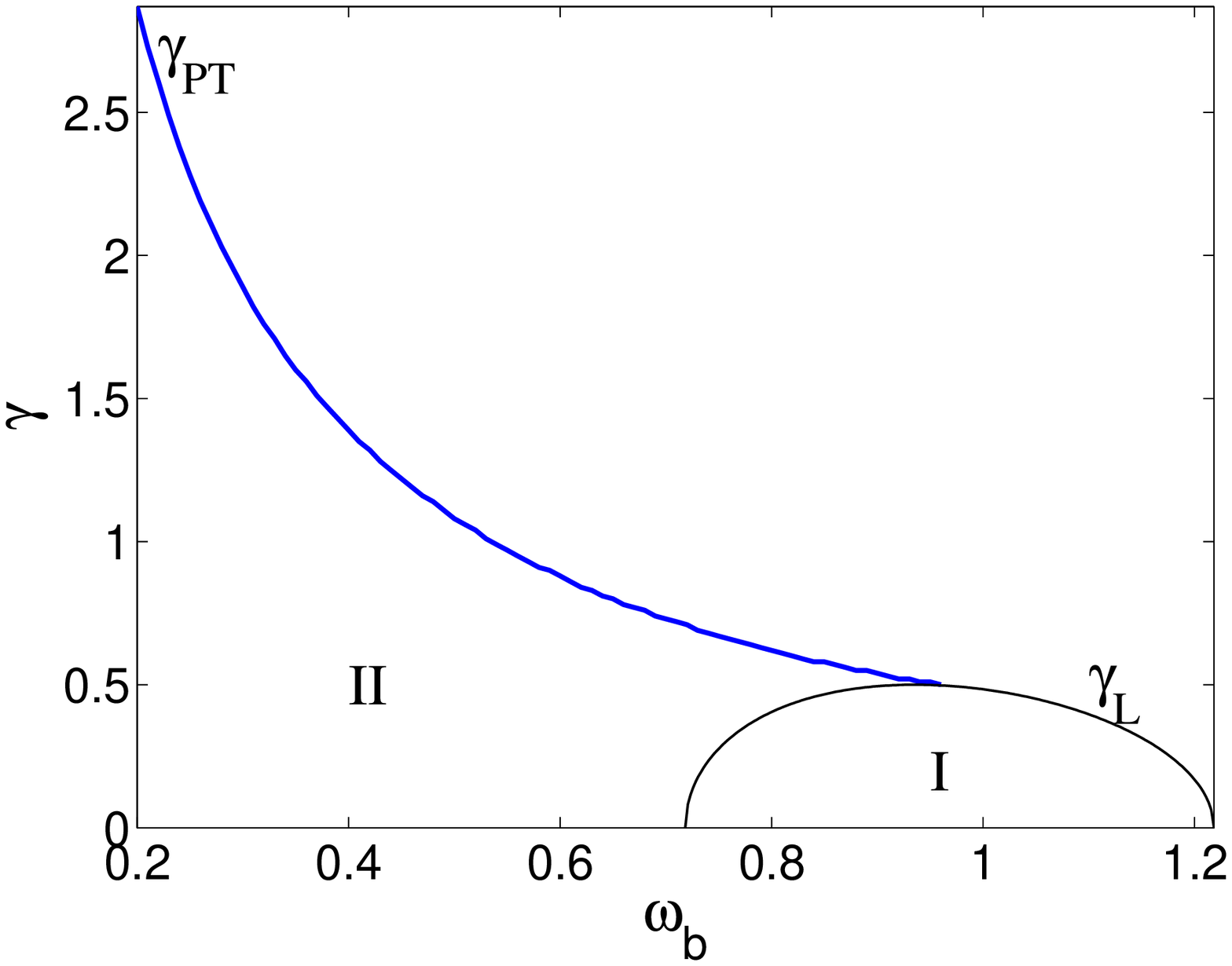} &
\includegraphics[width=6cm]{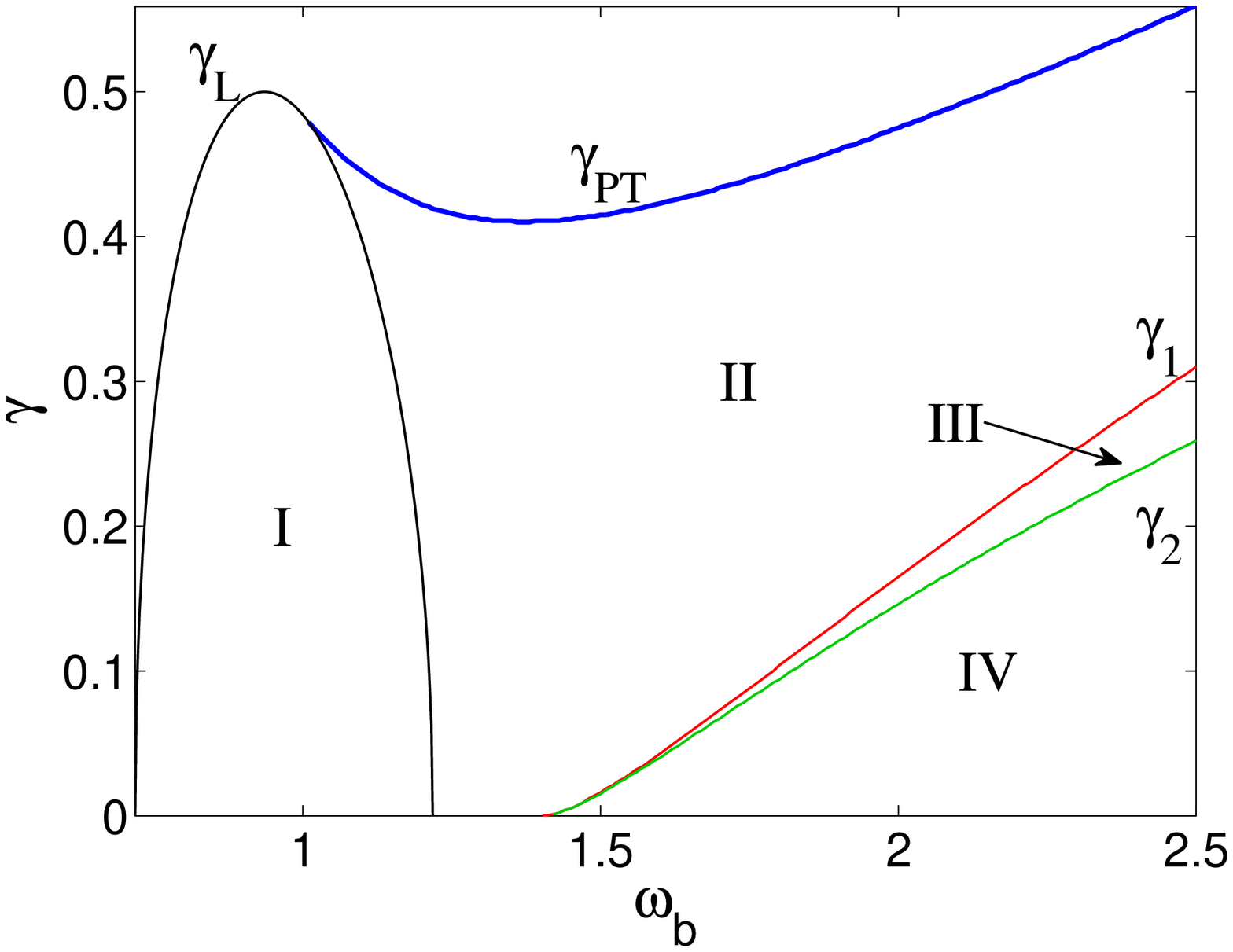} \\
$\epsilon=1$, $\delta=3\epsilon/2$ &
$\epsilon=-1$, $\delta=3\epsilon/2$ \\
\includegraphics[width=6cm]{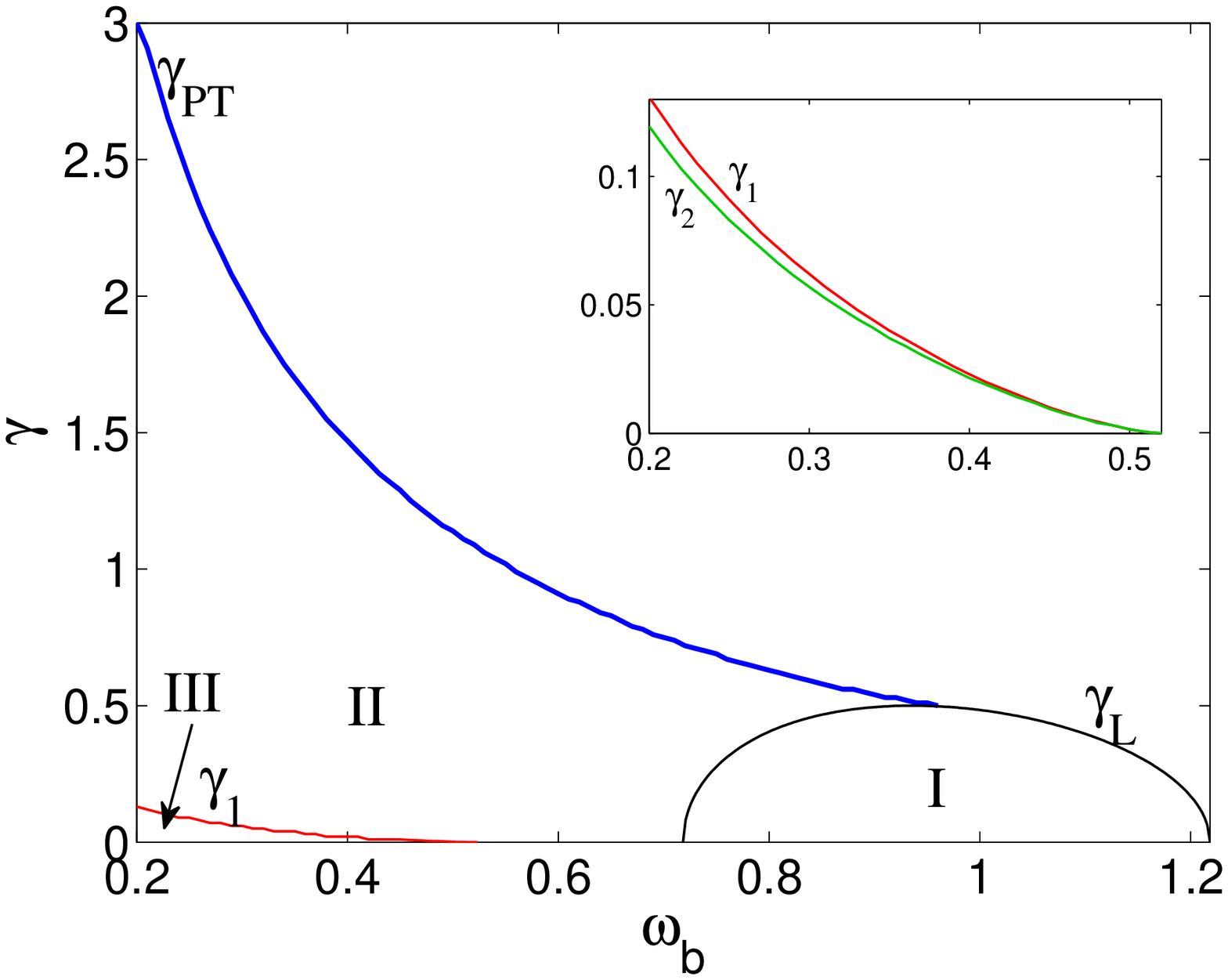} &
\includegraphics[width=6cm]{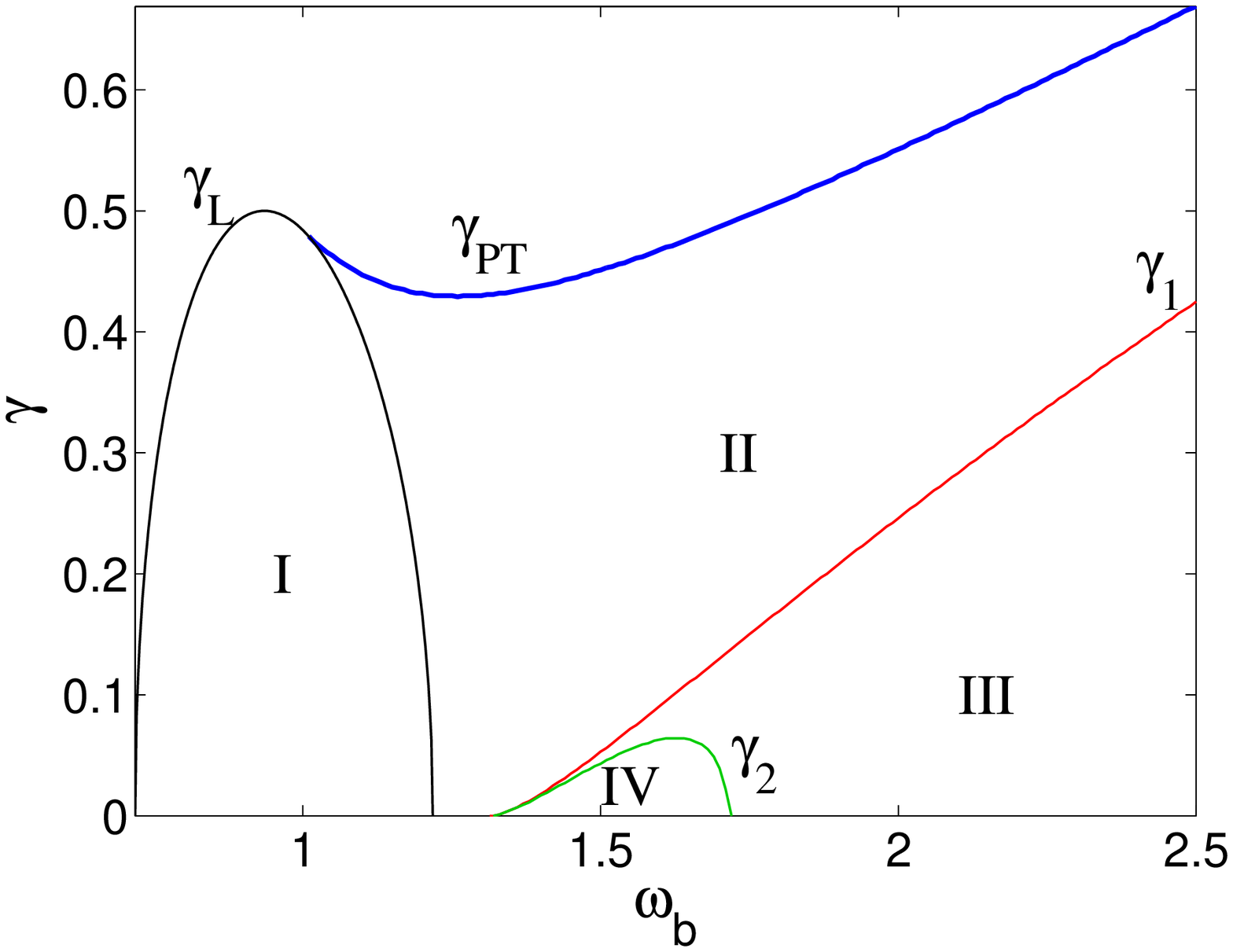} \\
$\epsilon=1$, $\delta=3\epsilon$ &
$\epsilon=-1$, $\delta=3\epsilon$ \\
\includegraphics[width=6cm]{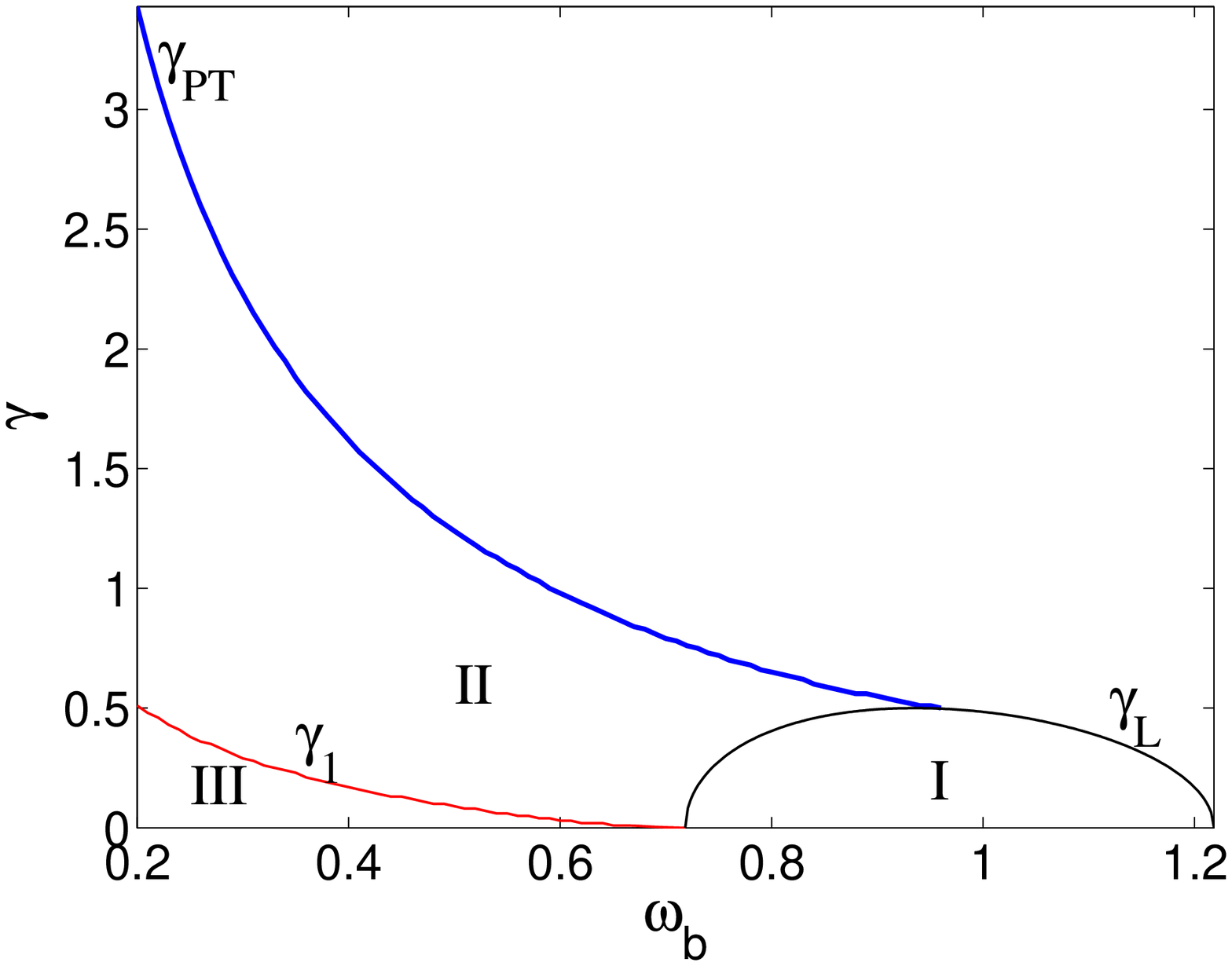} &
\includegraphics[width=6cm]{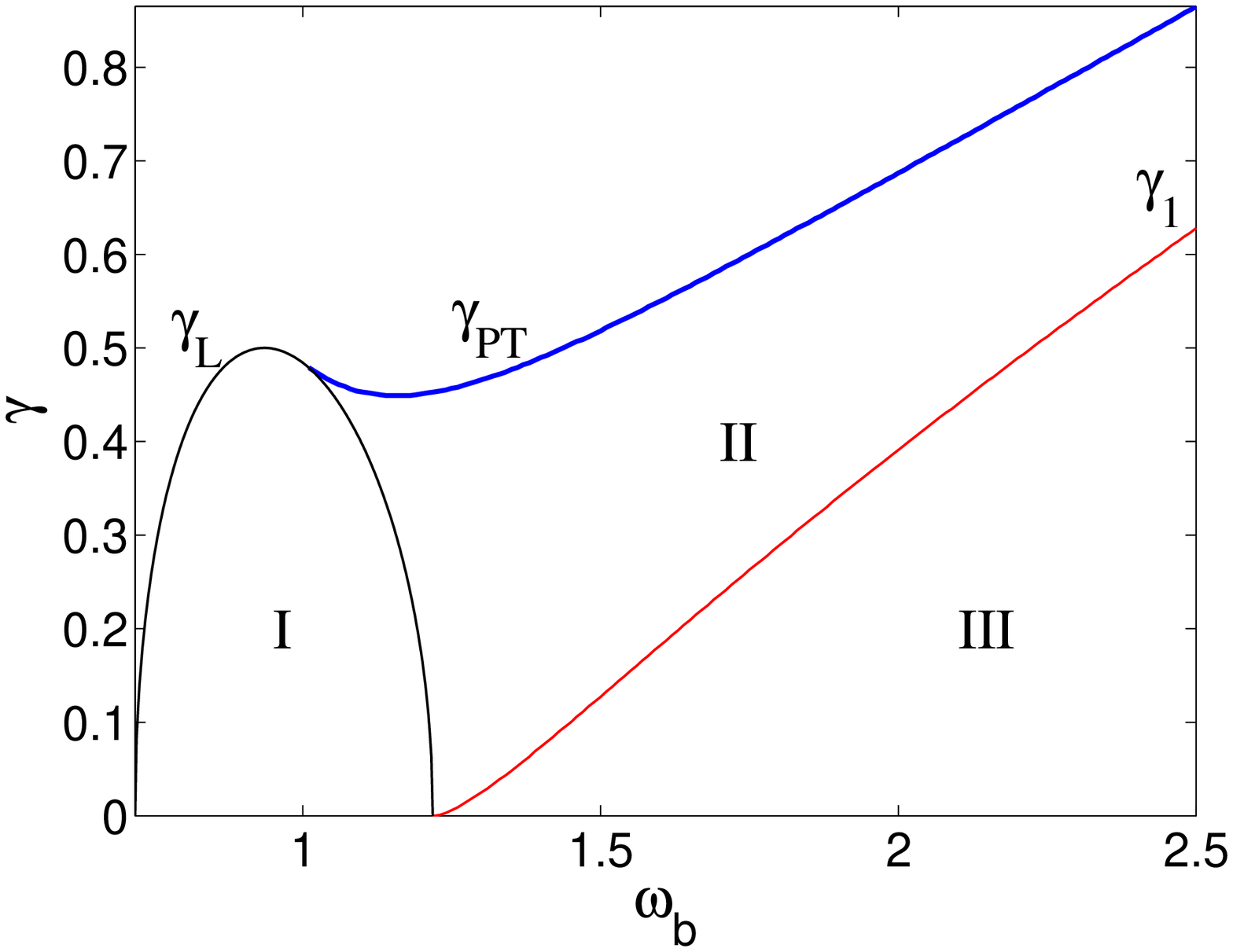} \\
\end{tabular}
\caption{$\gamma$-$\omega_b$ plane for $k=\sqrt{15}/8$. Details on the meaning of each curve and region can be found in the text. The linear limit
of the oscillator system is denoted by $\gamma_L$, while the
upper $\mathcal{PT}$-symmetric threshold of solution existence is
denoted by $\gamma_{\mathcal{PT}}$. An additional delimiter
of the existence of further solutions \Ap and \Sp is also given by
$\gamma_1$. The existence regions of the different solutions are
encompassed by these curves both in the soft $\epsilon=1$
case (left panels) and in the hard $\epsilon=-1$ case (right panels).}
\label{fig:RWA}
\end{figure}

We show below the properties of the \Ao, \So, \Ap and \Sp solutions
in the cases $\delta=\epsilon$, $\delta=3\epsilon/2$ and $\delta=3\epsilon$ for both soft ($\epsilon=+1$) and hard ($\epsilon=-1$) potentials. A summary of the existence and stability regions is displayed in Fig. \ref{fig:RWA}, where
the panels depict the $\gamma$-$\wb$ planes. Notice that although \Ao, \So, \Ap and \Sp solutions are, strictly speaking, defined only at $\gamma=0$, we will use this notation for solutions at $\gamma\neq0$ that are obtained by continuation from the Hamiltonian ($\gamma=0$) limit.

Prior to starting the analysis for arbitrary $\gamma$, we will briefly show the properties of the asymmetric solutions at $\gamma=0$. As explained above, we will choose $k=\sqrt{15}/8$. Notice that for this parameter value, $\wass^2<0$ if $\delta=3\epsilon/2>0$ and consequently, there is no asymmetric solution for this regime. However, there are asymmetric solutions if $\delta=3\epsilon/2<0$ and $\wb>\wash\approx1.7136$. In fact, at $\wb=\wash$ there is a pitchfork bifurcation, as the \So solution is unstable for $\wb<\wash$ and becomes stable past the bifurcation point, where a pair of branches corresponding to unstable asymmetric solutions emerge. For $\delta=3\epsilon$, the situation is similar in the hard case in what regards the existence of solutions (now $\wash=\wA\approx1.2182$); for the soft case, the asymmetric solution exists for $\wb\leq\wass=\wS\approx0.7182$ and bifurcates from the \Ao solution. Notice that the \Ao (for the
soft case) and the \So (for the hard case) are all {\em stable}; furthermore,
the AS solution appears to be marginally stable and highly
degenerate as all the eigenfrequencies $\Omega$ are equal to zero;
recall also the special, completely integrable nature of this special limit.

We analyze now the properties of the {\em soft} potential when $\gamma\neq0$. In the $\delta=\epsilon$ case, there are two main regions: in region I, only \Ao solutions exist, as \So solutions bifurcate from the left arm of the $\gL(\wb)$ curve (i.e. $\omega_+$), which corresponds to the symmetric linear modes; at the right of region I, no solutions are found because \Ao solutions bifurcate
from  the right arm (i.e. $\omega_-$) of the linear dispersion relation.
In this soft case of $\epsilon>0$, the bifurcations occur to the left
of $\gL(\wb)$, whereas in the hard case of $\epsilon<0$,
they arise to the right
of $\gL(\wb)$.
It is easy to show that from (\ref{eq:linear}), $\gL$ is given by:

\begin{equation}
    \gL(\wb)=\frac{\sqrt{(k^2-1)+2\wb^2-\wb^4}}{\wb} .
\end{equation}

Consequently, region I is bounded between $\wb=\wS\approx0.7182$, $\wb=\wA\approx1.2182$ and $\gamma=\gamma_{PT,L}=0.5$. In region II, both \Ao and \So solutions exist, and experience the $\cP\cT$ phase transition at the
curve designated as $\gPT(\wb)$. Notice also that all the solutions existing in both regions I and II  are stable. In addition, \Ap and \Sp solutions can only be found for $\wb<\wps$, but their existence range is quite small as $\wps\approx0.1782$ and only exist for $\gamma<0.1$.

For $\delta=3\epsilon/2$, both regions I and II have the same properties as before. In addition, region III is included, which is below the curve $\gamma_1(\wb)$. In that region, all four solutions exist and are stable except for \Ao. This solution becomes stable only nearby i.e. {between the curves $\gamma_2(\wb)$ and $\gamma_1(\wb)$}. This small stability region can be observed between red and green curves of the inset of the corresponding panel (notice that this phenomenon was also observed in the $\delta=\epsilon$ case, but was not showcased due to the very small range of existence of \Sp and \Ap solutions therein). In region II {only two solutions exist}; for $\wb<\wps\approx0.5233$, {i.e. above
the curve $\gamma_1(\wb)$}, \So and \Sp solutions coexist and collide/disappear at $\gPT(\wb)$, whereas for $\wb>\wps$, the coexisting solutions are \Ao and \So. {As a side comment, the reason for the existence of \Sp for $\wb<\wps$ and of \Ao for $\wb>\wps$ within Region II has to do with the fact that these solutions effectively ``morph'' from one to the other (smoothly) as this frequency is crossed.}

For $\delta=3\epsilon$, the scenario is similar to the last one, except for two points: first, the $\gamma_1$ curve finishes at $\wb=\wS$ and encompasses
an accordingly broader region III; and second, the solutions in region III, \Ao and \Ap,  are stable for any value of $\gamma$ and $\wb$.

We focus now on the {\em hard} potential (i.e. $\epsilon=-1$) properties. In all the cases, we can find the region I, with the same properties as in the soft case (although now it is the \So solution that exists and the
\Ao that bifurcates into existence beyond the boundary of the region).
In addition, region II is present in every case, enclosed between curves $\gPT(\wb)$ and $\gamma_1(\wb)$; there are two solutions therein: the \Ao solution and the \So for $\wb<\wph$ and the \Ao and \Ap for $\wb>\wph$. Under the curve $\gamma_1(\wb)$, whose minimum value takes place at $\wb=\wph$ (so that $\wph=\wA$ for $\delta=3\epsilon$), the four kinds of solutions coexist, so that \So and \Sp collide and disappear at this line. {As in the soft case, there is a ``morphing'' from the \So to \Ap solutions when the frequency $\wph$ is crossed.}

Thus, the most significant difference between the three considered regimes
lies in the existence of region IV and curve $\gamma_2(\wb)$. Region III is characterized by the fact that all the solutions exist (as mentioned above) and are stable. However, below the curve $\gamma_2(\wb)$ (i.e. in region IV), solution \So becomes unstable. Notice that for $\delta=\epsilon$ this region exists for every $\wb>\wph\approx1.4029$. However, if $\delta=3\epsilon/2$, region IV is shrunk to the range $\wph\approx1.3138<\wb\lesssim1.71$. Finally, region IV has totally vanished at $\delta=3\epsilon$.

\section{Analysis of the oscillator dimer}\label{sec:KG}

In this section, we complete the description of the system by
returning to the original oscillator system and analyzing its
exact periodic orbits (that up to now we had only approximated
using the RWA).
This is done by numerically solving in the Fourier space the dynamical equations set (\ref{eq:dyn}) [cf. Appendix \ref{app:numerics}]. That is, we express the solution in the form:

\begin{equation}
    u(t)=\sum_n y_n\exp(\ii n \wb t),\qquad v(t)=\sum_n z_n\exp(\ii n \wb t) .
\end{equation}

We have considered the same cases as in Section \ref{sec:RWA}, namely, $\delta/\epsilon$ equal to 1, $3/2$ and 3, with $\epsilon=\pm1$.

Prior to showing the results, we want to remark that the Fourier coefficients of \So and \Ao solutions (due to their symmetry)
have the following property:

\begin{equation}\label{eq:Fourier_criterion}
    y_n=z_n^*\ \textrm{(\So)},  \qquad y_n=-z_n^*\ \textrm{(\Ao)}.
\end{equation}

In what follows, we will first show the properties of the solutions at the Hamiltonian limit $\gamma=0$. Afterwards, we will be focusing in the different cases of $\delta/\epsilon>0$ for $\gamma\neq0$. In most cases, results will be compared with the previously found results for the RWA.

\subsection{Solutions for $\gamma=0$}

We start by analyzing the modes that can be expressed analytically at the $\gamma=0$ limit. In fact, these can be expressed in terms of Jacobi elliptic functions. If $\epsilon>0$ (soft potential), the solutions are of the form:

\begin{equation}
    u(t) = A\ \sn[\beta\ t;m]\,,\quad v(t) = \pm A\ \sn[\beta\ t;m]\,,
\end{equation}
with
\begin{equation}
    A=\beta\sqrt{\frac{2m}{\epsilon+\delta}},\, ~~~\beta^2=\frac{1\mp k}{1+m},\ ~~~\wb=\frac{\pi\beta}{2K(m)}.
\end{equation}
with the upper (lower) sign corresponding to the \So (\Ao) solution, $K(m)$ the complete elliptic integral of the first kind with modulus $m$ \footnote{Here, we use the definition $K(m)=\int_0^1\!\frac{\mathrm{d}x}{\sqrt{(1-x^2)(1-mx^2)}}$.}, and $0<m<1$. As $K(m)>\pi/2$, it is easy to deduce that $\wb<\omega_S=\sqrt{1-k}$ for the \So solution and $\wb<\omega_A=\sqrt{1+k}$ for the \Ao solution, as within the RWA.

If $\epsilon<0$ (hard potential), these modes can be expressed as:

\begin{equation}
    u(t) = A\ \cn[\beta\ t;m]\,,\quad v(t) = \pm A\ \cn[\beta\ t;m]\,,
\end{equation}
with
\begin{equation}
    A=\beta\sqrt{-\frac{2m}{\epsilon+\delta}},\, ~~~\beta^2=\frac{1\mp k}{1-2m},\ ~~~\wb=\frac{\pi\beta}{2K(m)} ,
\end{equation}
where $0<m<1/2$. Similar to the soft case, $\wb>\omega_S$ for the \So solution and $\wb>\wA$ for the \Ao solution. Notice that for these solutions to
exist $\delta < -\epsilon$.

{At $\delta = \epsilon$ and for a hard potential, the \Sp and \Ap solutions are given by}:

\begin{equation}
    u(t)= A \sn(\beta t;m) +B \sqrt{m} \cn(\beta t;m)\,,\quad~~v(t) = \pm[A \sqrt{m} \sn(\beta t;m) -B \sqrt{m} \cn(\beta t;m)]\,,
\end{equation}
provided
\begin{equation}
    A = \sqrt{\frac{(3m-2)\beta^2+2}{4\epsilon}},\,B=\sqrt{\frac{1-(2m+1)\beta^2}{2\epsilon}},\, \beta^2 = 2k/m,\ ~~~\wb=\frac{\pi\beta}{2K(m)} .
\end{equation}

The AS solution can be analytically expressed whenever $\delta=3\epsilon$. If $\epsilon>0$, it is given by:

\begin{equation}
    u(t) = A_+\ \sn[\beta_+\ t;m_+]+A_-\ \sn[\beta_-\ t;m_-]\,,\quad v(t) = A_+\ \sn[\beta_+\ t;m_+]-A_-\ \sn[\beta_-\ t;m_-]\,,
\end{equation}
with
\begin{equation}
    A_\pm=\beta_\pm\sqrt{\frac{2m_\pm}{\epsilon}},\, ~~~\beta_\pm^2=\frac{1\mp k}{1+m_\pm},\ ~~~\wb=\frac{\pi\beta_+}{2K(m_+)}=\frac{\pi\beta_-}{2K(m_-)},\,
\end{equation}
whereas if $\epsilon<0$, the AS solution is:

\begin{equation}
    u(t) = A_+\ \cn[\beta_+\ t;m_+]+A_-\ \cn[\beta_-\ t;m_-]\,,\quad v(t) = A_+\ \cn[\beta_+\ t;m_+]-A_-\ \cn[\beta_-\ t;m_-]\,,
\end{equation}
with
\begin{equation}
    A_\pm=\beta_\pm\sqrt{-\frac{2m_\pm}{\epsilon}},\, ~~~\beta_\pm^2=\frac{1\mp k}{1-2m_\pm},\ ~~~\wb=\frac{\pi\beta_+}{2K(m_+)}=\frac{\pi\beta_-}{2K(m_-)},\, .
\end{equation}

It can be {numerically} observed that \Ao, \So and AS solutions exist for every $\delta$ whereas \Ap and \Sp do not exist for $\delta\leq3\epsilon/2$ in the $\epsilon=+1$ case. On the contrary, a new solution denoted as \At exists for the soft potential and all of the considered values of $\delta$; this new solution, which was not found in the $\delta=0$ case, is characterized by a high increase of the third harmonic in the Fourier series, and, consequently, cannot be predicted by the RWA. The existence of this new solution can be caused by the hybridization of the \So mode with frequency $\wb$ and the \Ao mode with frequency $3\wb$; this symmetry breaking effect
could happen whenever $\wb<\wA/3\approx0.4061$. Notice that the \At mode bifurcates from the \So mode at $\omega_3$, which exactly coincides with $\wA/3$ when $\delta=3\epsilon$, but is smaller than this when $\delta<3\epsilon$ (e.g. for $\delta=3\epsilon/2$, $\omega_3\approx0.384$ whereas $\omega_3\approx0.365$ for $\delta=\epsilon$.). There is no stability change at this bifurcation.

The asymmetric (AS) solution preserves the properties of the RWA. That is, it bifurcates from the \Ao solution in soft potentials and from the \So solution for hard potentials.
The AS solution does not exist for $\epsilon=\delta$ and for $\delta=3\epsilon/2>0$. Besides, all the Floquet exponents are $\theta=0$ (or, equivalently, the Floquet multipliers are $+1$) for $\delta=3\epsilon$. For $\delta=3\epsilon/2<0$, the AS solution is unstable, as in the RWA. In addition, for $\delta=3\epsilon$, the AS, \Sp and \Ap solutions bifurcate from the \Ao solution at $\wb=\wS$ if $\epsilon=+1$, with the \Sp and \Ap solutions being stable and the \Ao stable (unstable) for $\wb>\wS$ ($\wb<\wS$); if $\epsilon=-1$, the AS, \Sp and \Ap solutions bifurcate from the \So mode at $\wb=\wA$, with the \So solution being
stable and the \Sp unstable, whereas  the \Ap is marginally stable as are
the AS solutions (all the Floquet exponents are zero). In the $\delta=3\epsilon/2<0$ case, the \Sp and \Ap solutions, which are stable, bifurcate from the \So solution at $\wb\approx1.306$ which is close to $\wph$; the \So solution is stable (unstable) for $\wb$ smaller (higher) than the bifurcation point. This latter bifurcation also occurs for $\delta=\epsilon=-1$, taking place in this case at $\wb\approx1.386$. In addition, the AS solution (which is unstable) bifurcates from the \So solution (which changes its stability) at $\wb\approx1.708$, a value which is close to (but not exactly at) $\wash$. {We must also mention that for the case analyzed in \cite{PRA}, i.e. $\delta=0$, the AS solution bifurcates from the \So (\Ao) solution in the soft (hard) potential. This situation is reversed in the present observations for sufficiently large $\delta\neq0$, which suggests the existence of a critical point.}

Finally, as within the RWA, the energy coincides for the AS, \Sp and \Ap solutions when $\delta=3\epsilon$.

\begin{figure}
\begin{tabular}{cc}
$\epsilon=1$, $\delta=\epsilon$, $\gamma=0$ &
$\epsilon=-1$, $\delta=\epsilon$, $\gamma=0$ \\
\includegraphics[width=6cm]{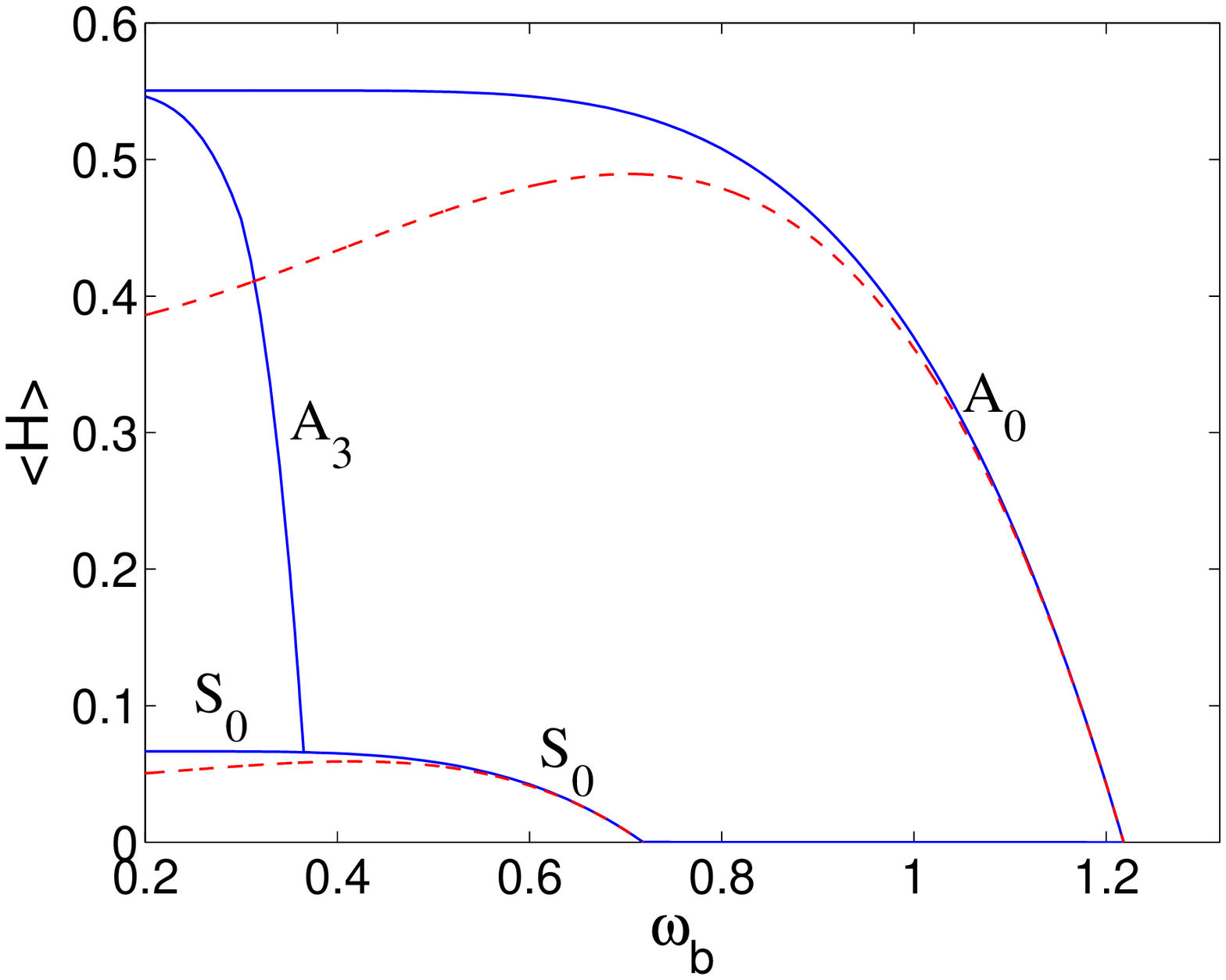} &
\includegraphics[width=6cm]{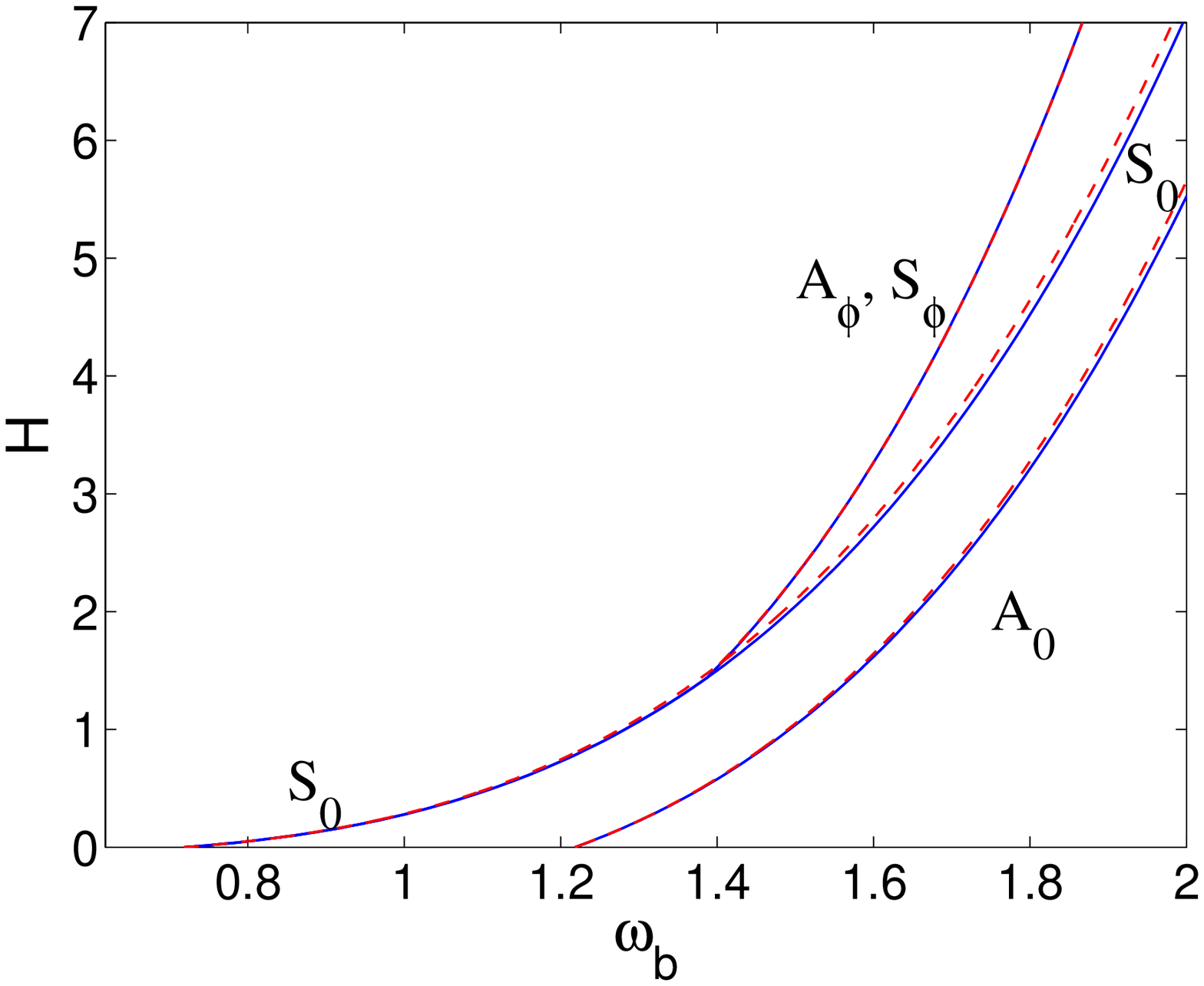} \\
$\epsilon=1$, $\delta=3\epsilon/2$, $\gamma=0$ &
$\epsilon=-1$, $\delta=3\epsilon/2$, $\gamma=0$ \\
\includegraphics[width=6cm]{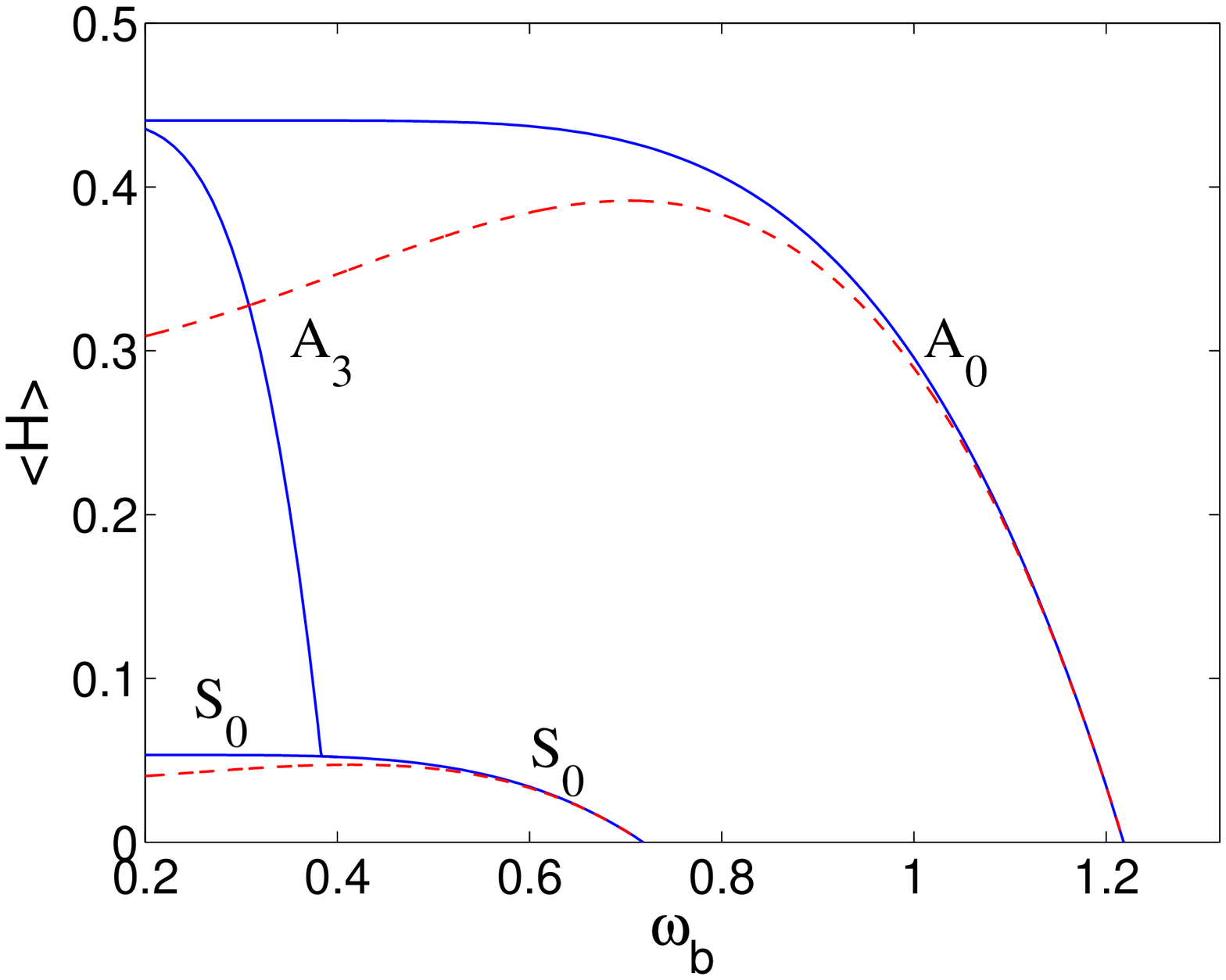} &
\includegraphics[width=6cm]{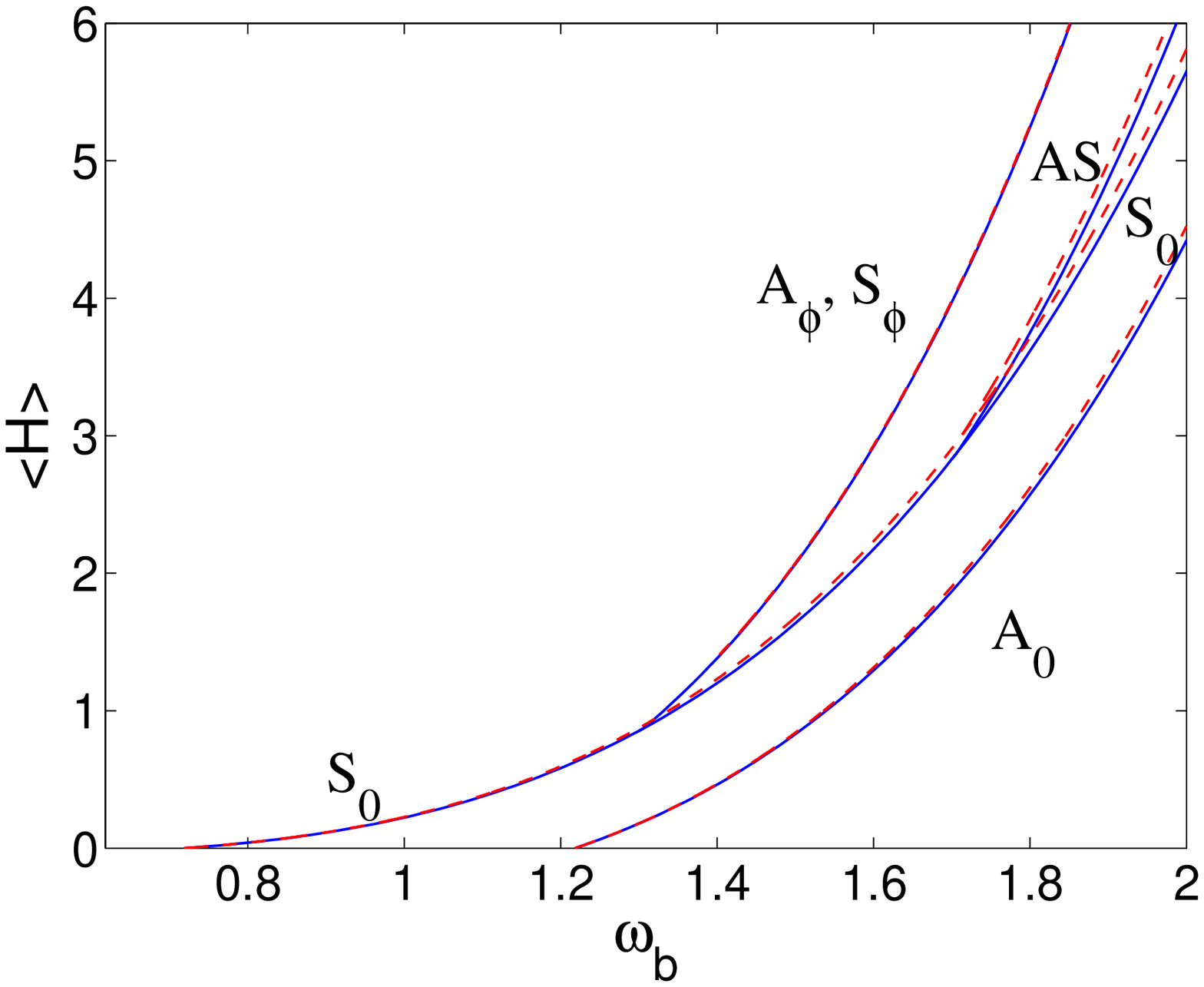} \\
$\epsilon=1$, $\delta=3\epsilon$, $\gamma=0$ &
$\epsilon=-1$, $\delta=3\epsilon$, $\gamma=0$ \\
\includegraphics[width=6cm]{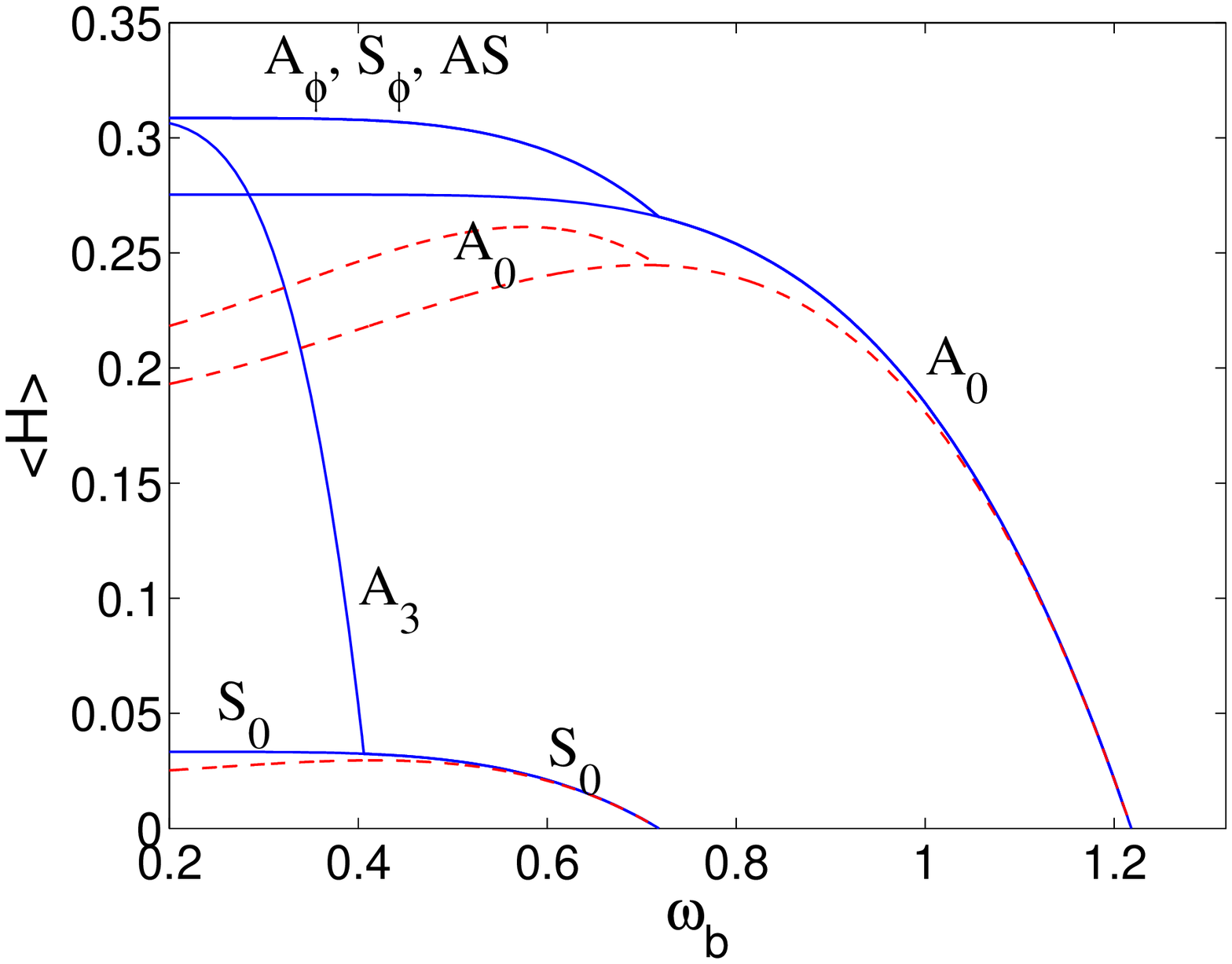} &
\includegraphics[width=6cm]{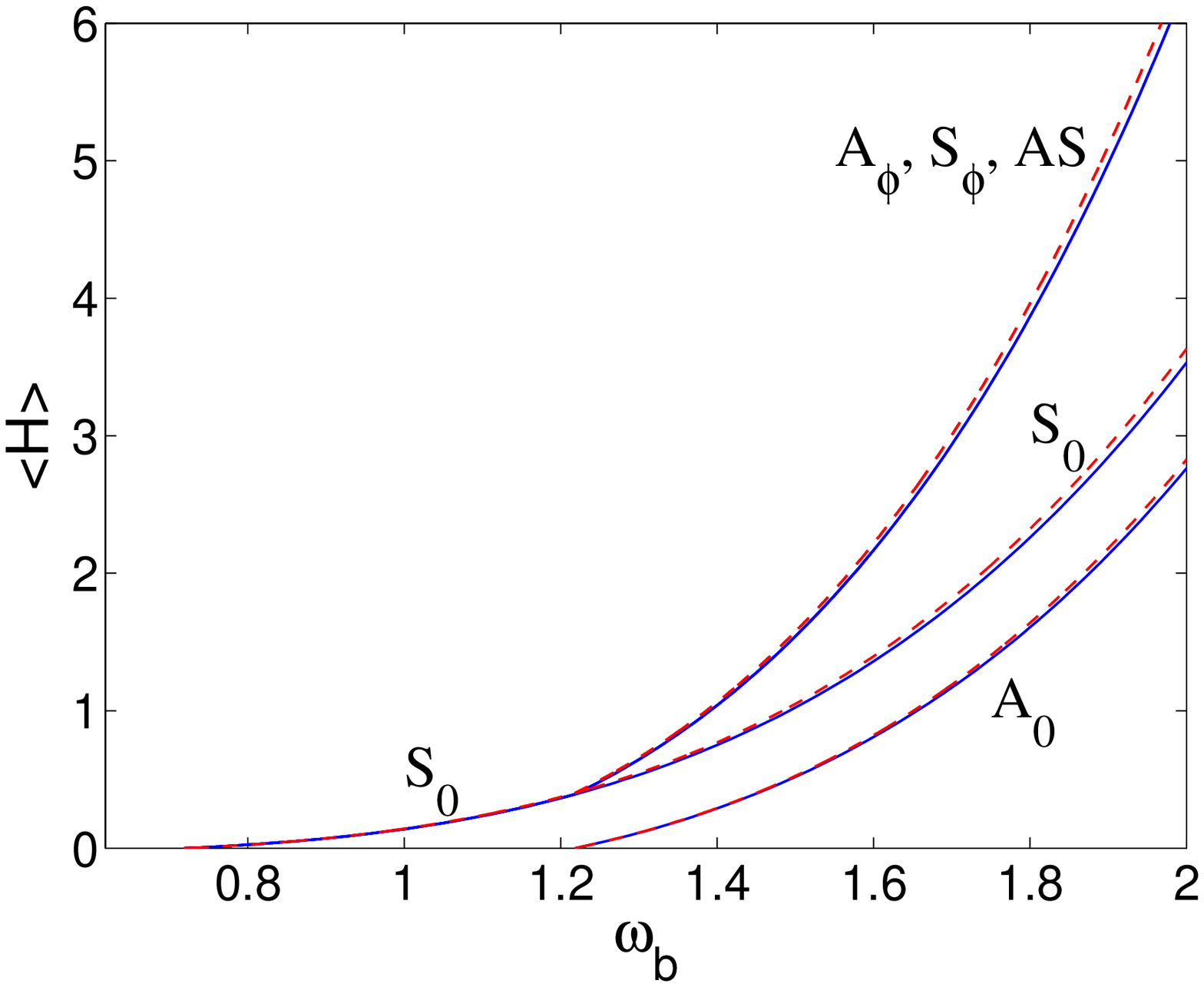} \\
\end{tabular}
\caption{Energy versus $\wb$ at $\gamma=0$. Full blue lines correspond to the Hamiltonian (\ref{eq:Ham}) of full Klein-Gordon dimer whereas the dashed red lines represent the averaged energy within the RWA (\ref{eq:RWAHgamma0}).}
\label{fig:gamma0}
\end{figure}

\begin{figure}
\begin{tabular}{cc}
\So solution &
\Ao solution \\
\includegraphics[width=6cm]{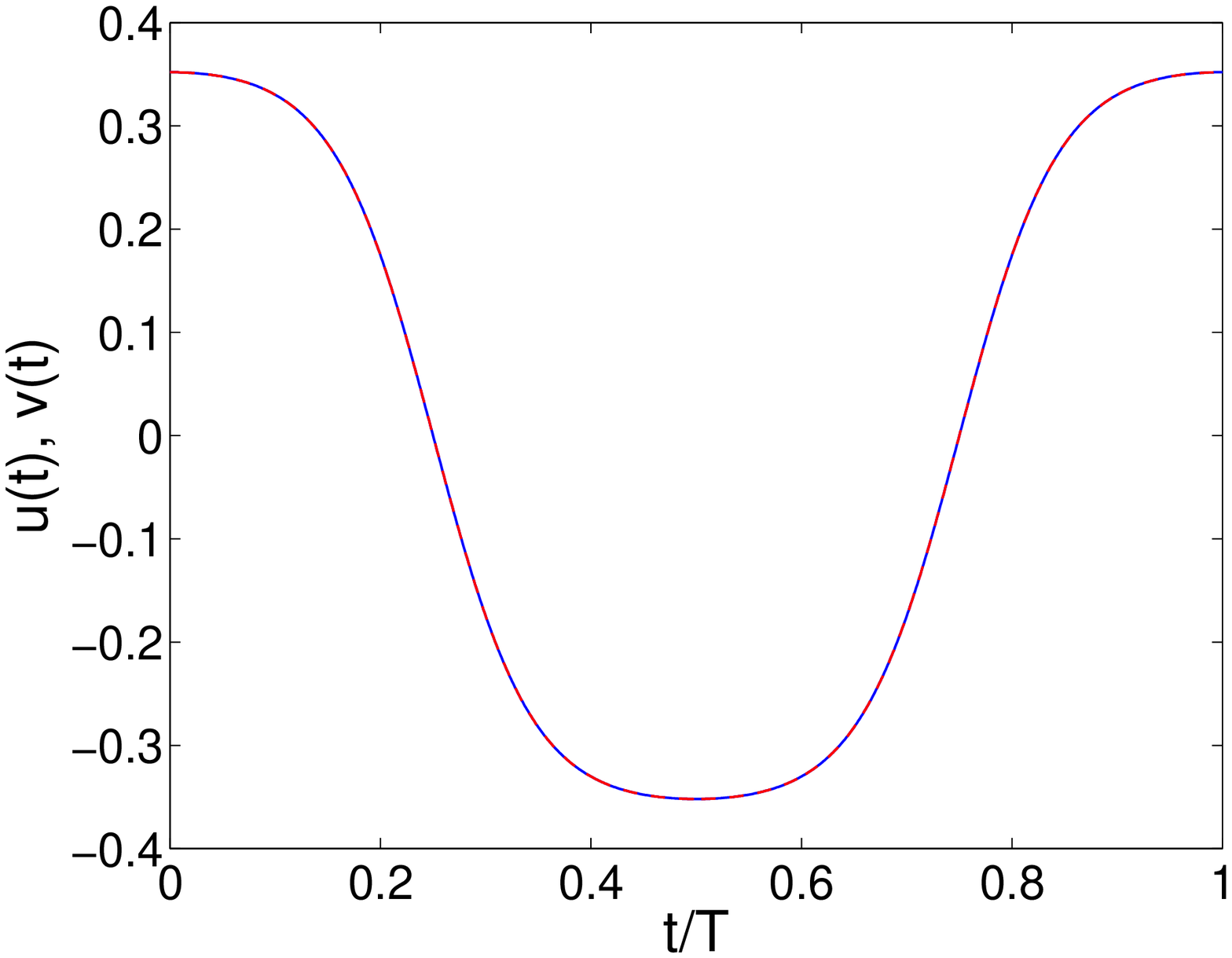} &
\includegraphics[width=6cm]{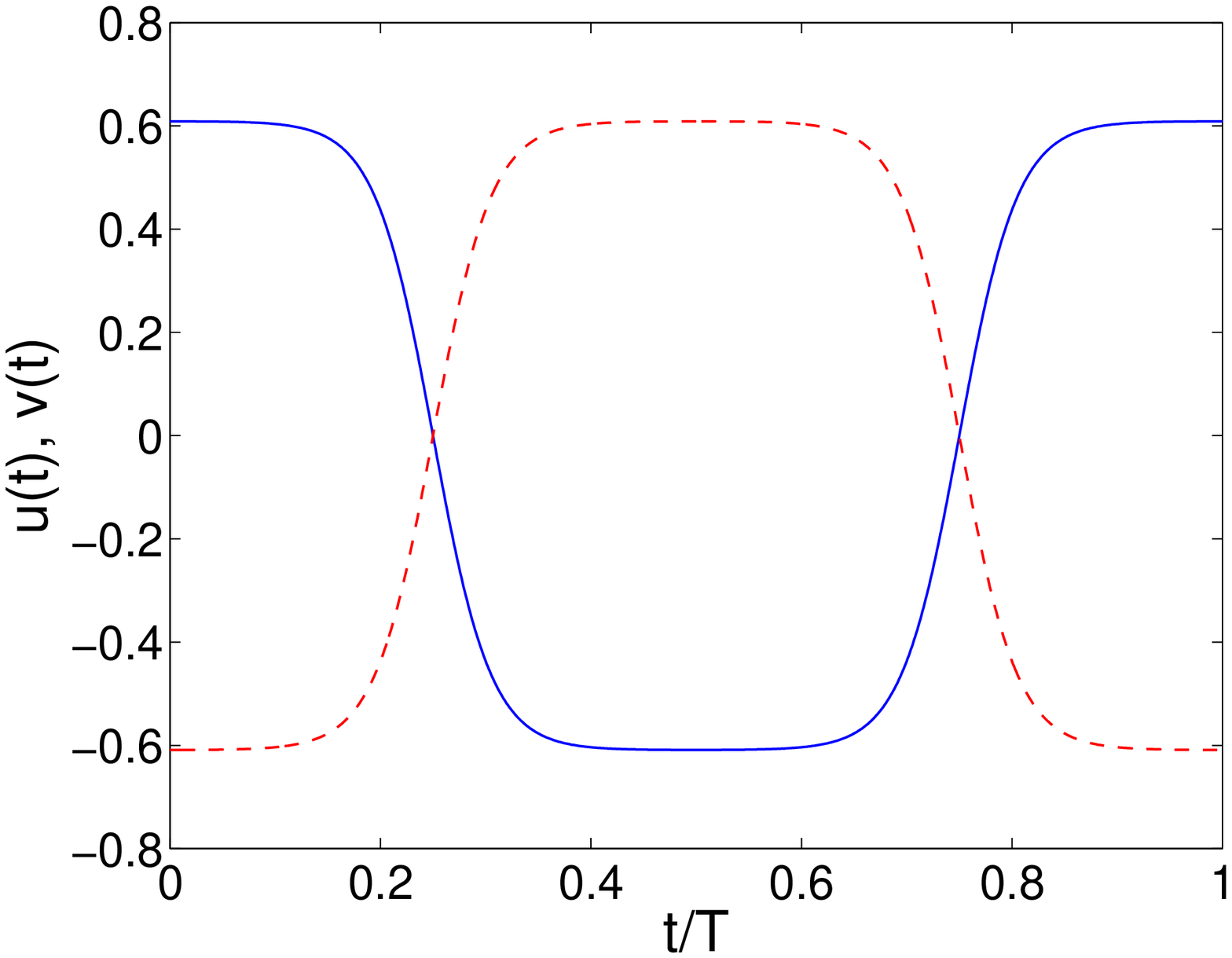} \\
\Sp solution &
\Ap solution \\
\includegraphics[width=6cm]{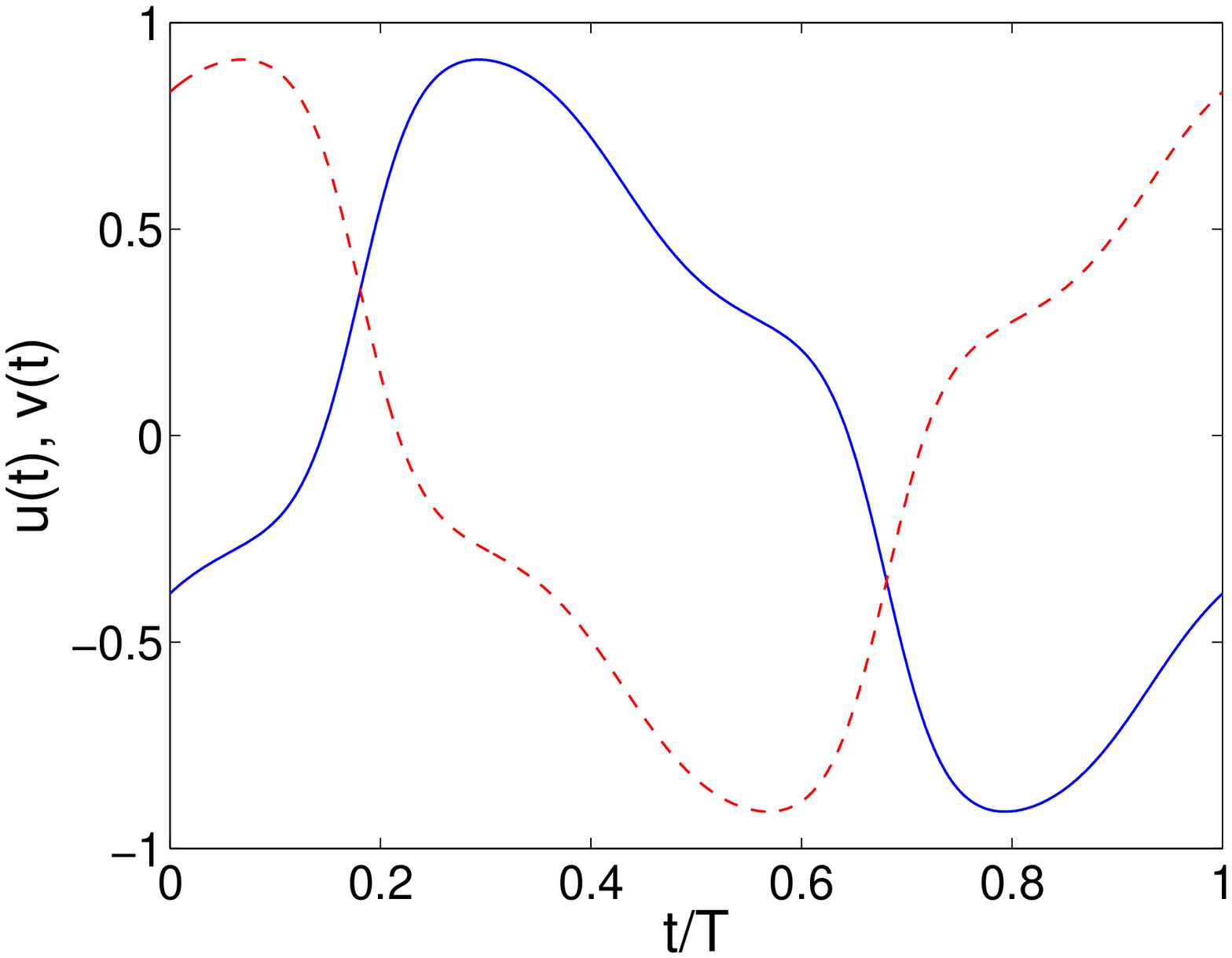} &
\includegraphics[width=6cm]{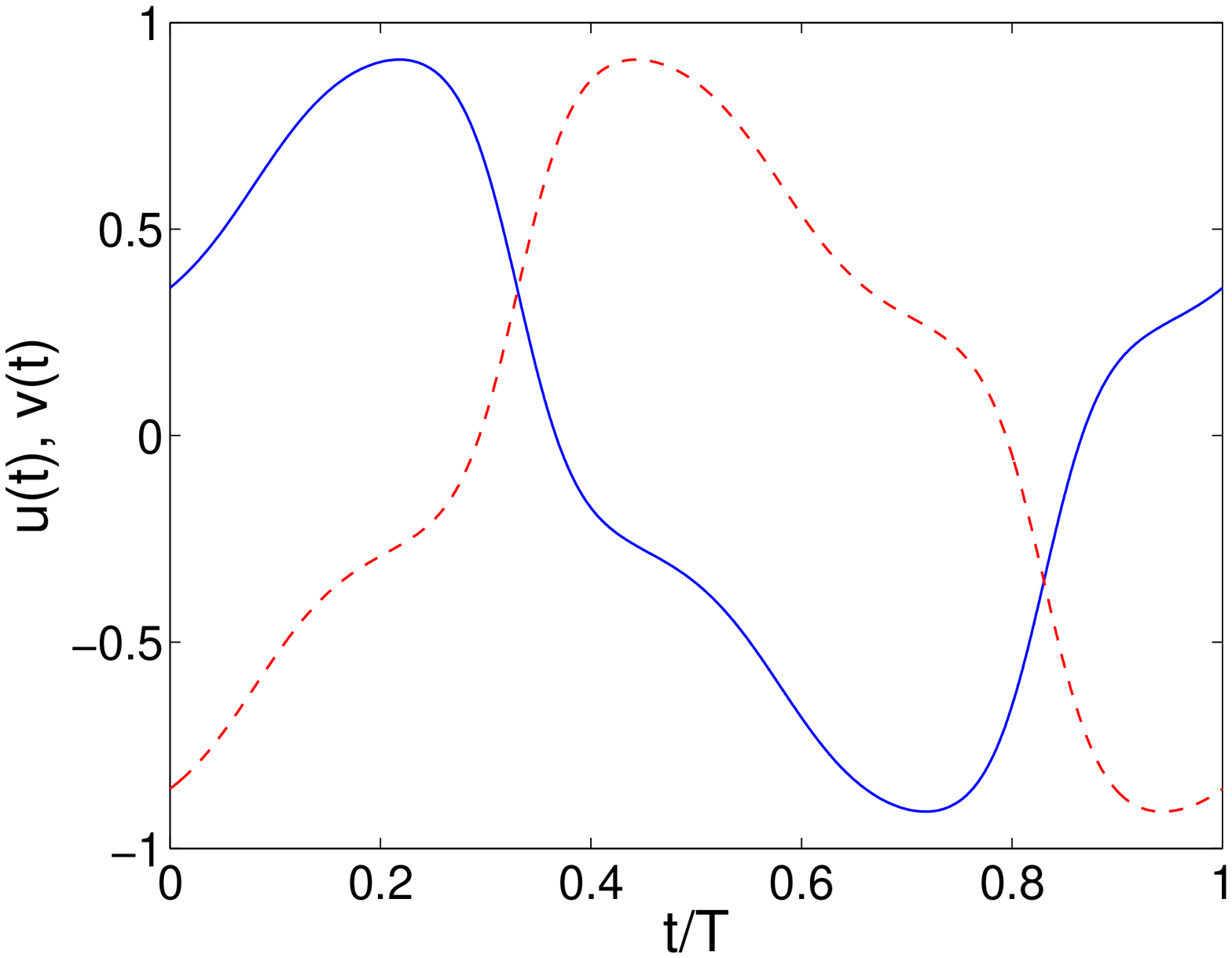} \\
AS solution &
\At solution \\
\includegraphics[width=6cm]{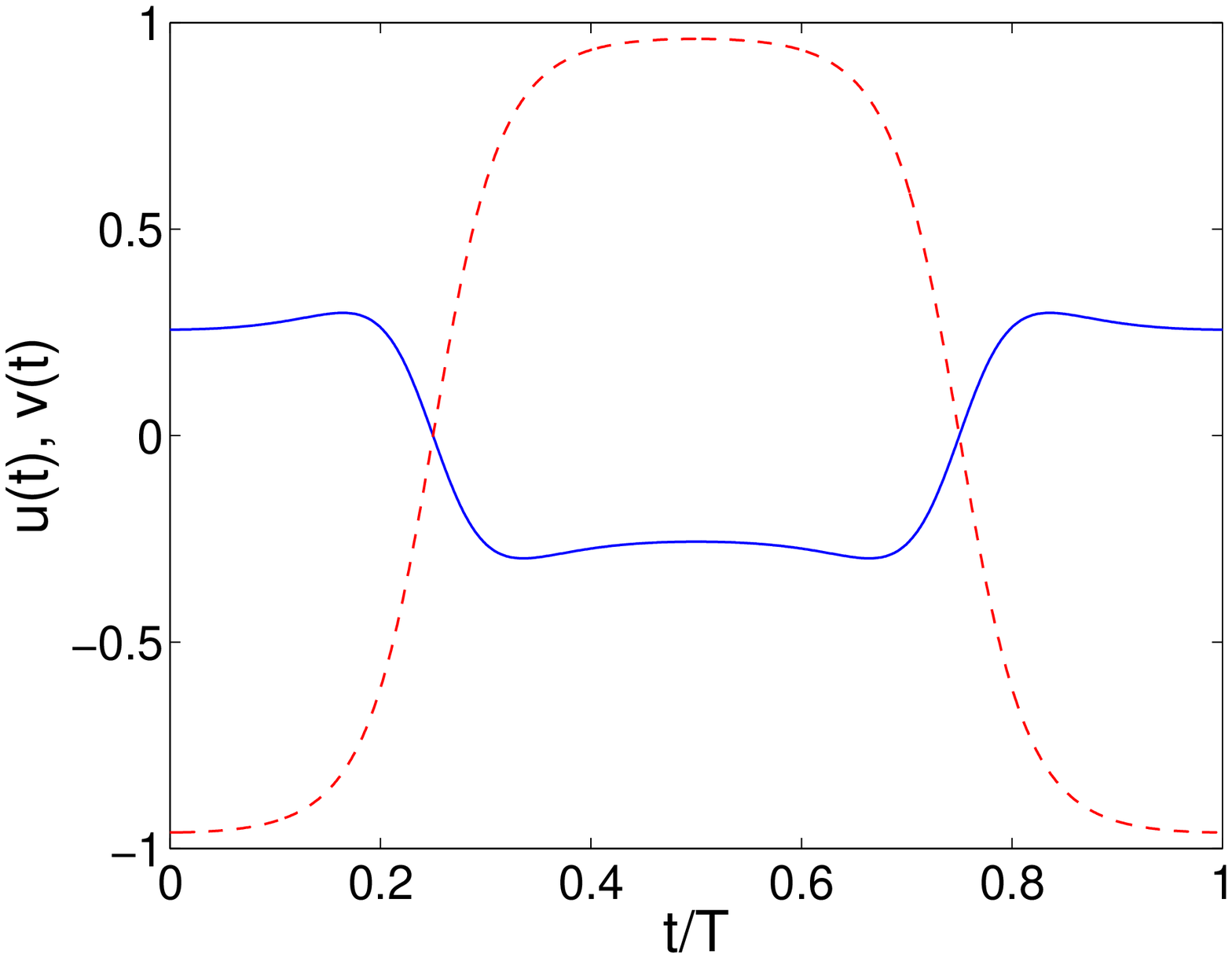} &
\includegraphics[width=6cm]{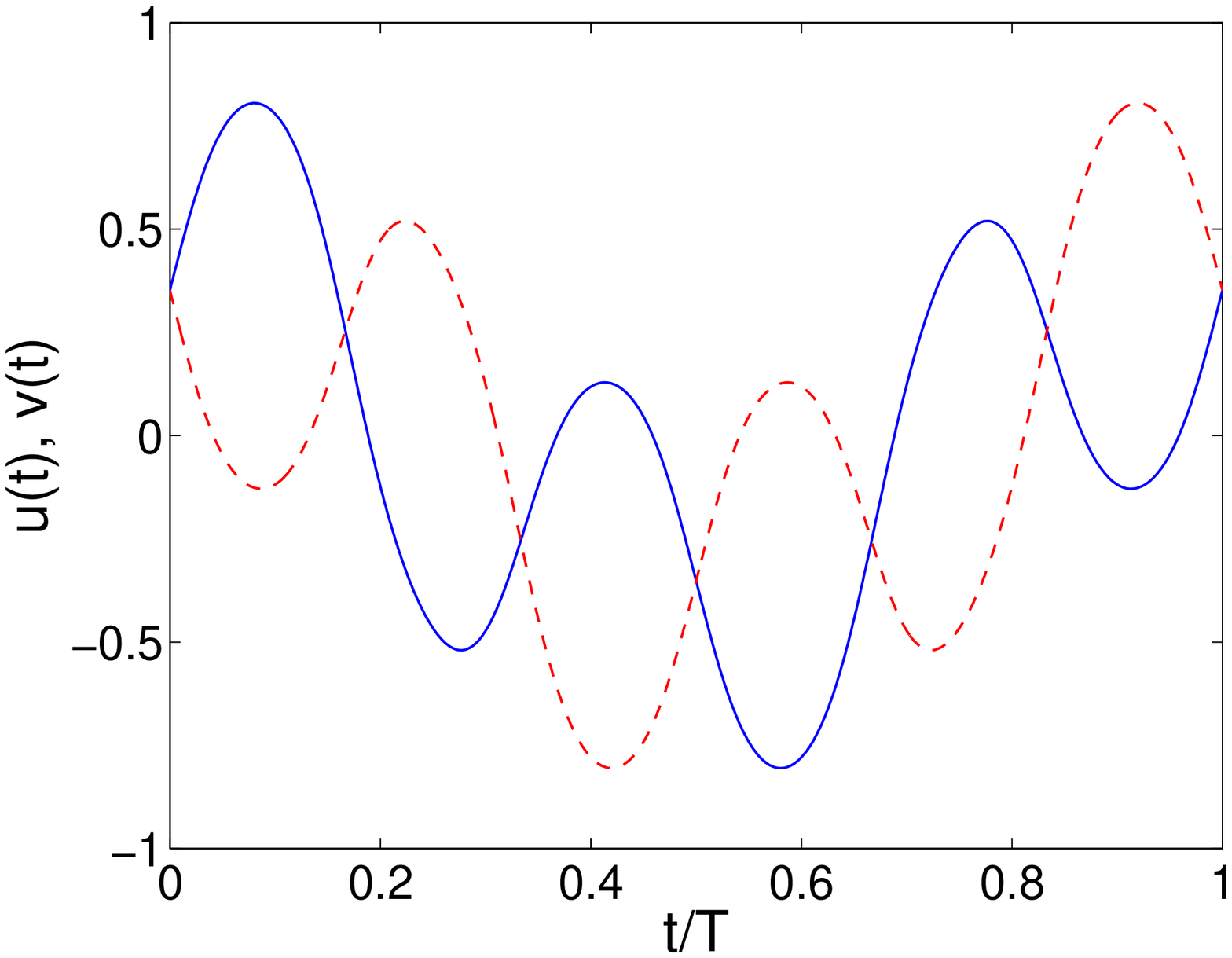} \\
\end{tabular}
\caption{Time evolution of all the different solutions
considered at $\delta=3\epsilon$ for the soft nonlinearity case
of $\epsilon=1$ and $\wb=0.3$; here
$\gamma=0$.}
\label{fig:gamma0_soft_evol}
\end{figure}

\begin{figure}
\begin{tabular}{cc}
\So solution &
\Ao solution \\
\includegraphics[width=6cm]{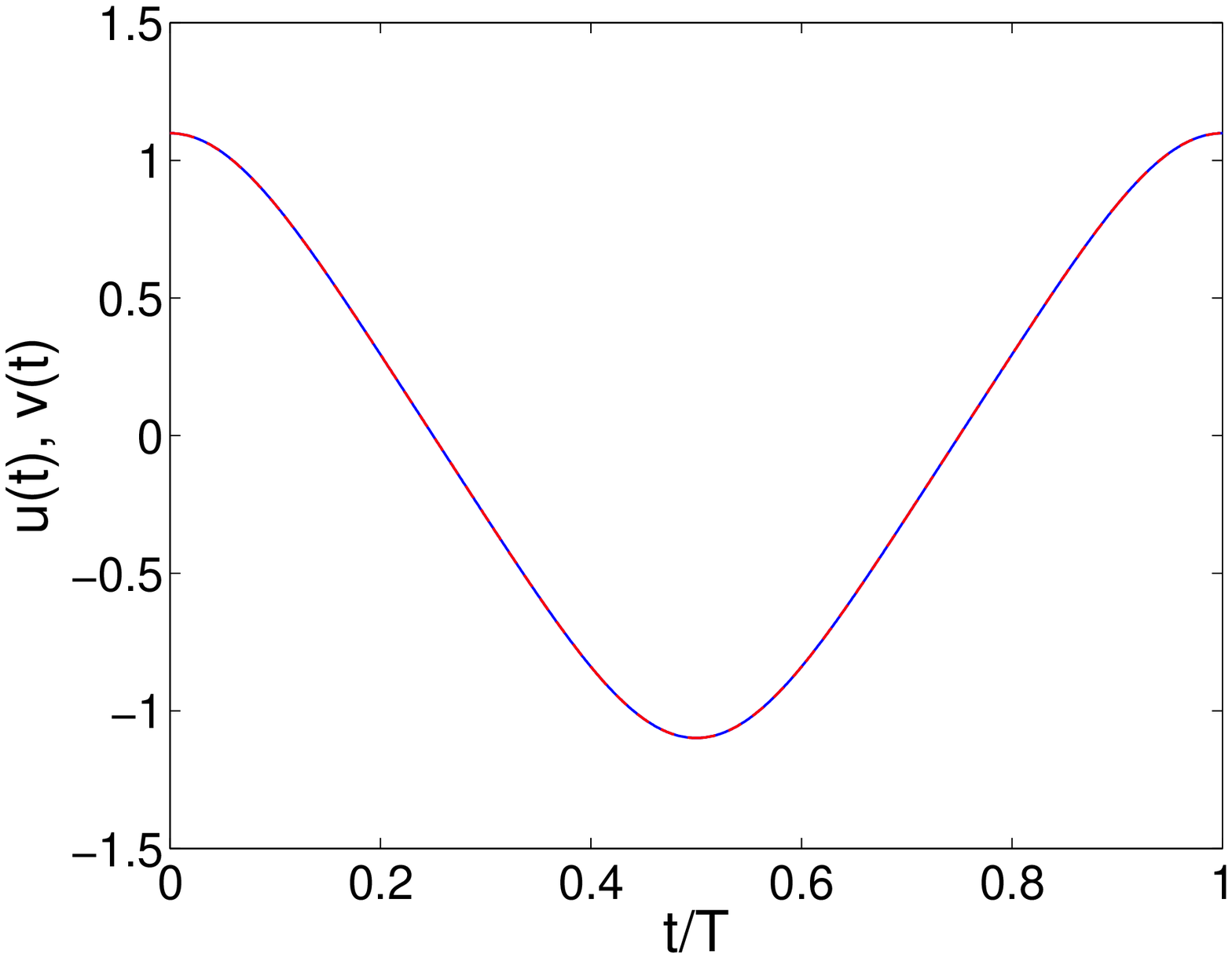} &
\includegraphics[width=6cm]{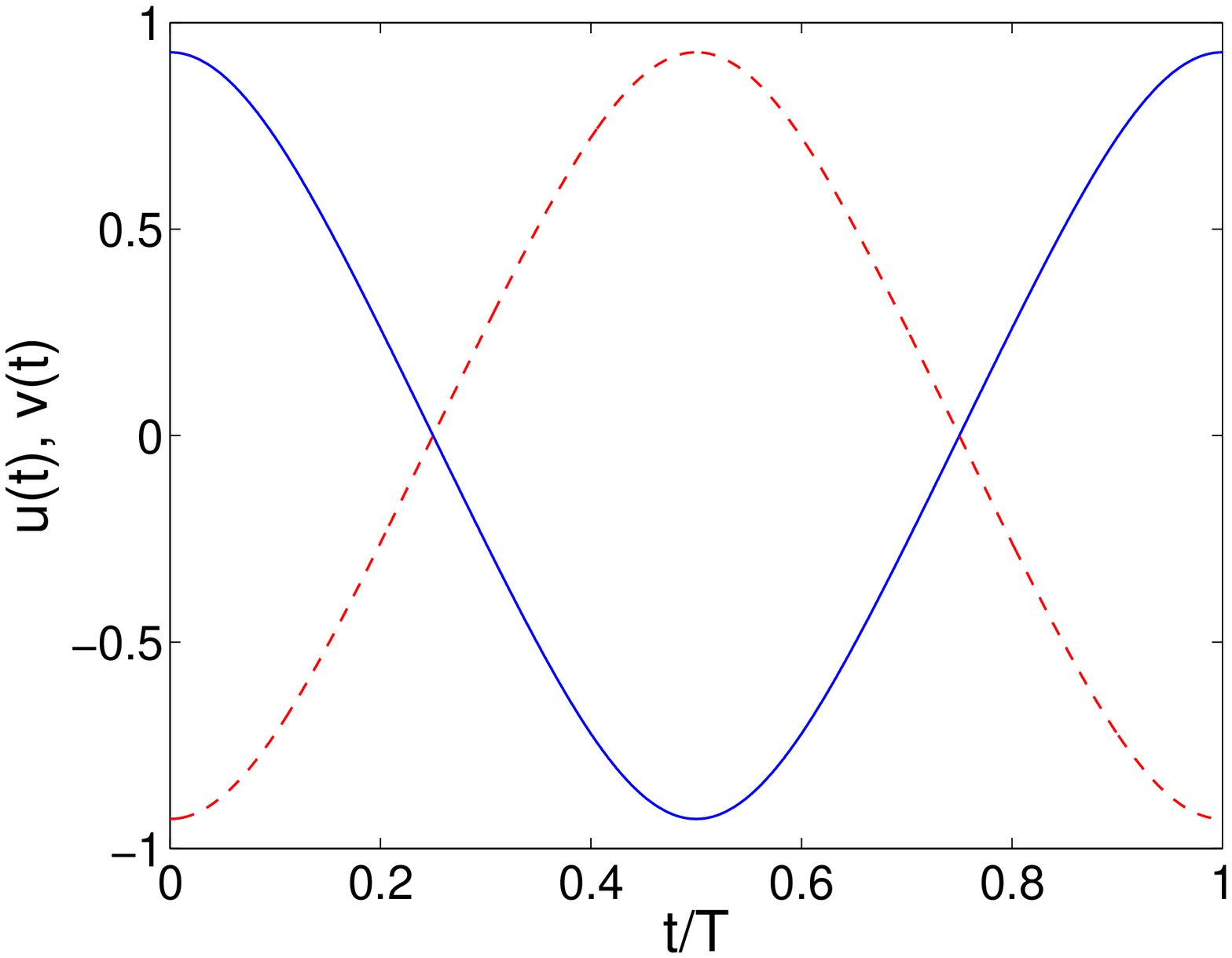} \\
\Sp solution &
\Ap solution \\
\includegraphics[width=6cm]{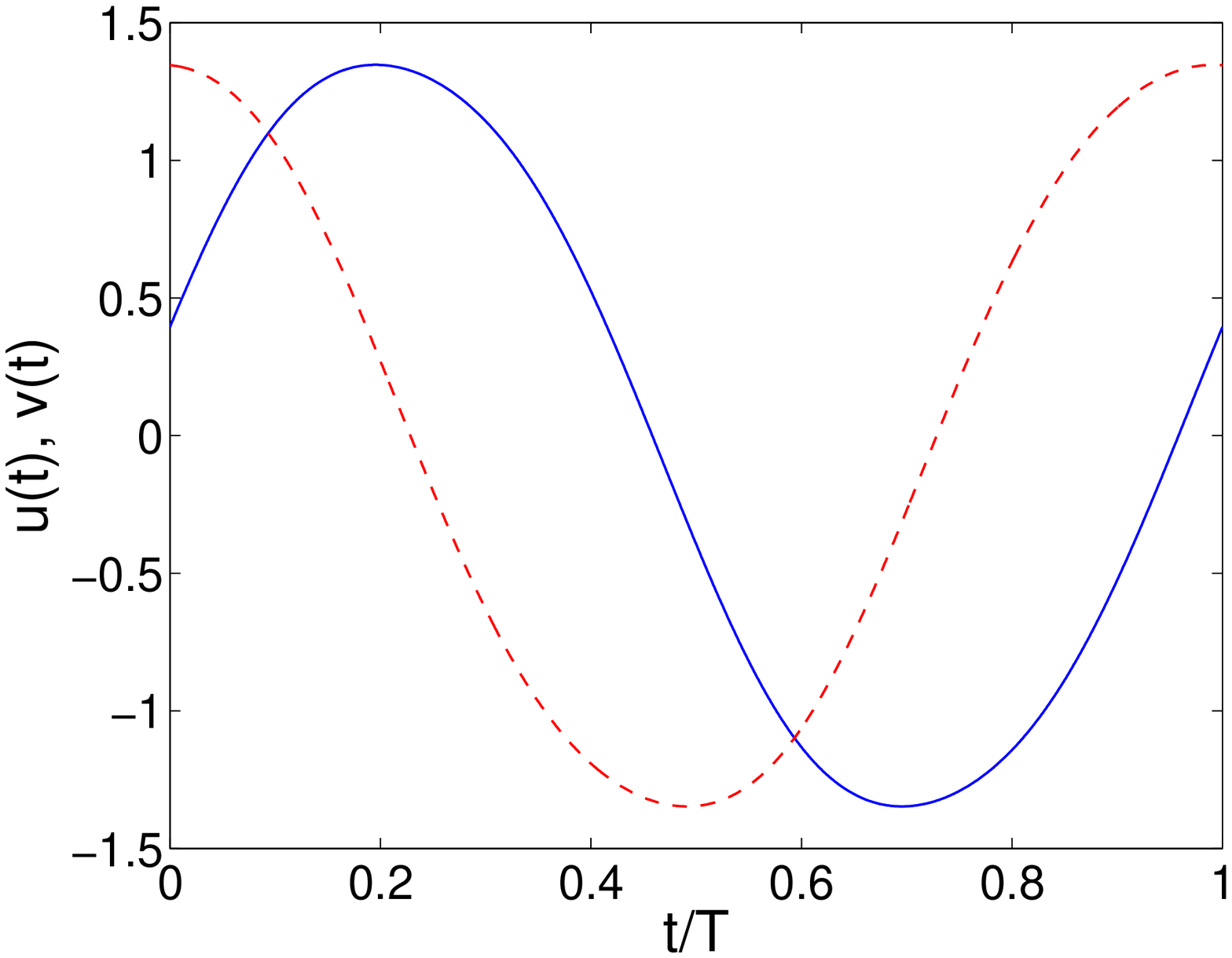} &
\includegraphics[width=6cm]{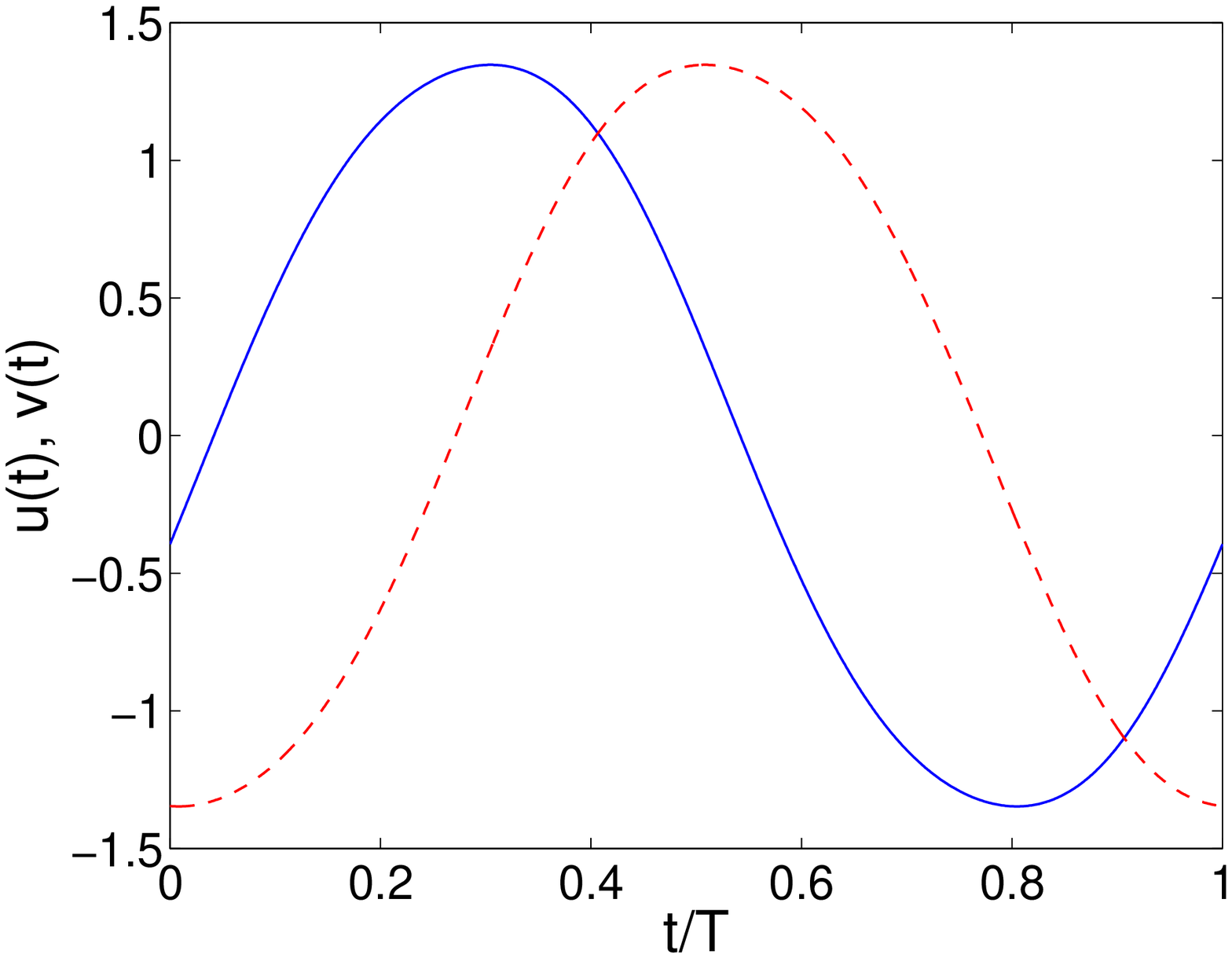} \\
AS solution & \\
\includegraphics[width=6cm]{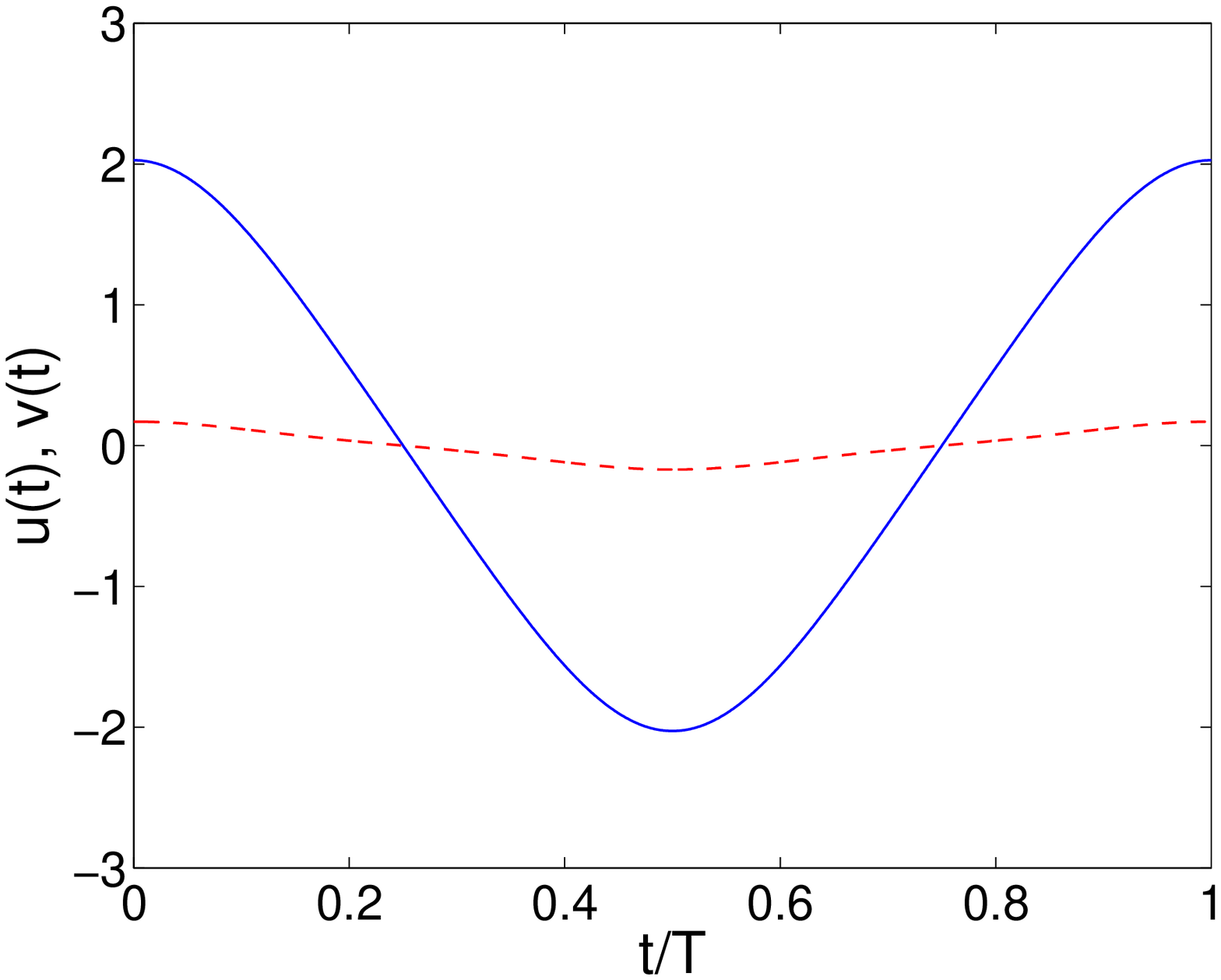} & \\
\end{tabular}
\caption{Time evolution of the solutions at $\delta=3\epsilon$ for
the hard nonlinearity case of $\epsilon=-1$ and $\wb=2$.
Again here, $\gamma=0$.}
\label{fig:gamma0_hard_evol}
\end{figure}

All the previous properties are summarized in Fig. \ref{fig:gamma0} where the Hamiltonian energy is depicted versus $\wb$ and compared with the averaged Hamiltonian for the RWA. This figure is complemented by Figs. \ref{fig:gamma0_soft_evol} and \ref{fig:gamma0_hard_evol} where the time evolution of the different solutions are displayed. Importantly, we should point out here
that it is evident that the approximations involved in the
RWA become demonstrably less accurate especially in the soft nonlinearity
case and particularly as the frequency $\omega$ decreases away from the
linear limit (and hence nonlinear terms become more significant).
Nevertheless, the qualitative agreement of the features of
 Fig. \ref{fig:gamma0} is still fairly satisfactory for the regime
of parameters considered herein. On the other hand, for the hard
nonlinearity case, the agreement seems to be even quantitatively
accurate for the frequency range considered.

\subsection{$\delta=\epsilon$ case: existence of exact solutions}

One of the main features of this case is the existence of two exact periodic solutions to (\ref{eq:dyn}):

\begin{equation}\label{eq:exactsol}
    u(t)=A\sin(\wt),\qquad v(t)=\pm A\cos(\wt)
\end{equation}
fulfilling that:
\begin{equation}\label{eq:exactcond}
    k=\mp \gamma\wb,\ ~~~A=\sqrt{\frac{1-\omega_b^2}{\epsilon}}\ ,
\end{equation}
with the upper sign corresponding to the symmetric solution and the lower one to the anti-symmetric solution. It is important to note that for a given
$k$, the frequency is proportional to $1/\gamma$. Thus, the two solutions
collide as $\gamma \rightarrow \infty$, when $\wb \rightarrow 0$.
That is, contrary to the ``standard'' model of $\delta=0$, since for the case
considered herein there exist nonlinear solutions {\it for all} values
of $\gamma$ that are not subject to the relevant transition~\footnote{We acknowledge here that Igor Barashenkov
in his recent talk at the SIAM conference on Nonlinear Waves and Coherent
Structures (Cambridge, August 2014)
reported an apparently similar feature as part of ongoing work with Dimitry Pelinovsky}.
This solution can actually be cast as $y_n=z_n=0\ \forall |n|>1$ and $\phi=\pm\pi/2$. If we fix the value of $\epsilon$, it is clear from Eq. (\ref{eq:exactcond}) that the properties of the solutions only depend on two parameters, as $k=k(\wb,\gamma)$.
That is, contrary to what we have discussed so far, here we do not fix
$k$ and vary $\gamma$ and $\wb$, but rather than varying $\gamma$ and $\wb$,
we fix a value of $k$ associated with them through Eq.~(\ref{eq:exactcond}).
Thus, we will consider the effect on the stability of varying parameters $\wb$ and $\gamma$ in the case $\epsilon=1$ (soft potential) and $\epsilon=-1$ (hard potential). Notice also that given the restrictions formulated in (\ref{eq:exactcond}) and the symmetry properties of the dynamical equations, the Floquet spectrum for a given set of parameters is the same for both solutions.

\begin{figure}[t]
\begin{tabular}{cc}
$\delta=\epsilon=+1$, $|\phi|=\pi/2$ &
$\delta=\epsilon=-1$, $|\phi|=\pi/2$ \\
\includegraphics[width=6cm]{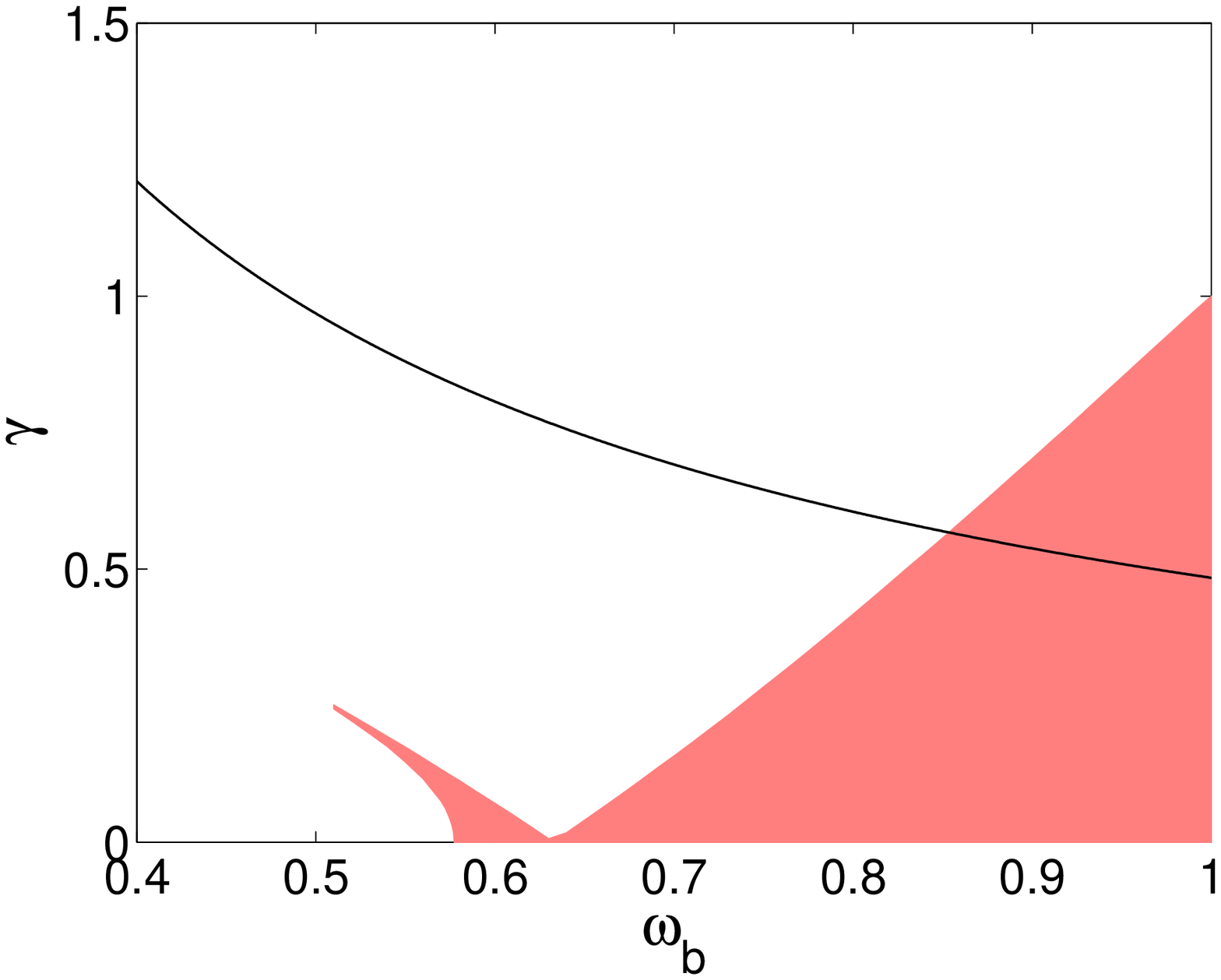} &
\includegraphics[width=6cm]{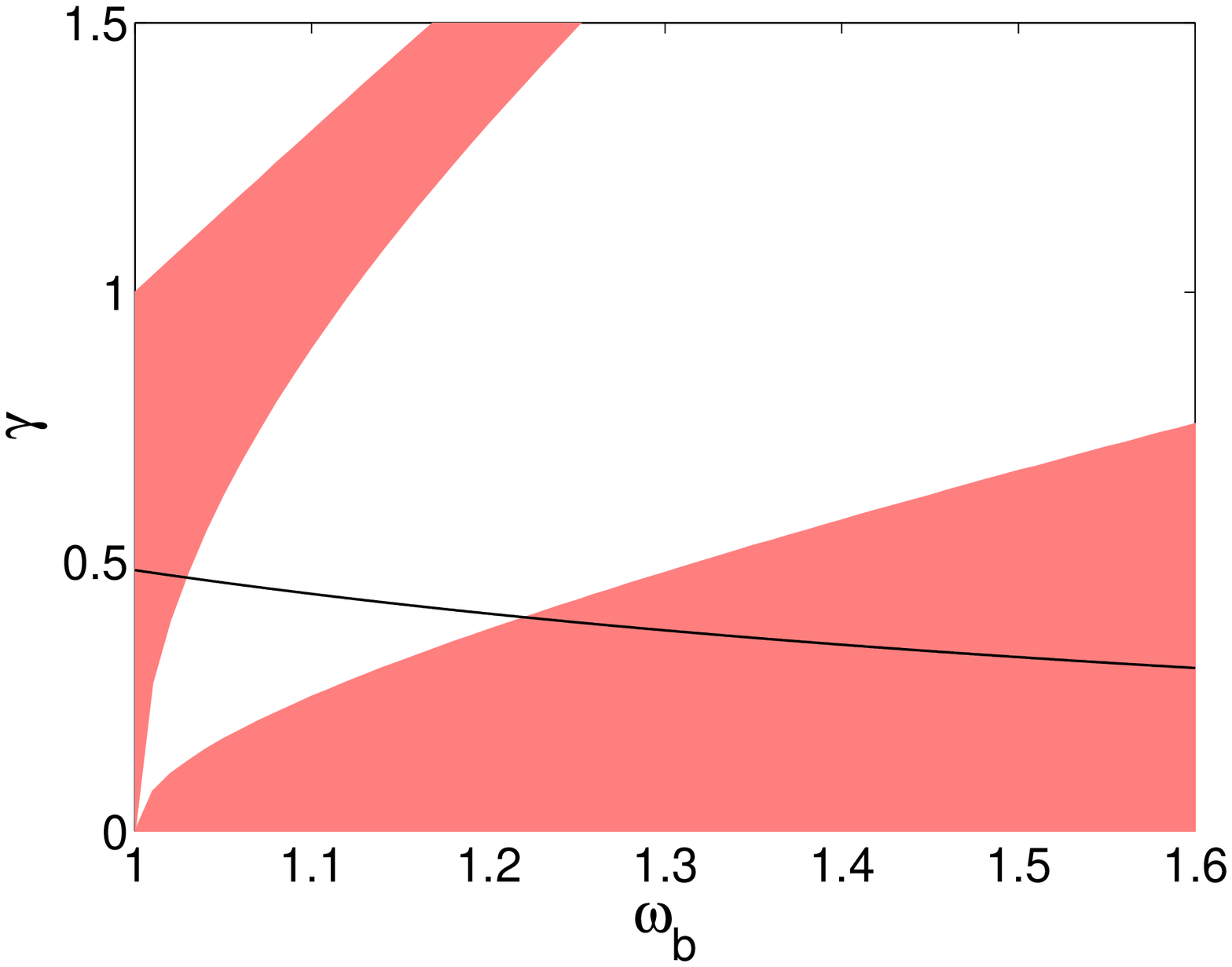} \\
\end{tabular}
\caption{Stability (shaded) and instability (blank) regions for the exact solutions (i.e. those with $\phi=\pm\pi/2$) in the Klein--Gordon dimer arising for $\delta=\epsilon$. The black line corresponds to $k=\sqrt{15}/8$
according to Eq.~(\ref{eq:exactcond}), i.e.,
the value generally considered throughout the paper.}
\label{fig:exactplane}
\end{figure}

Fig. \ref{fig:exactplane} shows the stability/instability regions for these solutions. Shaded areas correspond to stable solutions.
The  black line therein indicates the locus in the $\gamma$-$\wb$ plane
where $k=\sqrt{15}/8$ (i.e., the value used for other results in the
present work). From this line, it can be deduced that solutions with $\phi=\pm\pi/2$ when $\delta=\epsilon$ are stable in the range $\wb\in[0.8535,1]$ if $\epsilon=1$ and in $\wb\in[1,1.029] \cup [1.2206,\infty)$ if $\epsilon=-1$.

The averaged energy is, for both solutions, $<H>=\frac{1+2\omega^2-3\omega^4}{3\epsilon}$ which, for $\epsilon=1$ has a maximum at $\wb=3^{-1/2}\approx0.5774$; for $\epsilon=-1$, this function is monotonically decreasing. It is worth mentioning that for $\gamma=0$, where the averaged energy coincides with the Hamiltonian, there is a stability change at $\omega=3^{-1/2}$, the value at which $\partial H/\partial \omega$ changes its slope. This correlation between energy maximum and stability changes, which resembles the Vakhitov-Kolokolov criterion for NLS systems, is not observed for solutions that do not fulfill condition (\ref{eq:exactcond}) and, consequently, possess more than one harmonic in their Fourier series.

\begin{figure}[t]
\begin{center}
\begin{tabular}{cc}
$\delta=\epsilon=+1$ &
$\delta=\epsilon=-1$ \\
\includegraphics[width=6cm]{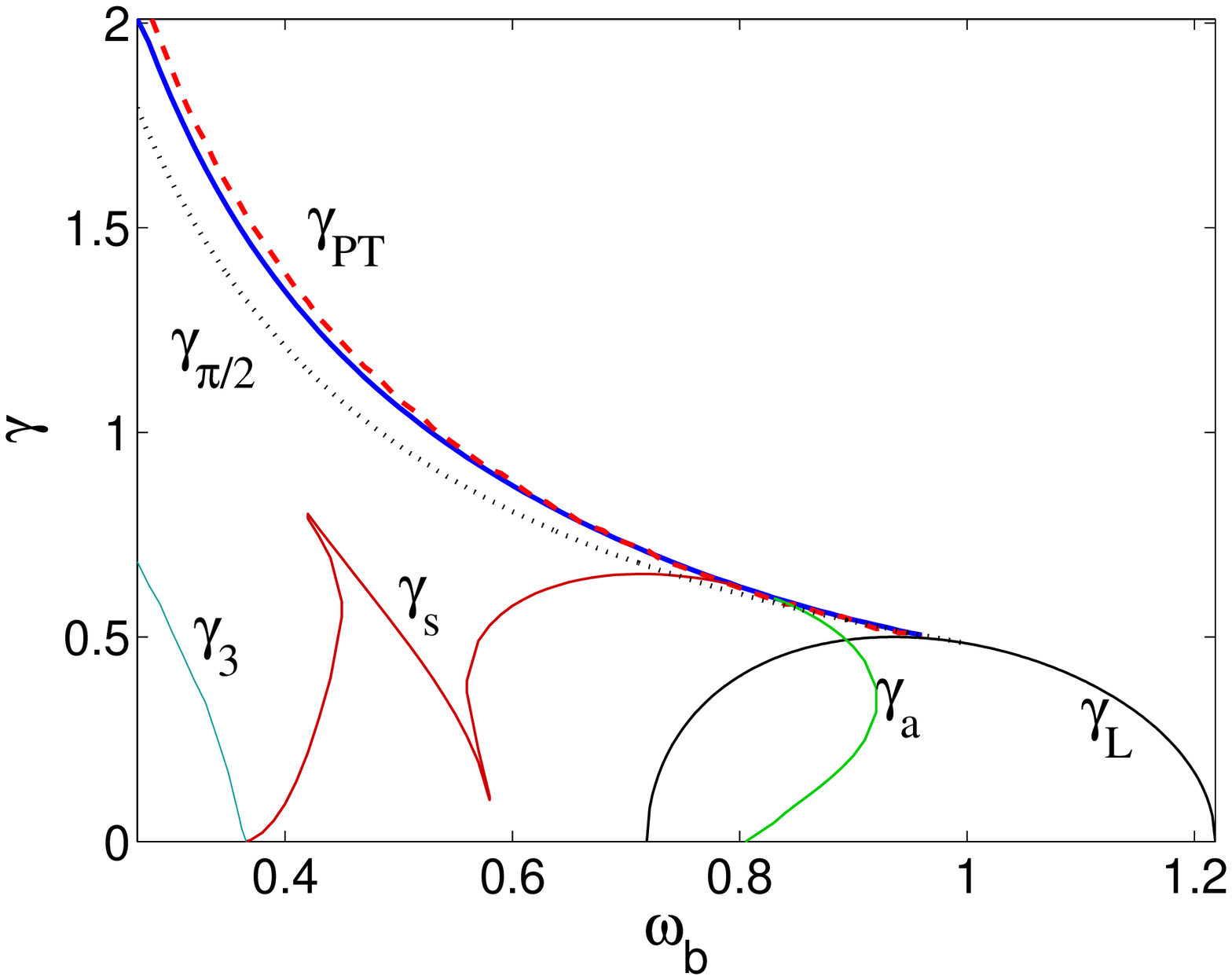} &
\includegraphics[width=6cm]{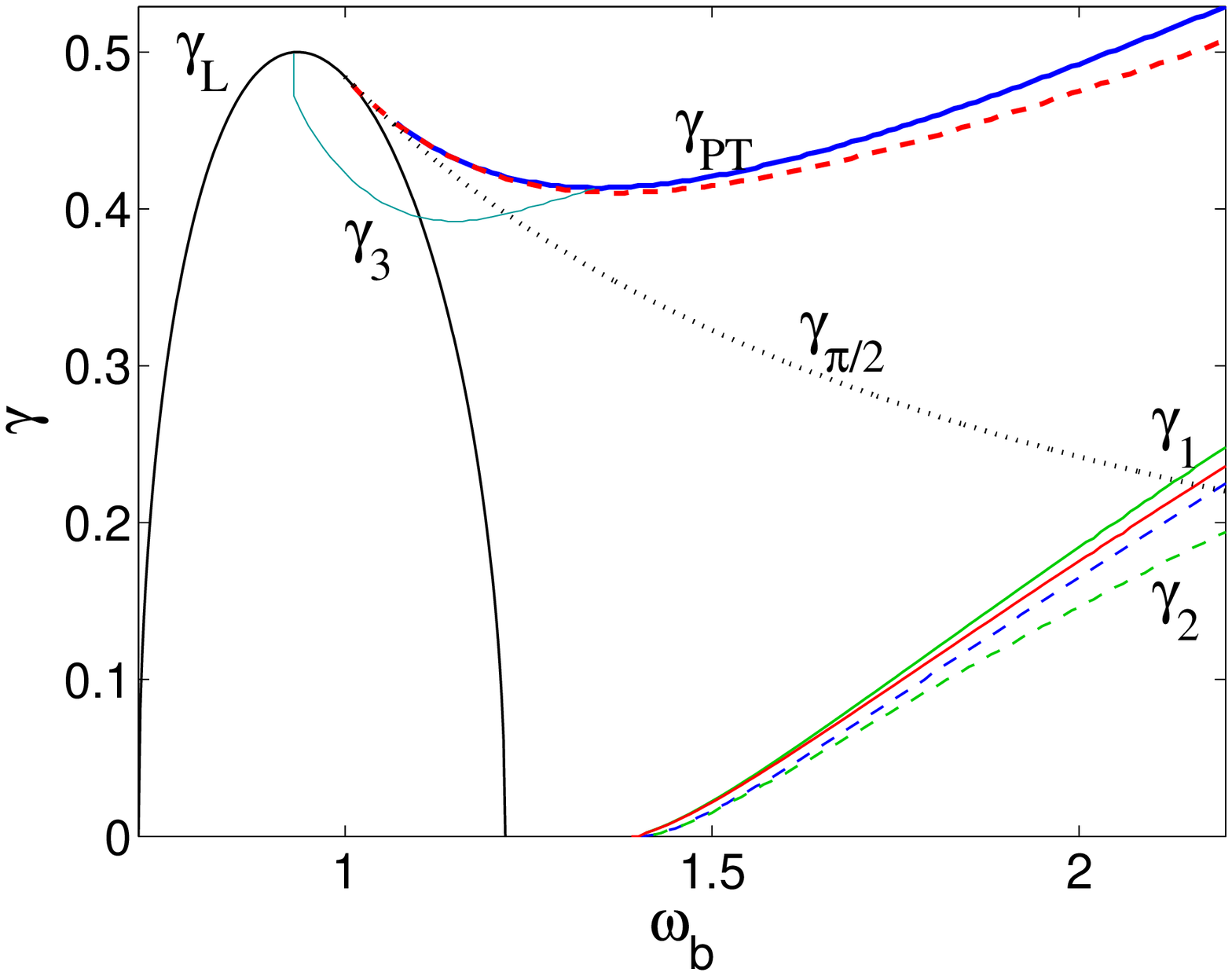} \\
\end{tabular}
\caption{Planes with curves separating regions of solutions that share the same properties when $\epsilon=1$ and $s=\sqrt{15}/8$ (see text). Dashed lines correspond to the RWA predictions and the dotted line $\gamma_{\pi/2}$ corresponds to the exact solutions with phase $\pi/2$ described in Fig. \ref{fig:exactplane}. Notice that the colors of the dashed lines are inverted with respect to that of the numerical results for a better visualization. This inversion pattern is followed
also in all figures comparing theory and numerical computations from here on.} \label{fig:delta1}
\end{center}
\end{figure}

The above mentioned exact solutions constitute only a subset of the whole $\gamma-\wb$ plane, which is depicted in Fig. \ref{fig:delta1}. This figure shows the existence range of the different solutions that arise for $\gamma\neq0$ and $\delta=\epsilon$. We explain below the different regions and curves.

\begin{figure}[t]
\begin{center}
\begin{tabular}{ccc}
    \includegraphics[width=6cm]{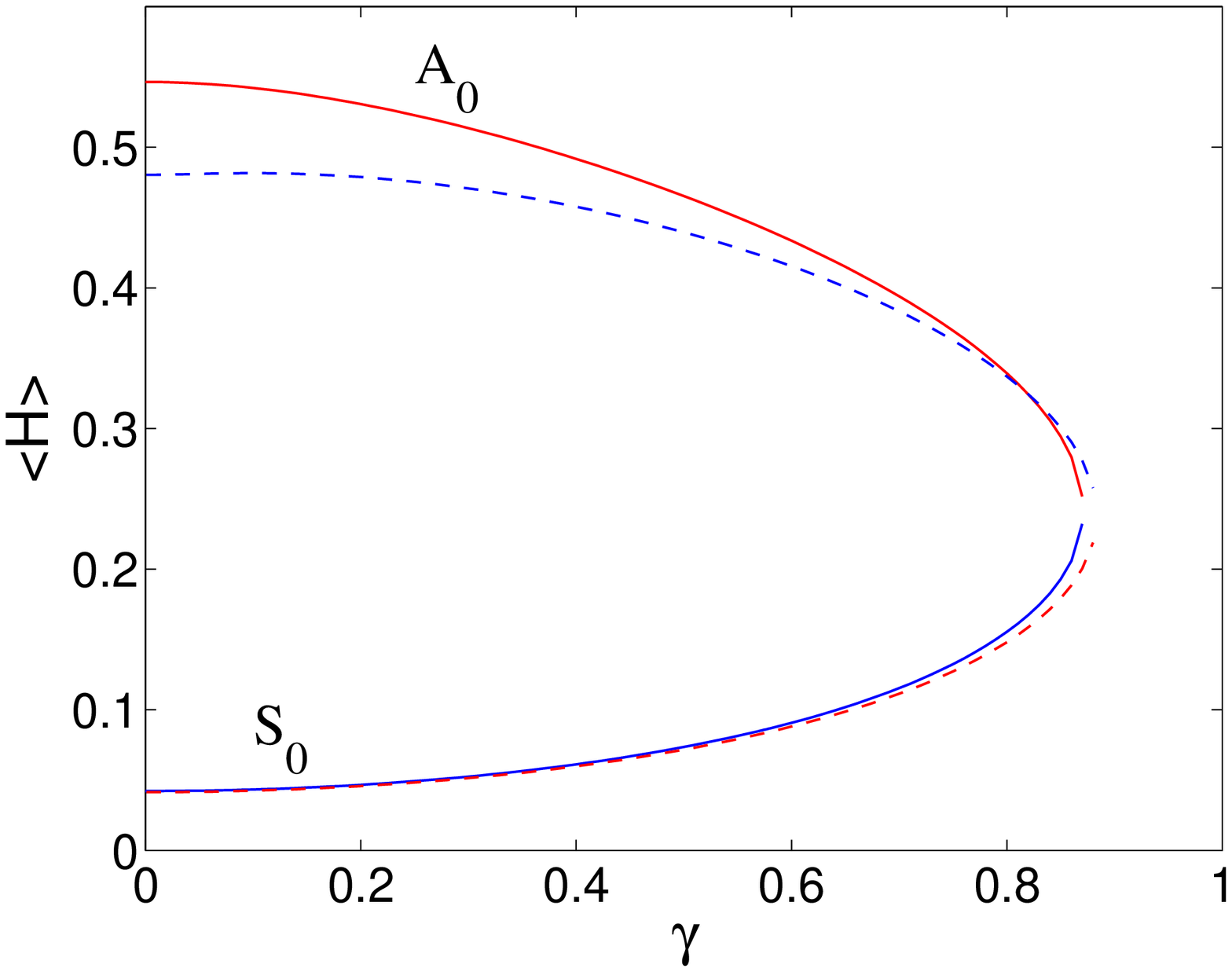} &
    \includegraphics[width=6cm]{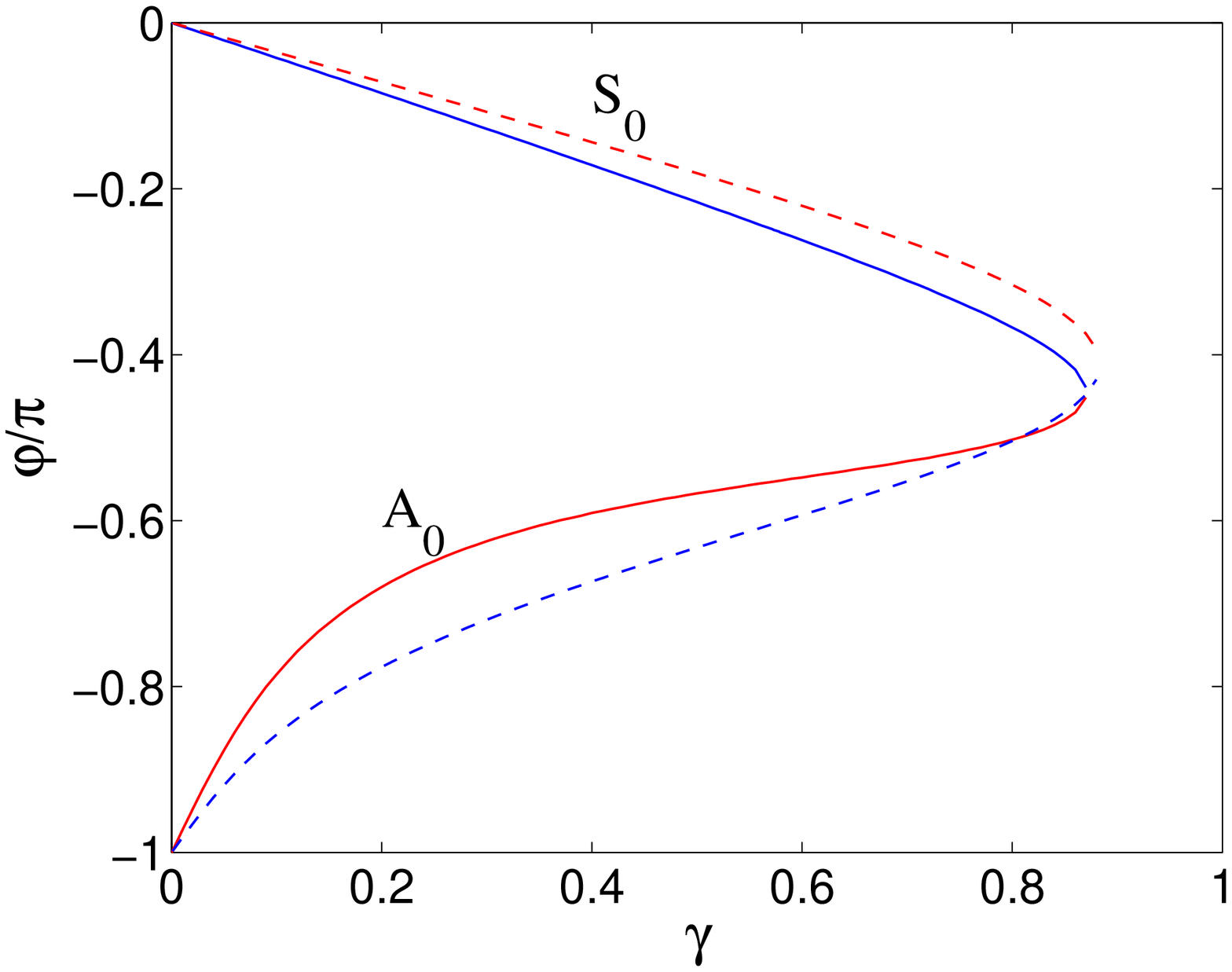} &
    \includegraphics[width=6cm]{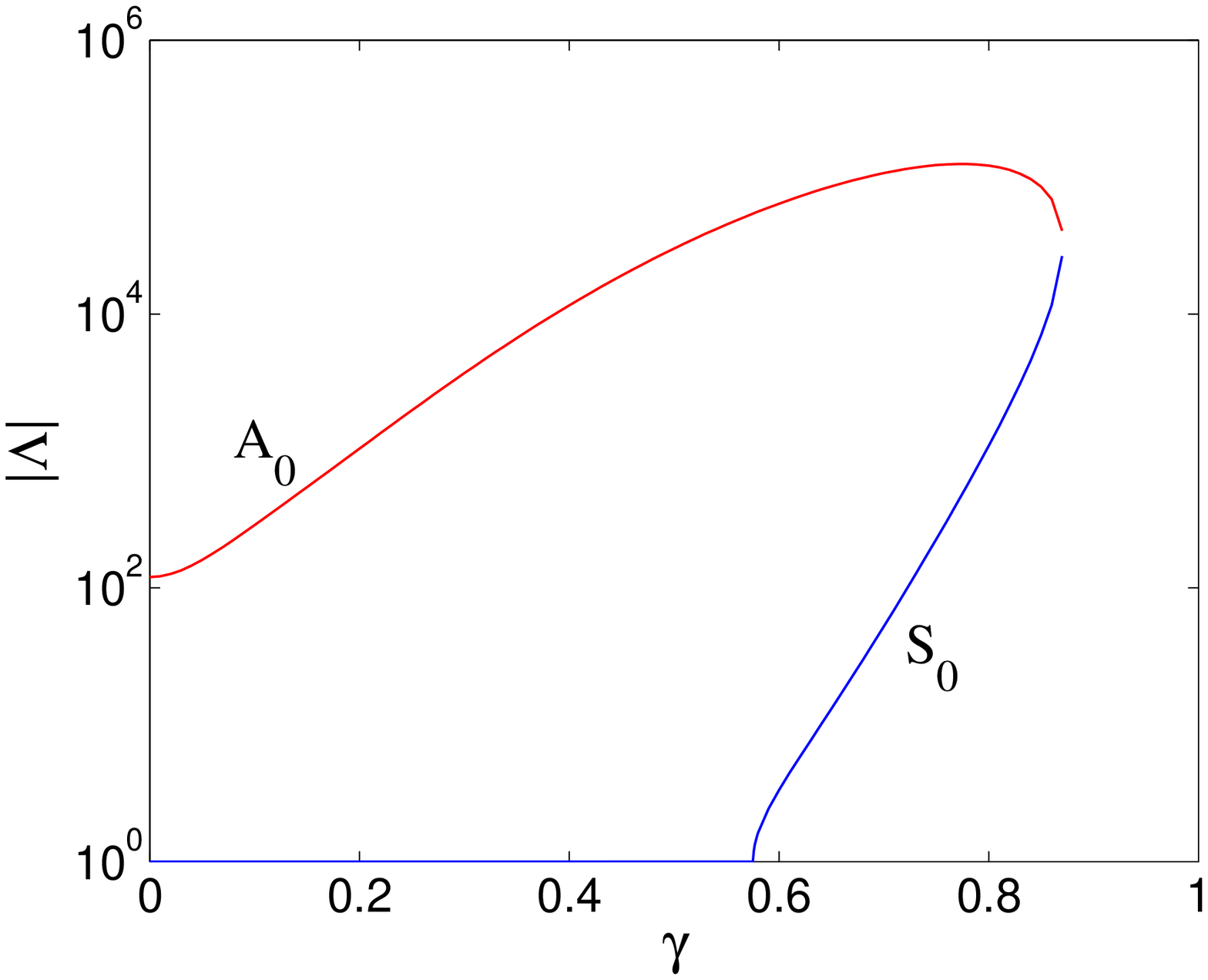} \\
\end{tabular}
\caption{Averaged energy (left), phase difference (center) and modulus of the Floquet multipliers
(right) [only the multipliers with moduli higher than one are shown] as a function of the gain/loss parameter $\gamma$ for $\epsilon=\delta=1$, $\wb=0.6$,
and $s=\sqrt{15}/8$. Notice the logarithmic scale in the $y$-axis of the latter graph. In the left and central panels, blue (red) solid line corresponds to the \So (\Ao) solution, while red (blue), i.e., reversed colors, dashed lines correspond to the \So (\Ao) branch from the RWA predictions. The figure on the right does not show the RWA prediction as they only qualitatively match
the results for the oscillator dimer for this value of $\wb$.}
\label{fig:softbif_1a}
\end{center}
\end{figure}

\begin{figure}[t]
\begin{center}
\begin{tabular}{cc}
    \includegraphics[width=6cm]{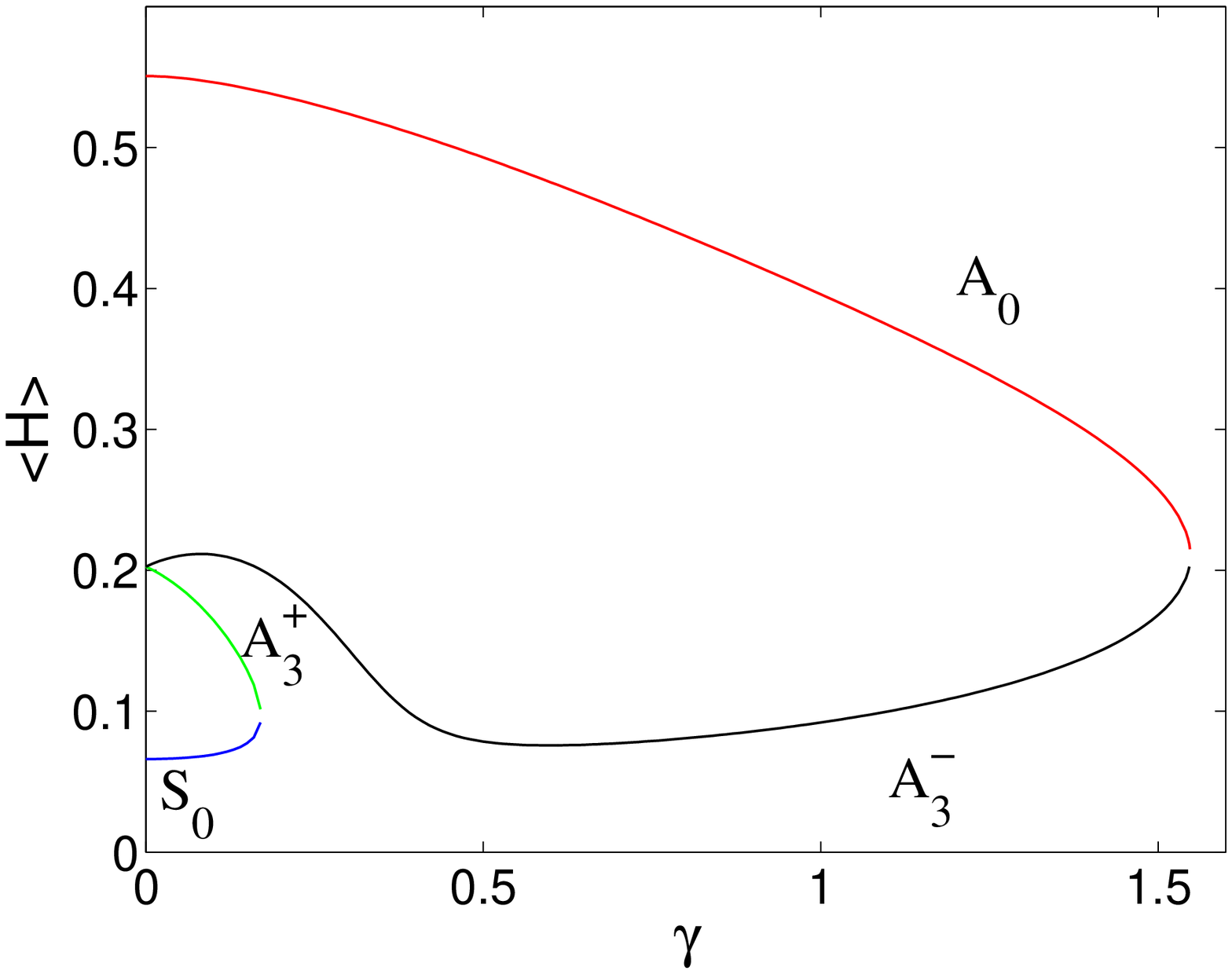} &
    \includegraphics[width=6cm]{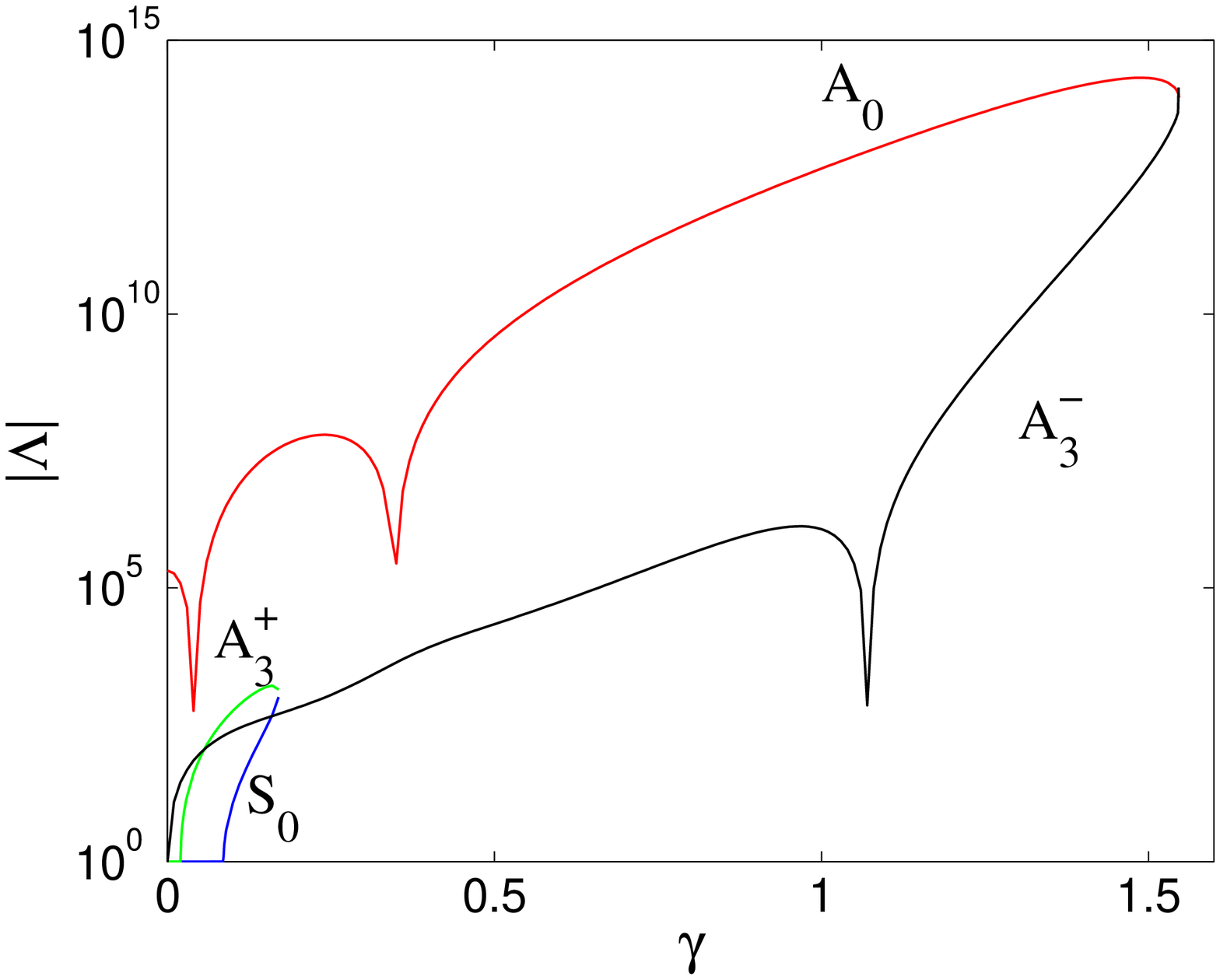} \\
    \includegraphics[width=6cm]{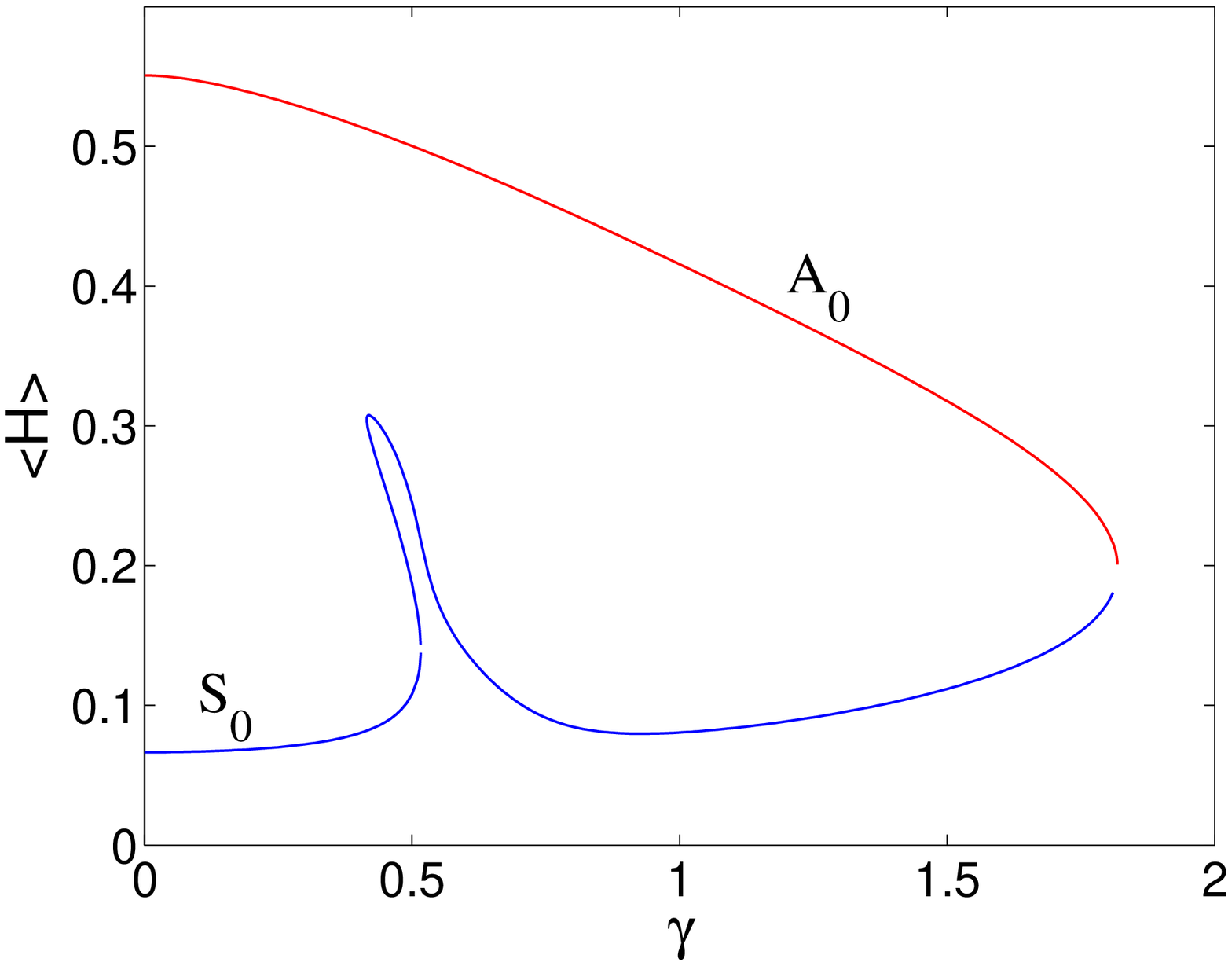} &
    \includegraphics[width=6cm]{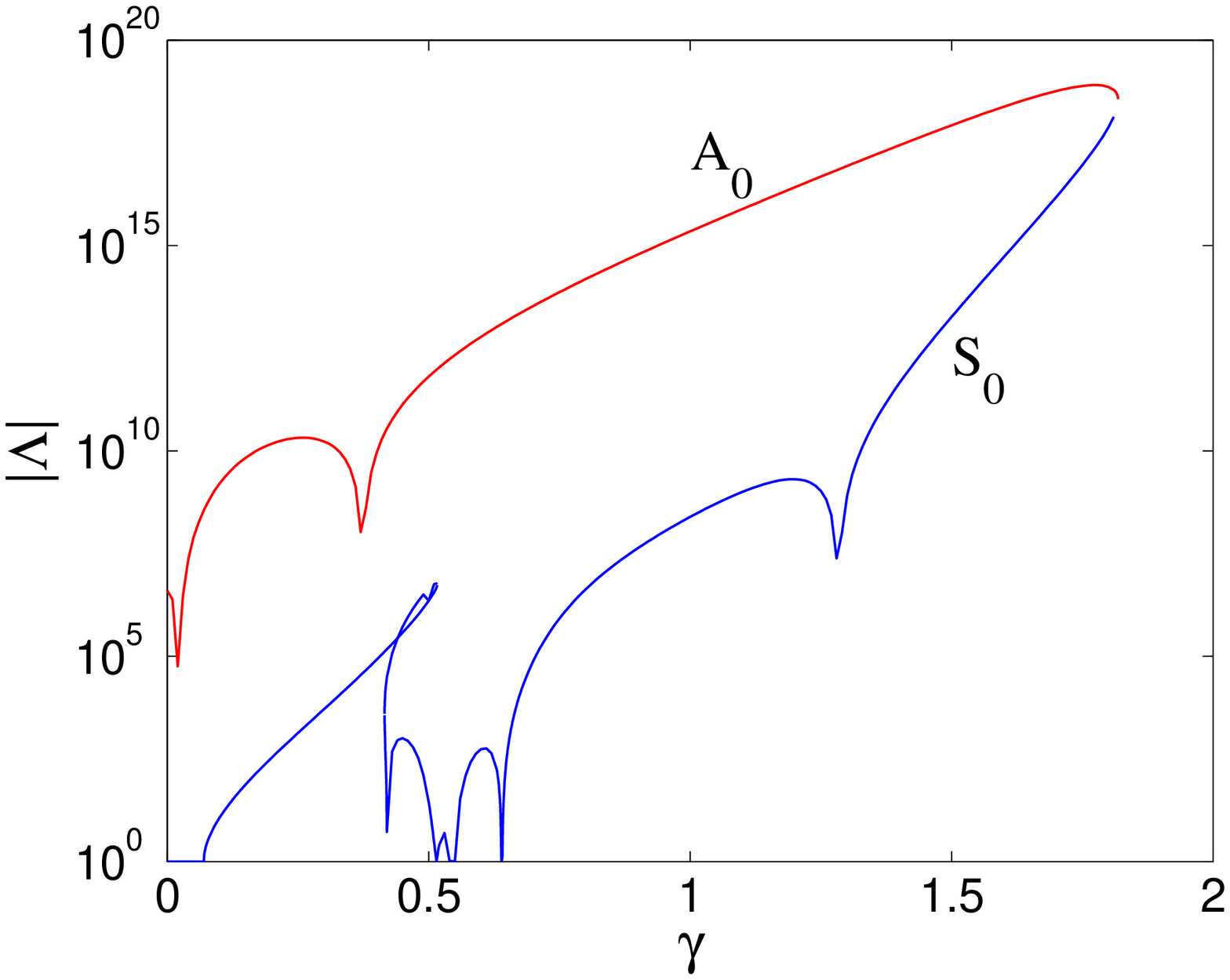} \\
\end{tabular}
\caption{Averaged energy (left) and modulus of the Floquet multipliers
(right) as a function of the gain/loss parameter $\gamma$ for $\epsilon=\delta=1$ and $\wb=0.35$ (top) or $\wb=0.3$ (bottom). Blue (red) line corresponds to the \So (\Ao) solution, while in the top panels, black and green lines represent
the two \At solutions discussed in the text, \Atp and \Atm.}
\label{fig:softbif_1b}
\end{center}
\end{figure}

In the case of {\em soft} potential, we observe, as expected, the curve $\gL$ which encloses a region with only \Ao solutions, as \So bifurcates from the left arm of the curve. In addition, above the curve $\gPT$, which indicates the $\cP\cT$ transition and is very close to the value predicted by RWA, there are no periodic orbits. {This transition is caused by the {\em collision} of \Ao and \So solutions whenever $\wb>\omega_3\approx0.365$}. Contrary to the expectation from RWA, there are three more curves in the considered range. At the right of curve $\ga$, \Ao solutions are stable; similarly, below
to the right of the curve denoted by $\gs$, \So solutions are stable. Consequently, for $\wb\lesssim0.8$ the $\cP\cT$ phase transition takes place between the
unstable \So and \Ao solutions. This behavior is similar to the one observed for the $\delta=0$ case~\cite{PRA}. Notice the existence of a third curve $\gamma_3$, which terminates at $\wb=\omega_3$. This curve corresponds to the loci
for the occurrence of the saddle-node bifurcation
between the \So and \Atp solutions.  With this notation we remark that this solution is a \At mode whose phase difference $\vphi$ is in the first quadrant for $\gamma\approx0$. Remarkably, there is a different behavior regarding \So in the regions between the curves $\gs$ and $\gamma_3$; in the former, the phase of the mode is in the fourth quadrant whereas in the latter, the phase lies in the first quadrant. Additionally, for $\wb<\omega_3$, the \Ao mode collides
and disappears at $\gPT$ with the \Atm mode; contrary to the \Atp case, the phase of this mode lies in the fourth quadrant. Notice also that in the region below curve $\gamma_3$, the stability description is not trivial; despite this, we can say that for small $\gamma$, both \Atp and \So solutions are stable. Finally, for $\wb\lesssim0.31$,
we observe that both \At solutions coalesce into the \Ao solution with frequency $3\wb$ and the \So solution transforms into a new solution that
collides and disappears with the \Ao solution at $\gPT$. A summary of the bifurcations for $\wb>\omega_3$, together with energy, phases, Floquet multipliers and comparisons with RWA are shown in Fig. \ref{fig:softbif_1a}. Figure \ref{fig:softbif_1b} shows the bifurcation diagrams and Floquet multipliers for $\wb=0.35$ and $\wb=0.3$.

\begin{figure}[t]
\begin{center}
\begin{tabular}{ccc}
    \includegraphics[width=6cm]{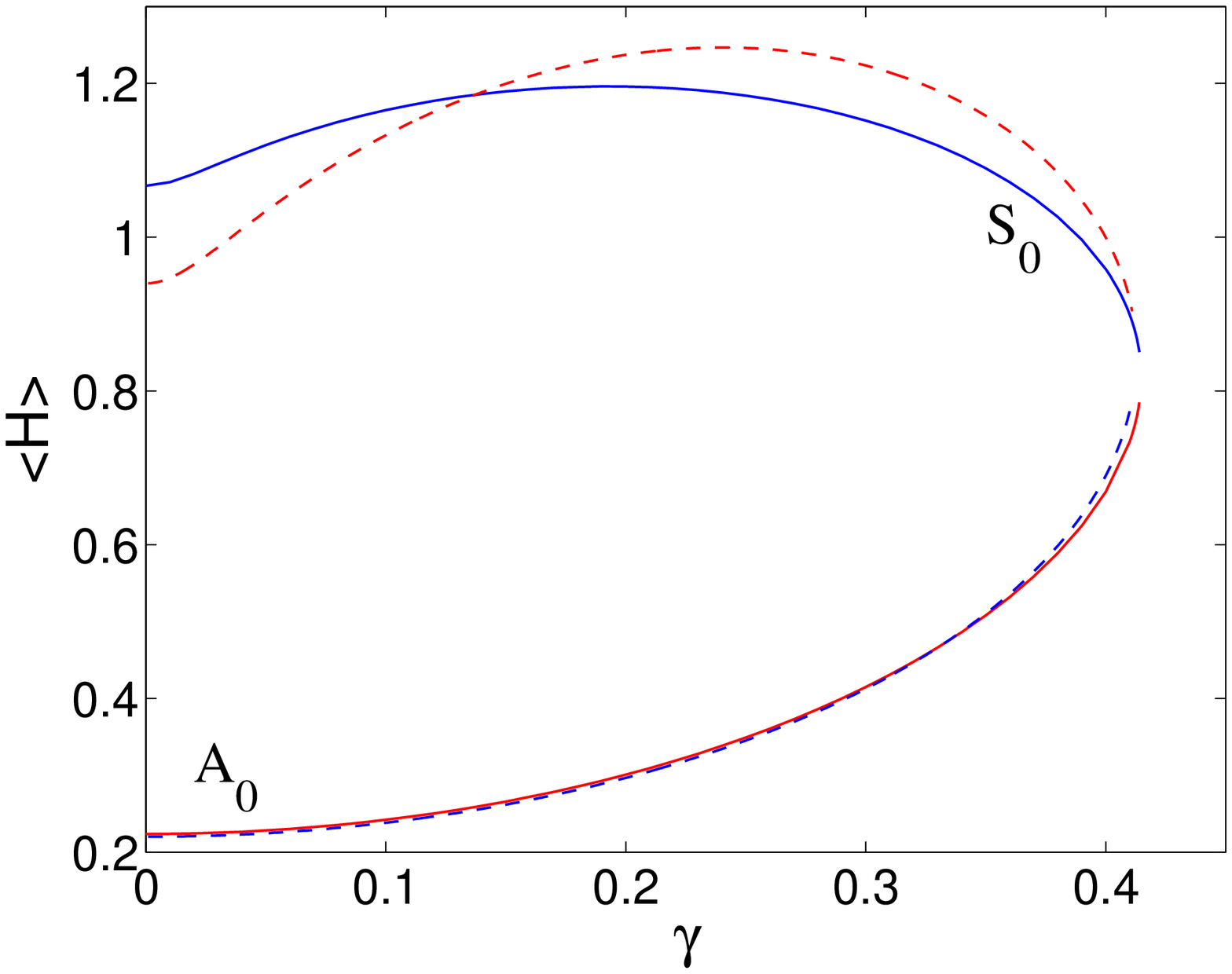} &
    \includegraphics[width=6cm]{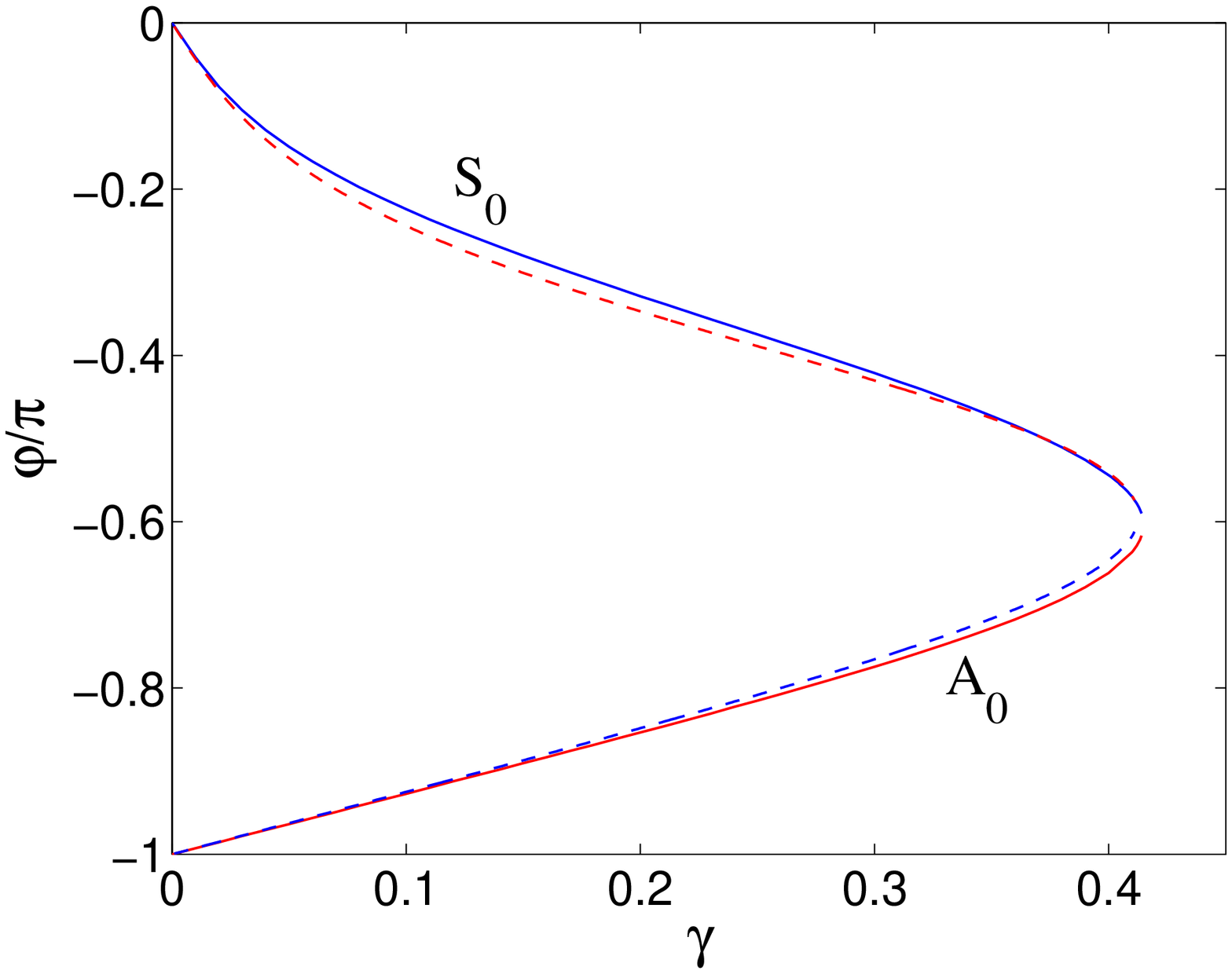} &
    \includegraphics[width=6cm]{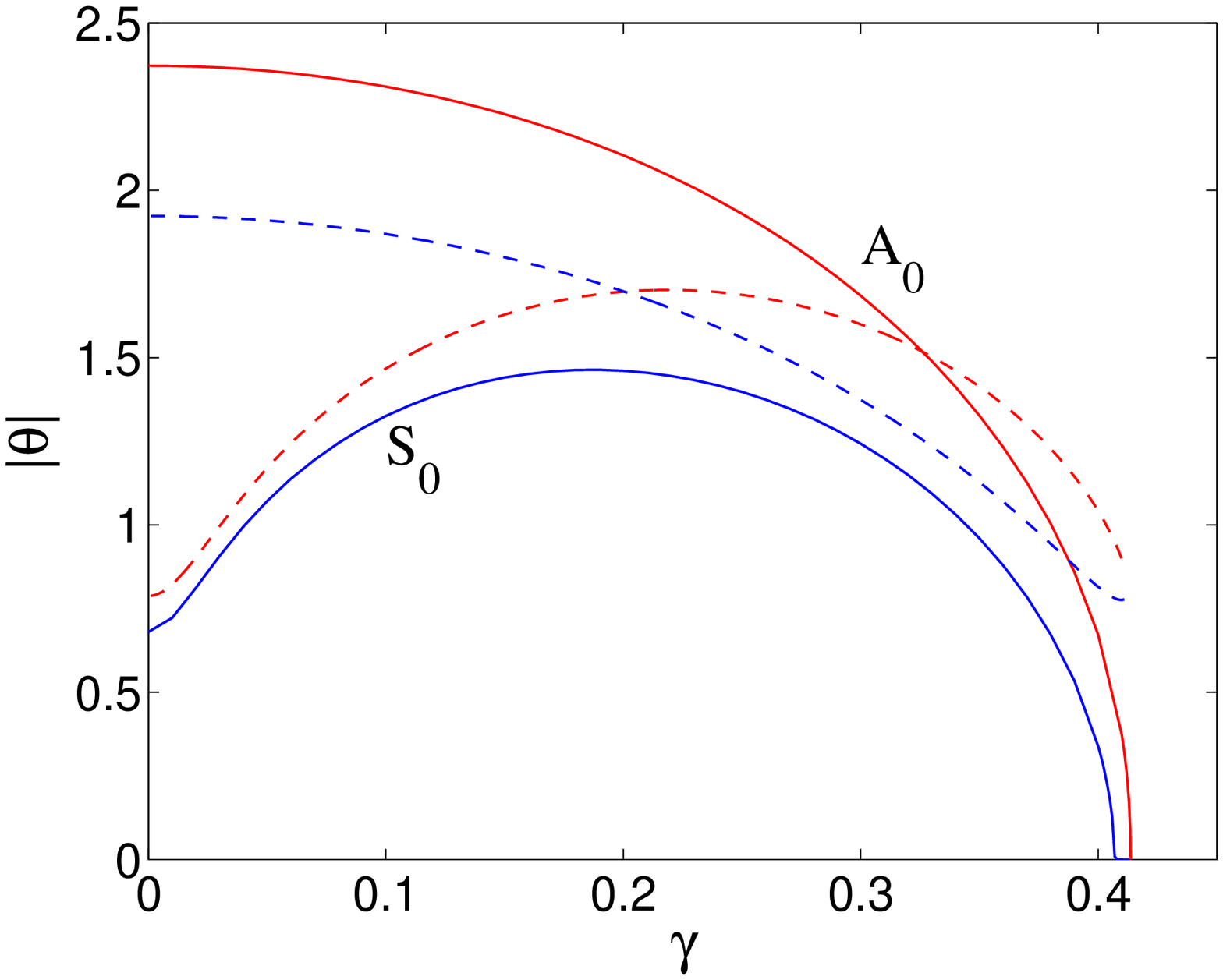} \\
    \includegraphics[width=6cm]{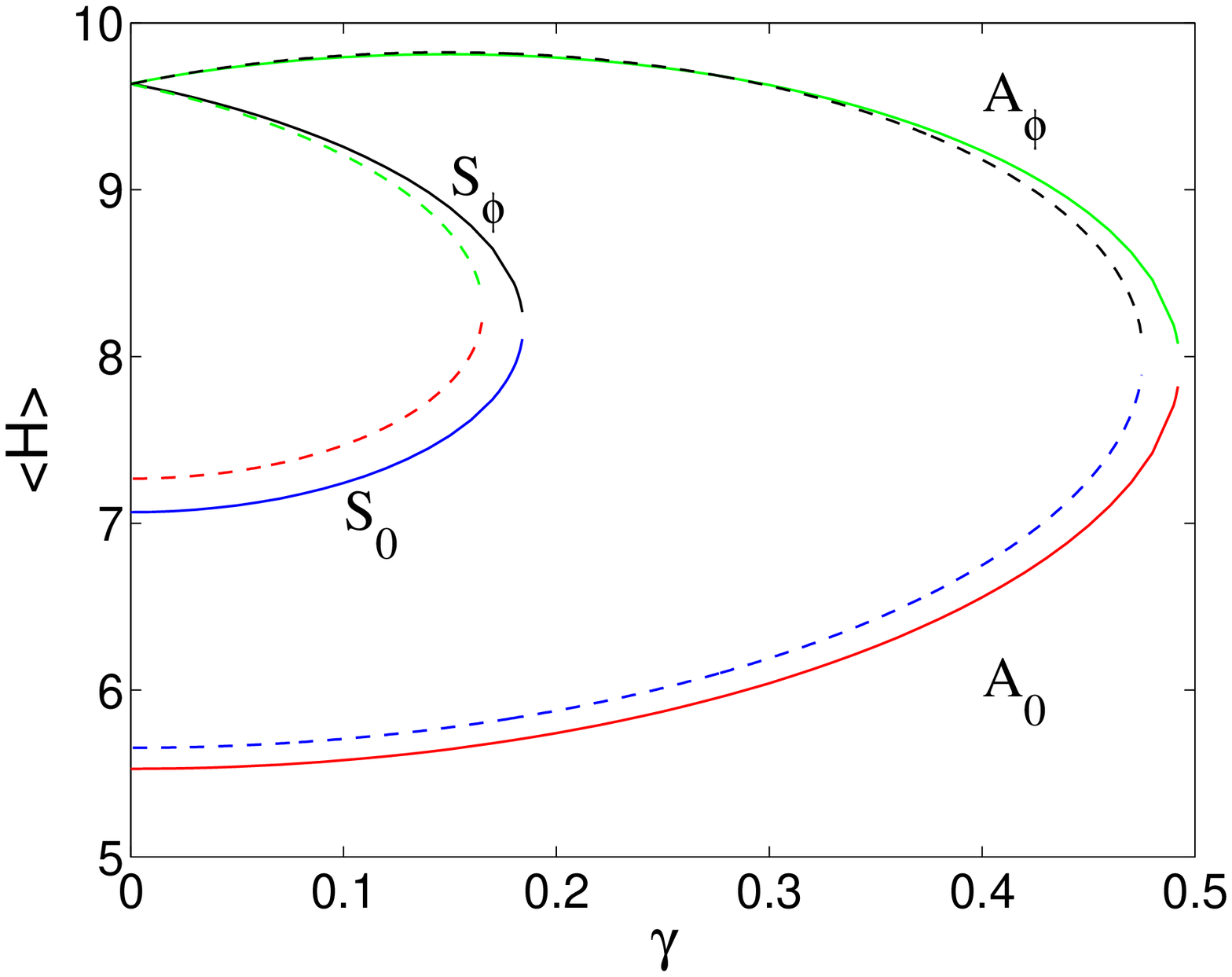} &
    \includegraphics[width=6cm]{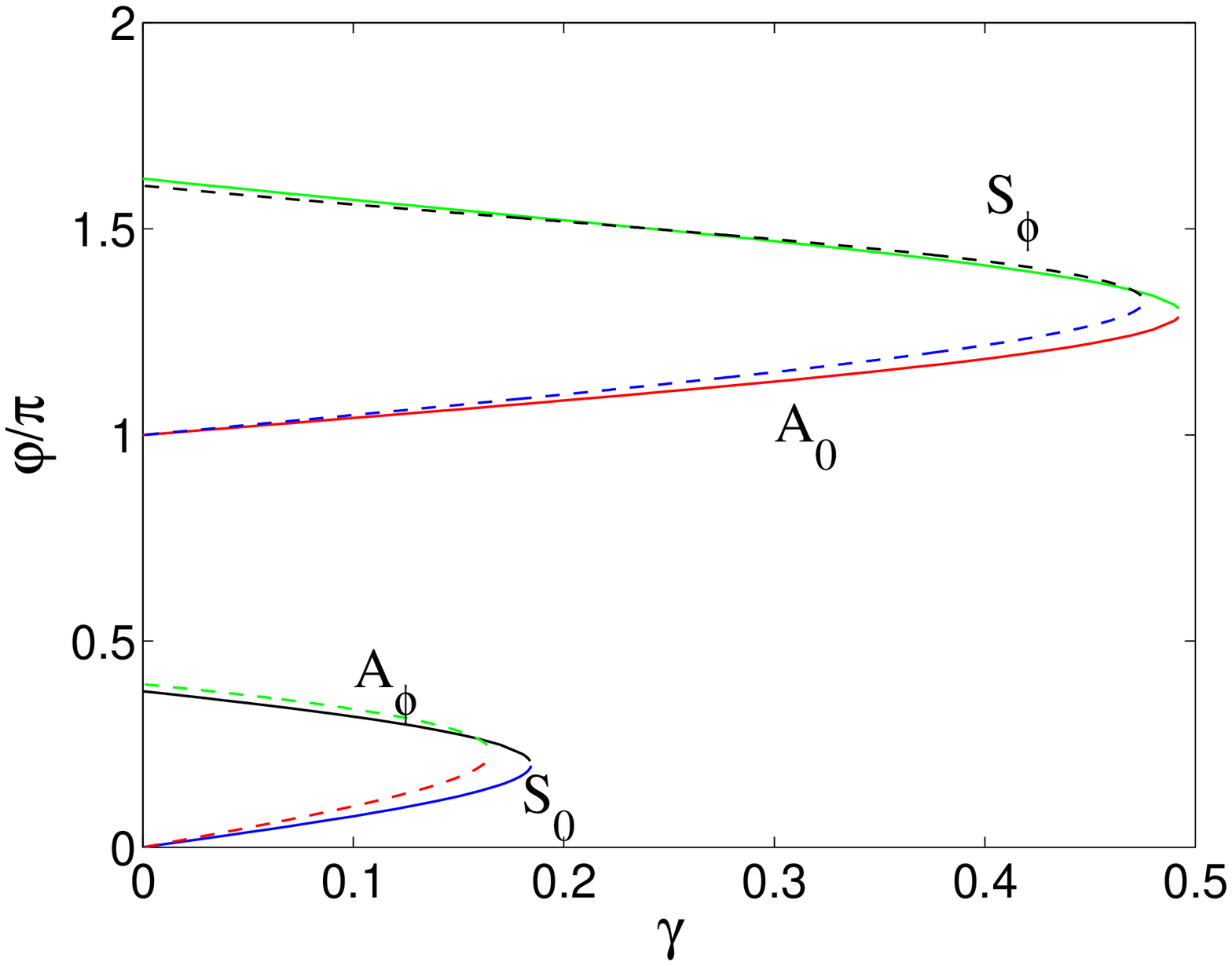} &
    \includegraphics[width=6cm]{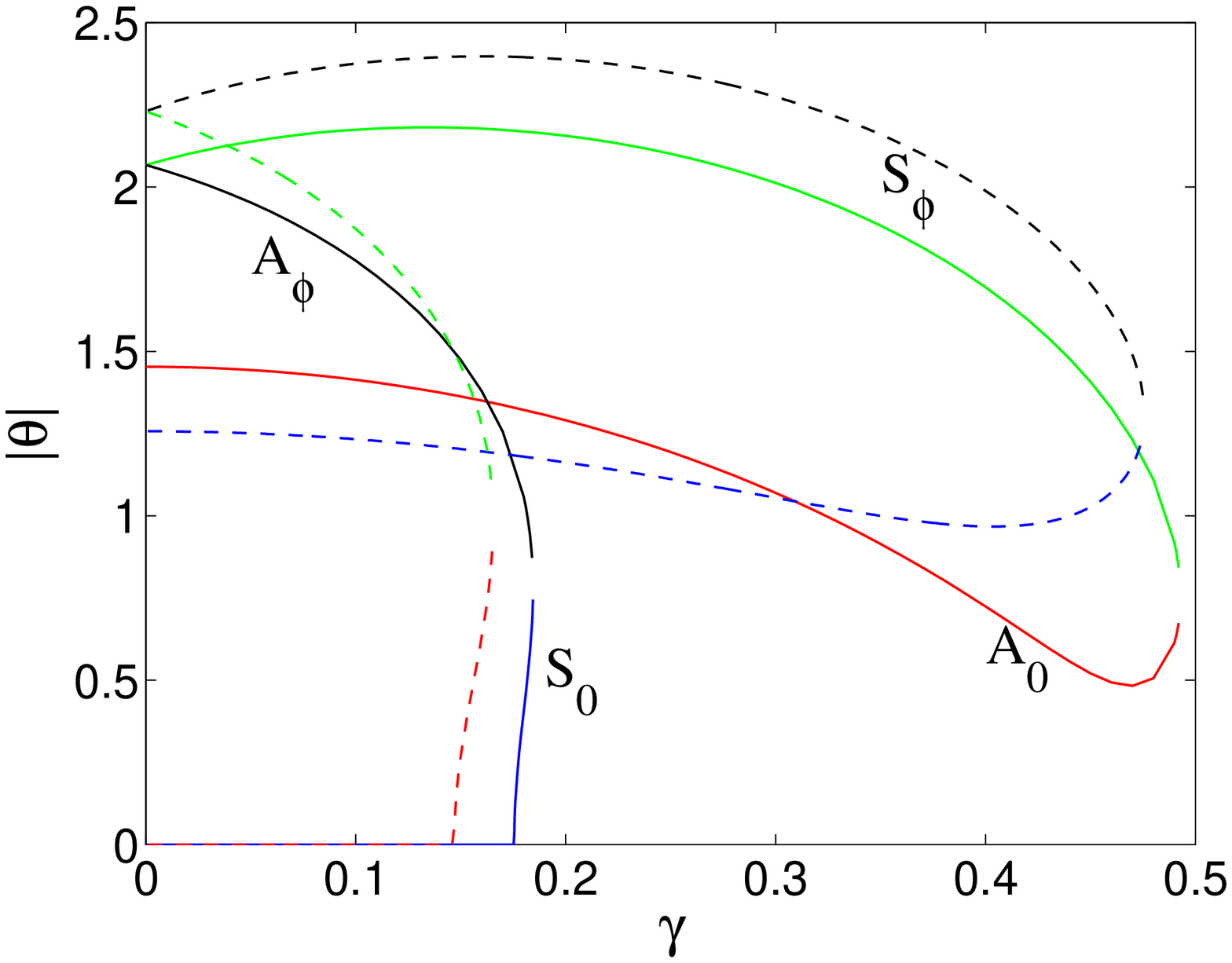} \\
\end{tabular}
\caption{Same as Fig. \ref{fig:softbif_1a} but for the case $\delta=\epsilon=-1$ and $\wb=1.3$ (top) and $\wb=2$ (bottom). Notice that,  as solutions are mostly stable, the Floquet argument is displayed instead of its modulus.}
\label{fig:hardbif_1}
\end{center}
\end{figure}

The case of hard potential ($\delta=\epsilon=-1$) is also illustrated in Fig. \ref{fig:delta1}. The curves and regions are equivalent to the RWA case, except for one fact: there is a region between curves $\gamma_3$ and $\gPT$ where the \So solution is unstable, similar to the $\delta=0$ case \cite{PRA}.
Since this is the only feature not captured by the RWA, it must
be directly connected with the emergence/role of higher harmonics
in the system.
Figure \ref{fig:hardbif_1} shows the averaged
energy, relative phase and Floquet exponents for the different solutions
 and compares them with the corresponding
RWA results for $\wb=1.3$ and $\wb=2$, identifying accurate
semi-quantitative agreement, as expected from the discussion
above.

\subsection{$\delta=3\epsilon/2$ case: Manakov-like coupling}

\begin{figure}[t]
\begin{center}
\begin{tabular}{cc}
$\delta=3\epsilon/2$. $\epsilon=+1$ &
$\delta=3\epsilon/2$. $\epsilon=-1$ \\
\includegraphics[width=6cm]{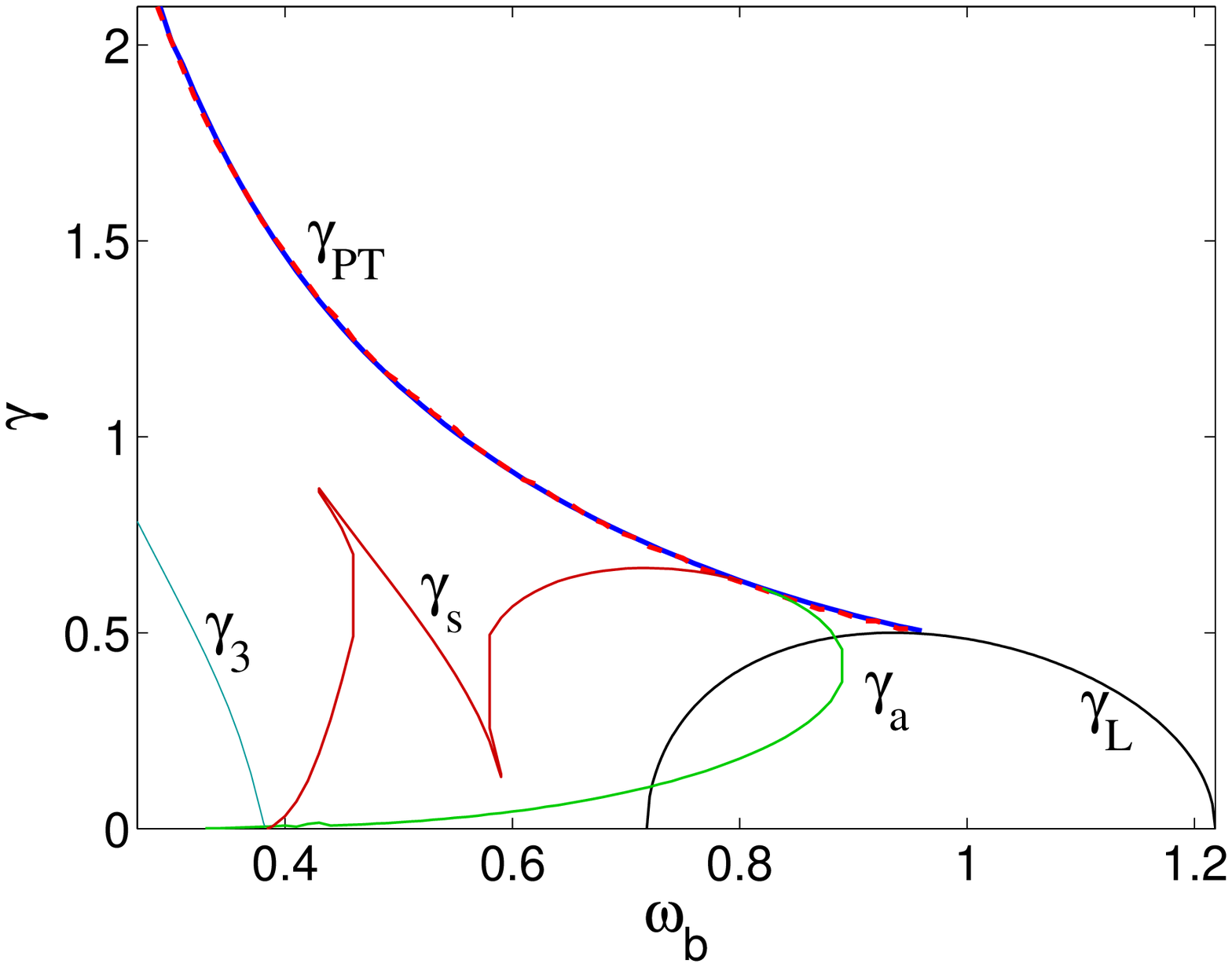} &
\includegraphics[width=6cm]{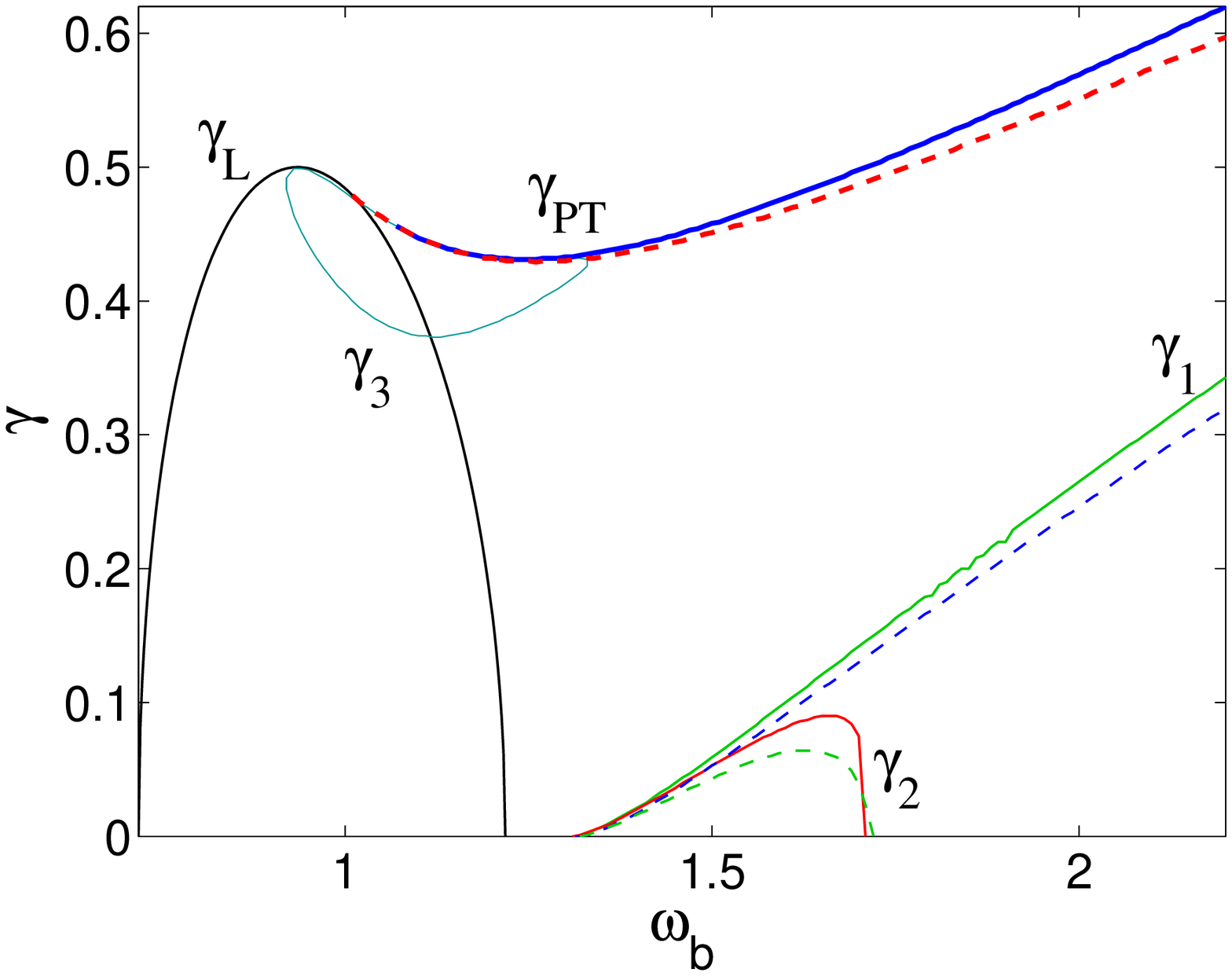} \\
\end{tabular}
\caption{Planes with curves separating regions of solutions that share the same properties when $\delta=3\epsilon/2$ and $s=\sqrt{15}/8$ (see text). } \label{fig:delta1_5}
\end{center}
\end{figure}

The interest of this case lies in the fact that, in the RWA, the coupling between fields, is similar to the Manakov equation i.e., bearing equal
self- and cross- interaction among the complex nonlinear variables
$\phi_{1,2}$. Figure \ref{fig:delta1_5} shows the different regions for the soft and hard case. In the soft case, the behavior is similar to that of $\delta=\epsilon$, even though the RWA predicted the existence of \Ap and \Sp solutions for this case. For $\wb\lesssim0.29$, the bifurcation diagram becomes very complex, similar to the $\delta=\epsilon$ case. For the hard case, the phenomenology is similar to the $\delta=\epsilon$ case; i.e. there is a good agreement with the RWA except for an additional region for which the \So solution is unstable. Because of the greater similarity with the $\delta=\epsilon$ case, no bifurcation diagrams are included for the present case.

\subsection{$\delta=3\epsilon$ case: Integrability}

\begin{figure}[t]
\begin{center}
\begin{tabular}{cc}
$\delta=3\epsilon$. $\epsilon=+1$ &
$\delta=3\epsilon$. $\epsilon=-1$ \\
\includegraphics[width=6cm]{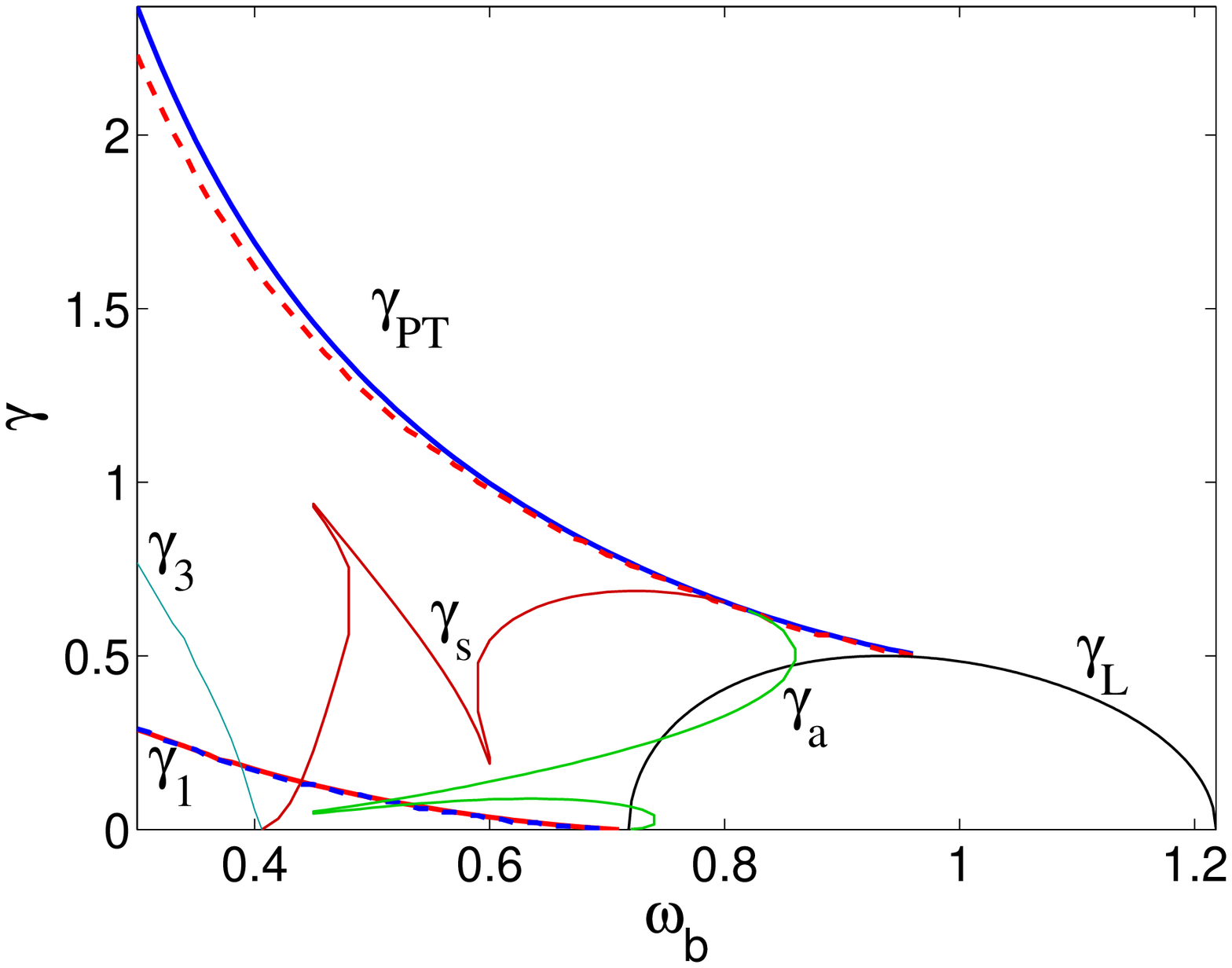} &
\includegraphics[width=6cm]{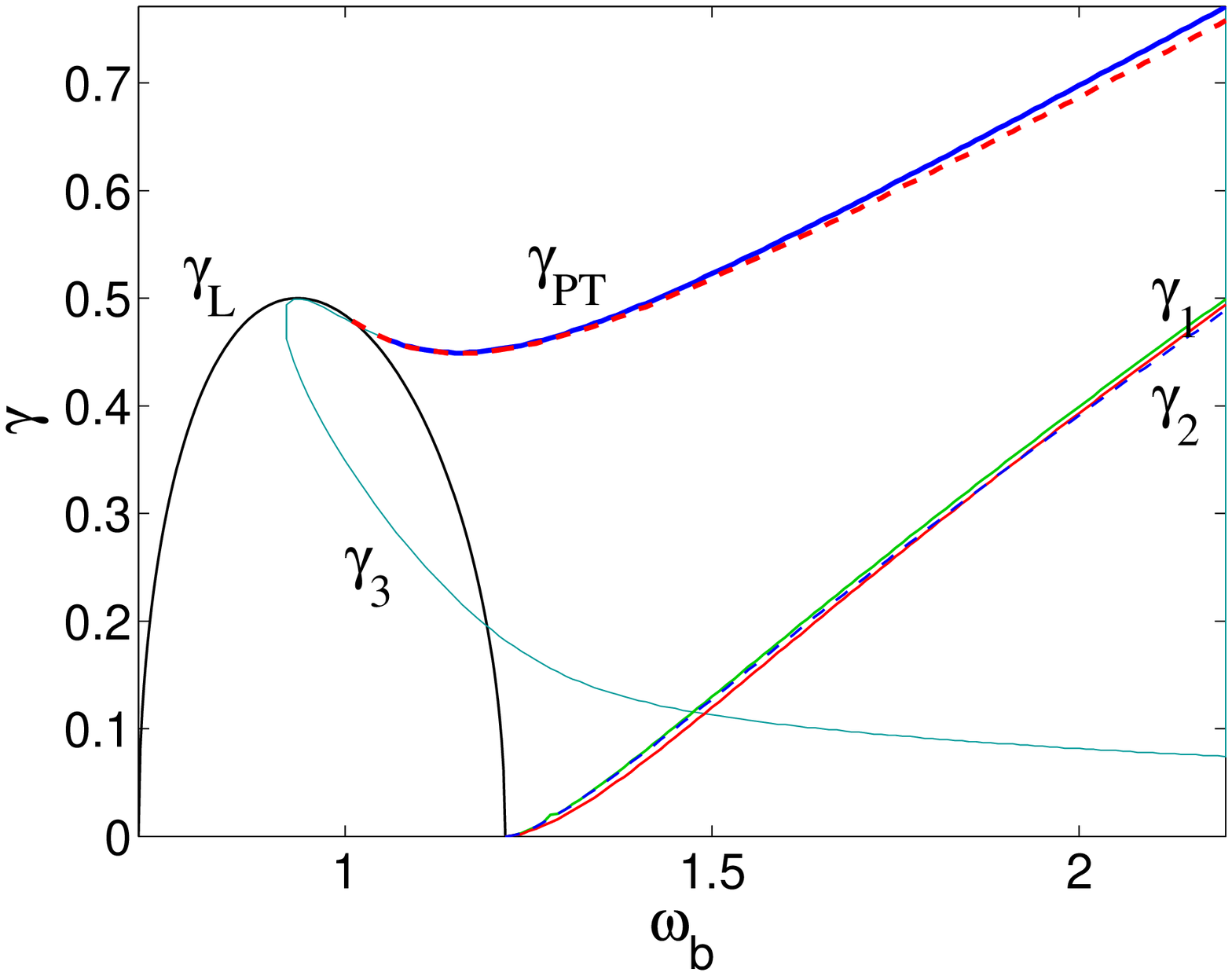} \\
\end{tabular}
\caption{Planes with curves separating regions of solutions that share the same properties when $\delta=3\epsilon$ and $s=\sqrt{15}/8$ (see text). } \label{fig:delta3}
\end{center}
\end{figure}

This case is arguably more interesting than the previous one, not only because of the existence of more solution families and also nontrivial
discrepancies with the RWA regimes, but also because of the integrability of
the dynamical equations, as they form the Hamiltonian $H_2$ [cf. Eq.~(\ref{eq:Ham2})]

Figure \ref{fig:delta3} illustrates the different regions in this case. For the soft case, the \Ap and \Sp modes do exist, as in the RWA. Contrary to the RWA predictions, however, the modes bifurcating at $\gPT$ are the \So and \Ap ones for $\wb>\omega_3=\wA/3$ and the \Atm and \Ap otherwise, whereas the modes \Ao and \Sp bifurcate at $\gamma_1$. The curves $\gs$ and $\gamma_3$ have a similar meaning as before, whereas at the right of curve $\ga$ it is the \Ap mode which is unstable. While the RWA predicted stability for modes below curve $\gamma_1$, here the \Ao and \Sp modes are stable for small $\gamma$ and unstable close to $\gamma_1$ (the change of stability curve is not shown in the figure in order not to make it even more complex).

The hard case is similar to the previous ones except for two facts: (i) the curve $\gamma_3$, above which the \So mode is unstable, extends now for every value of $\wb$, tending asymptotically to $\gamma=0$ for high $\wb$ (and, consequently, for $\wb>\wA$ the \Ap solution is unstable above the curve); (ii) below curve $\gamma_2$ (which does not exist within the RWA), the \Sp solution is unstable, contrary to the previous values of $\delta$ for which it was the \So mode that was unstable below the curve.

\begin{figure}[t]
\begin{center}
\begin{tabular}{ccc}
    $\delta=3\epsilon$, $\epsilon=1$, $\wb=0.6$ &
    $\delta=3\epsilon$, $\epsilon=1$, $\wb=0.3$ &
    $\delta=3\epsilon$, $\epsilon=-1$, $\wb=2$ \\
    \includegraphics[width=6cm]{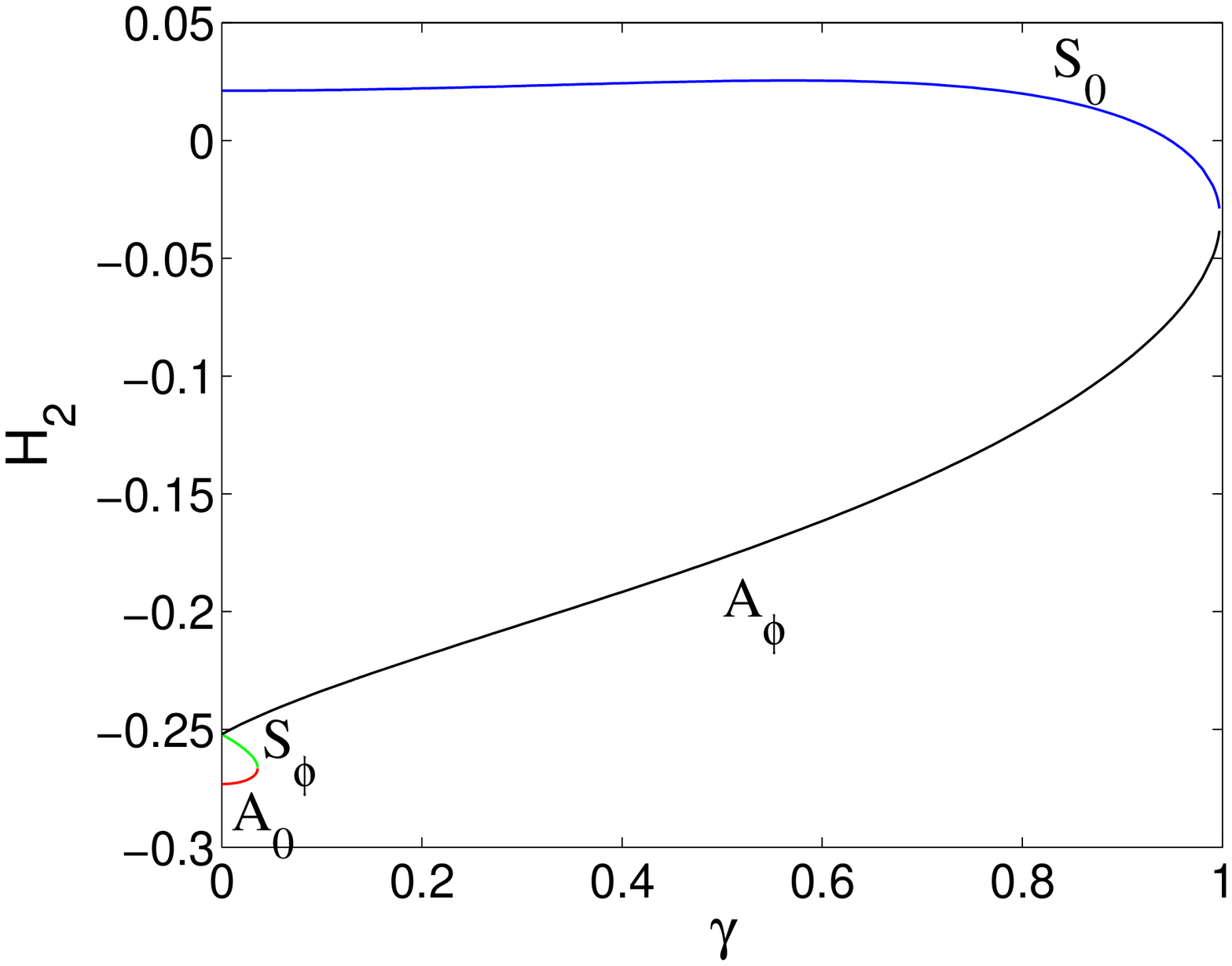} &
    \includegraphics[width=6cm]{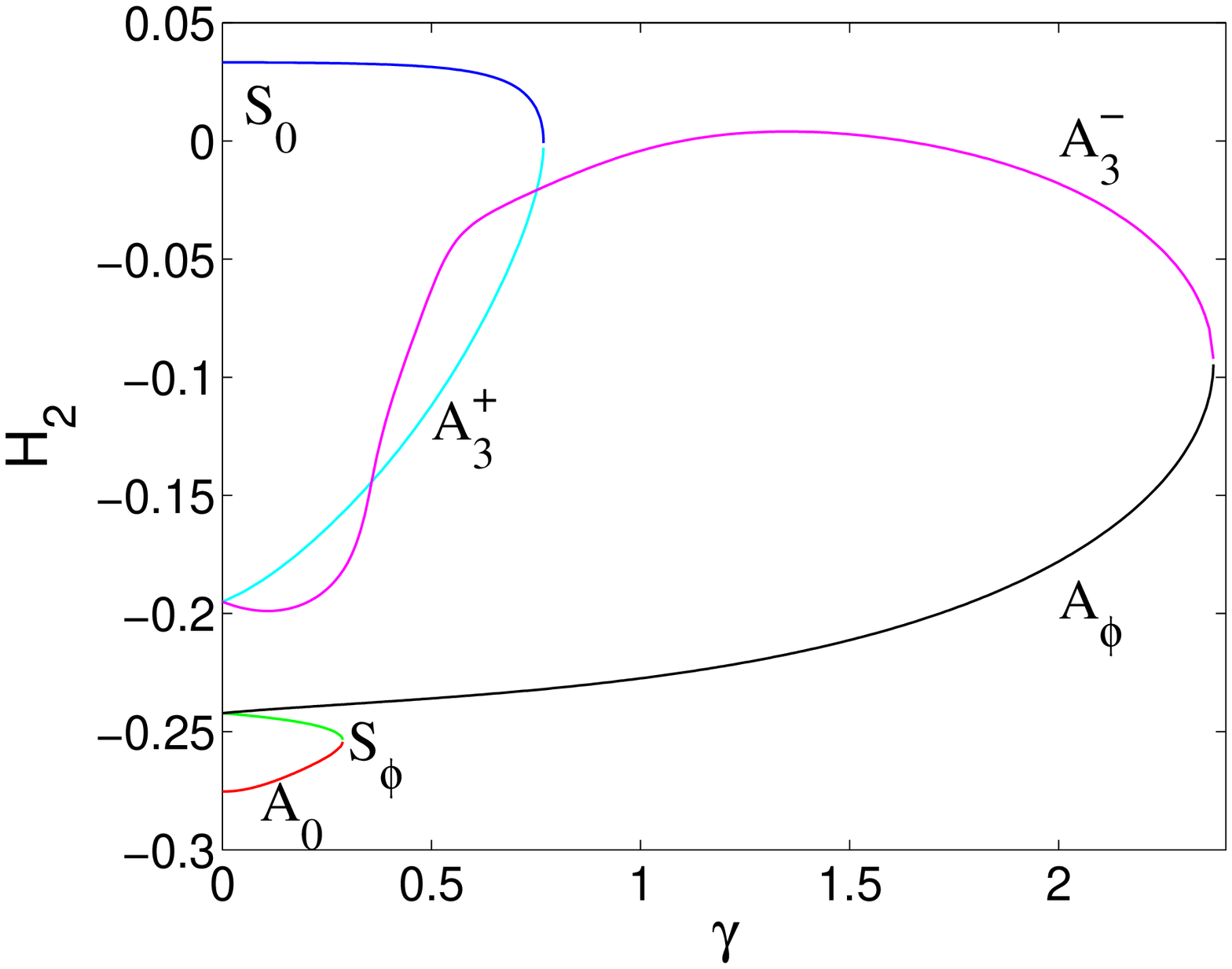} &
    \includegraphics[width=6cm]{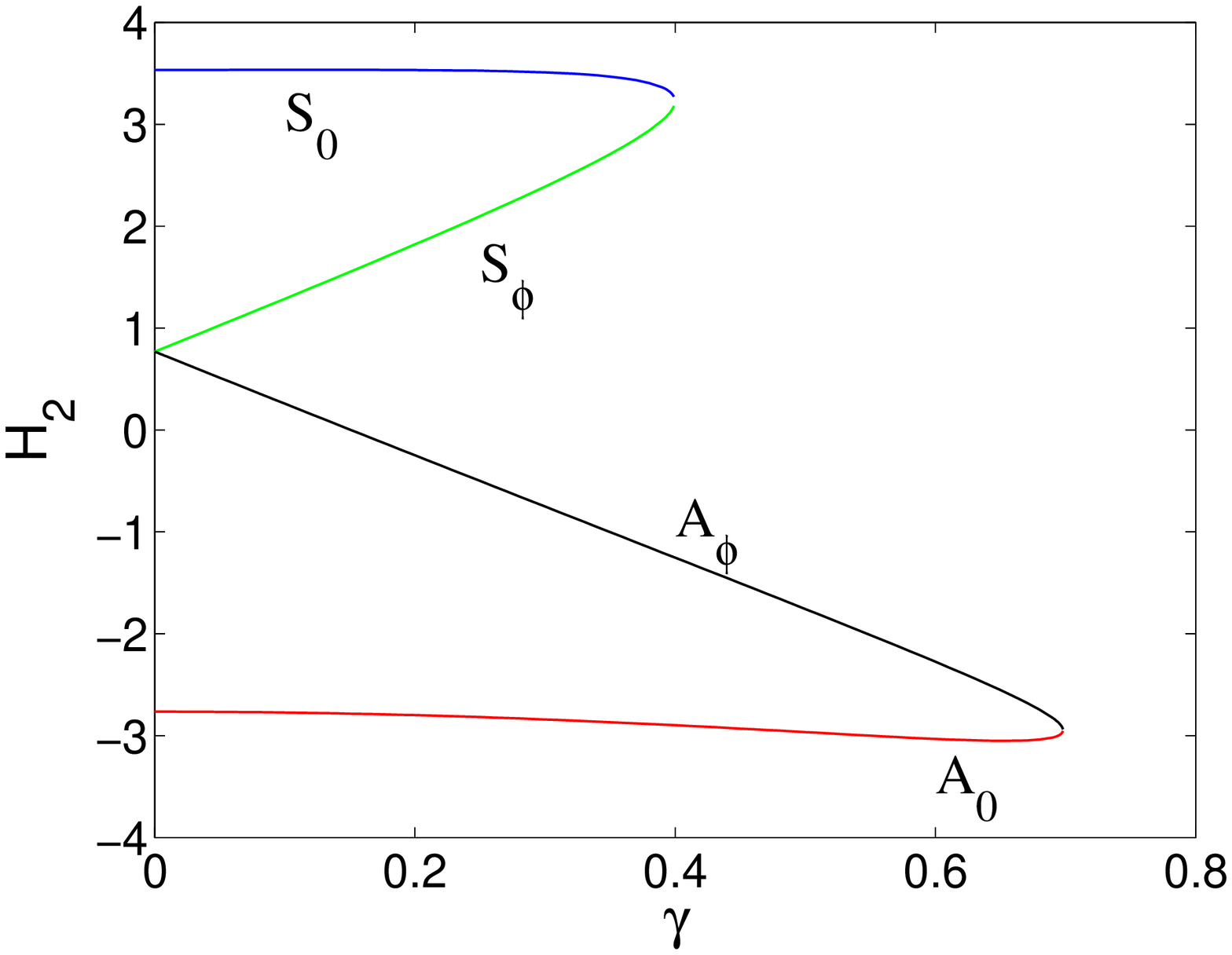} \\
\end{tabular}
\caption{Dependence of $H_2$ on $\gamma$ for each mode at different frequencies and $\epsilon$ for the case $\delta=3\epsilon$.}
\label{fig:bif_3}
\end{center}
\end{figure}

Figure \ref{fig:bif_3} illustrates the bifurcations mentioned above by means of the dependence of $H_2$ on $\gamma$ \footnote{{Notice that, contrary to the Hamiltonian $H$, it does not need to be averaged because $H_2$ is a constant of motion for $\delta=3\epsilon$}}. From the figure, it is
evident that depending on the particular value of the frequency, it is
possible that \So and \Ap, as well as \Sp and \Ao will collide
and disappear in pairwise saddle-center bifurcations (left); or,
\Atp and \So, as well as \Atm and \Ap may feature such collisions (middle);
or \So and \Sp, and \Ao and \Ap may collide and disappear hand-in-hand
(right panel).

\section{Dynamics of unstable solutions}

Finally,
in this section, we briefly touch upon some examples of the dynamical evolution of unstable modes. We are not aiming to be exhaustive; it should be
evident at this point that based on the bifurcation scenarios alone,
such a detailed study would warrant a separate paper. Instead, we aim
to present a few typical examples of dynamical outcomes observed
when evolving unstable configurations in this system.

\begin{figure}[t]
\begin{center}
\begin{tabular}{cc}
    $\delta=\epsilon$, $\epsilon=1$, $\wb=0.5$, $\gamma=0.6$ (\So) &
    $\delta=\epsilon$, $\epsilon=1$, $\wb=0.8$, $\gamma=0.5$ (\Ao) \\
    \includegraphics[width=6cm]{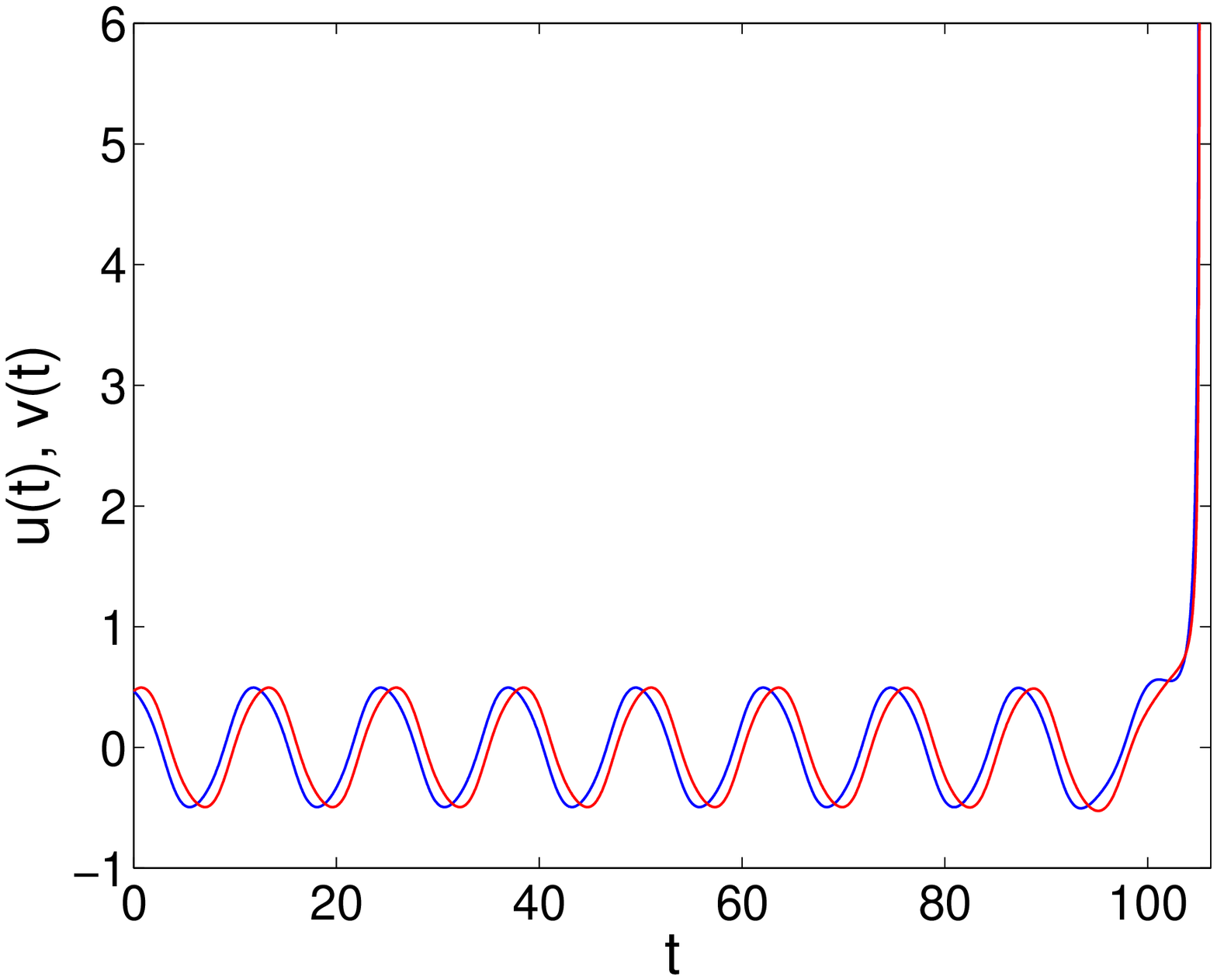} &
    \includegraphics[width=6cm]{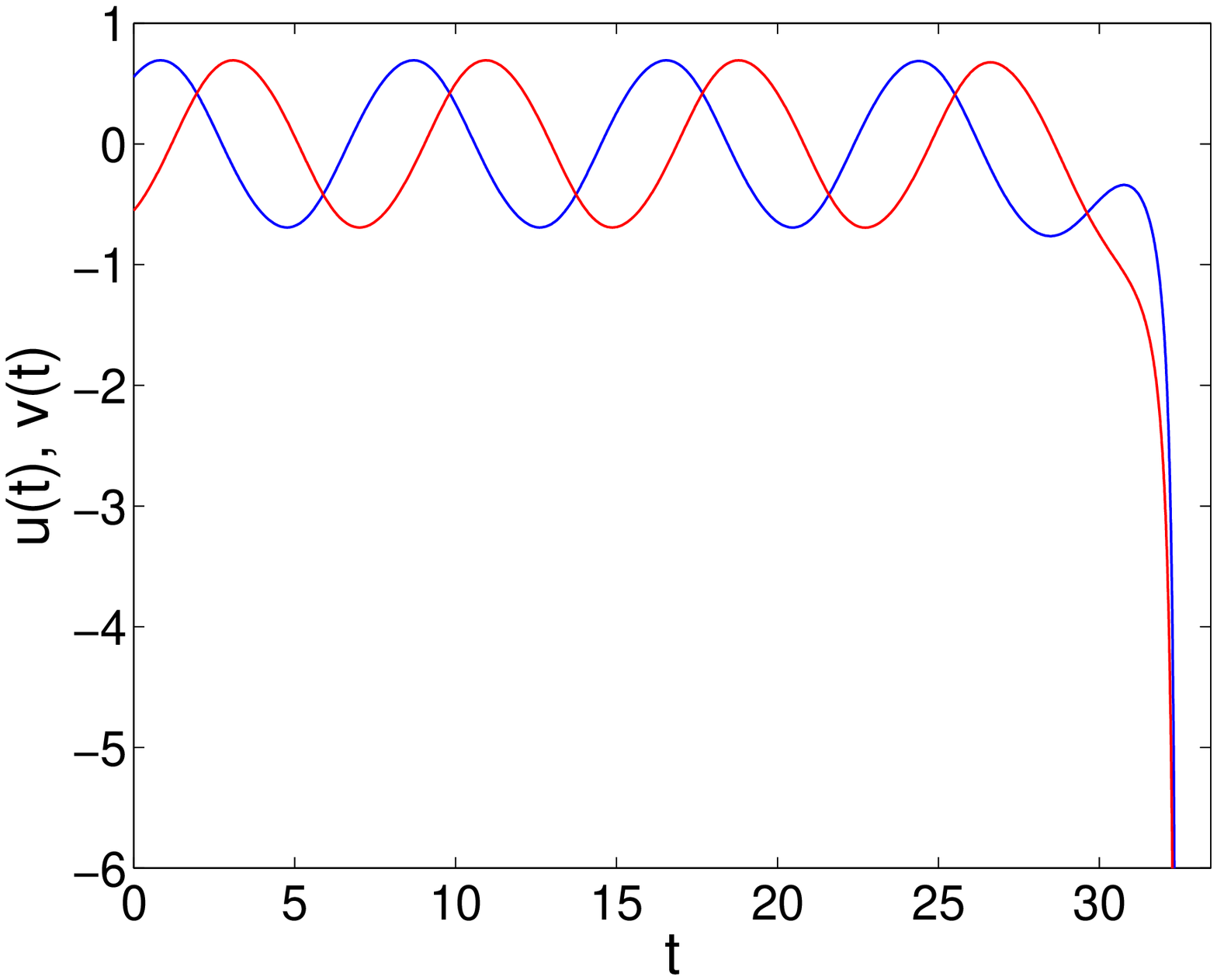} \\
    $\delta=3\epsilon$, $\epsilon=1$, $\wb=0.5$, $\gamma=0.05$ (\Ao) &
    $\delta=3\epsilon$, $\epsilon=1$, $\wb=0.7$, $\gamma=0.02$ (\Ap) \\
    \includegraphics[width=6cm]{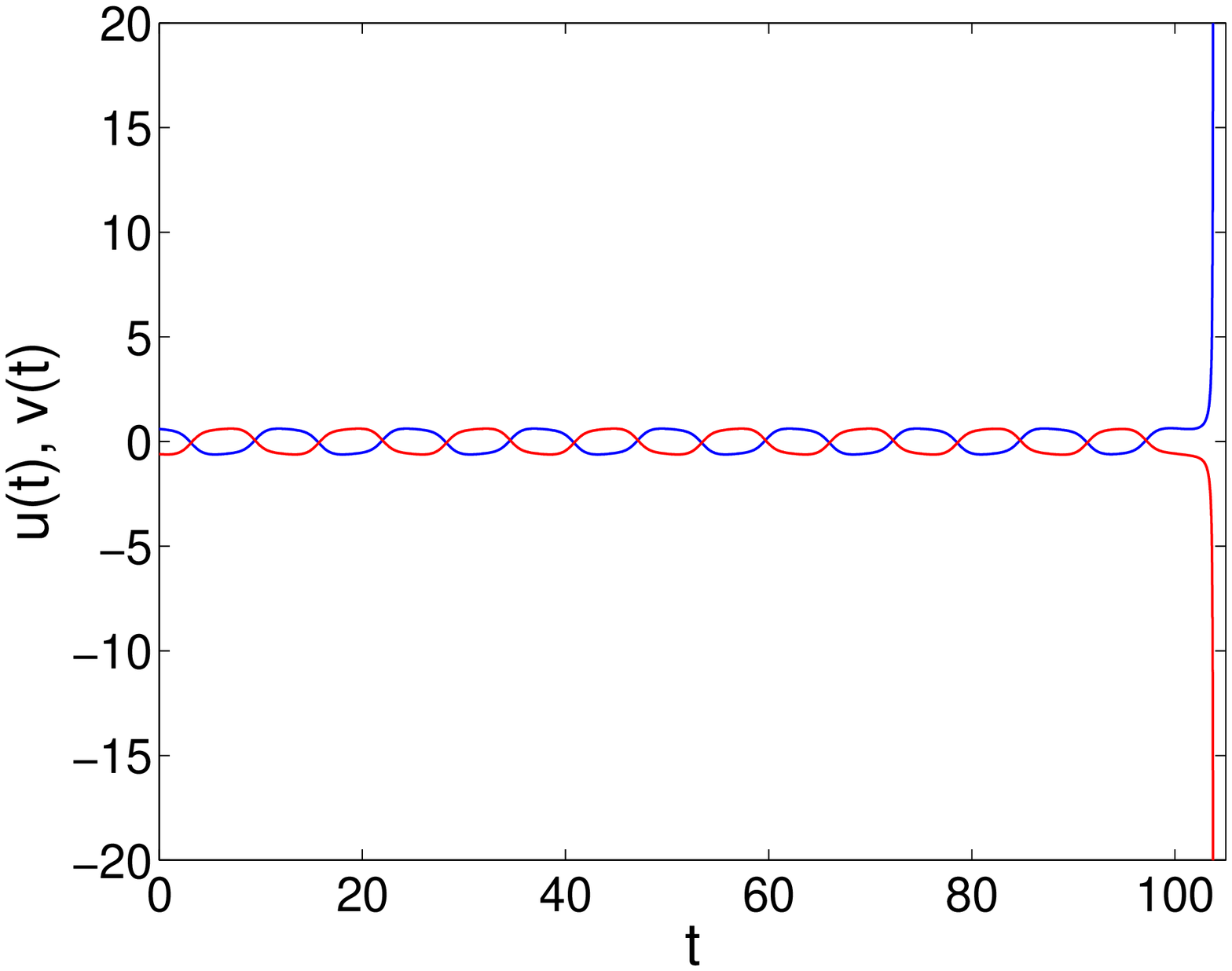} &
    \includegraphics[width=6cm]{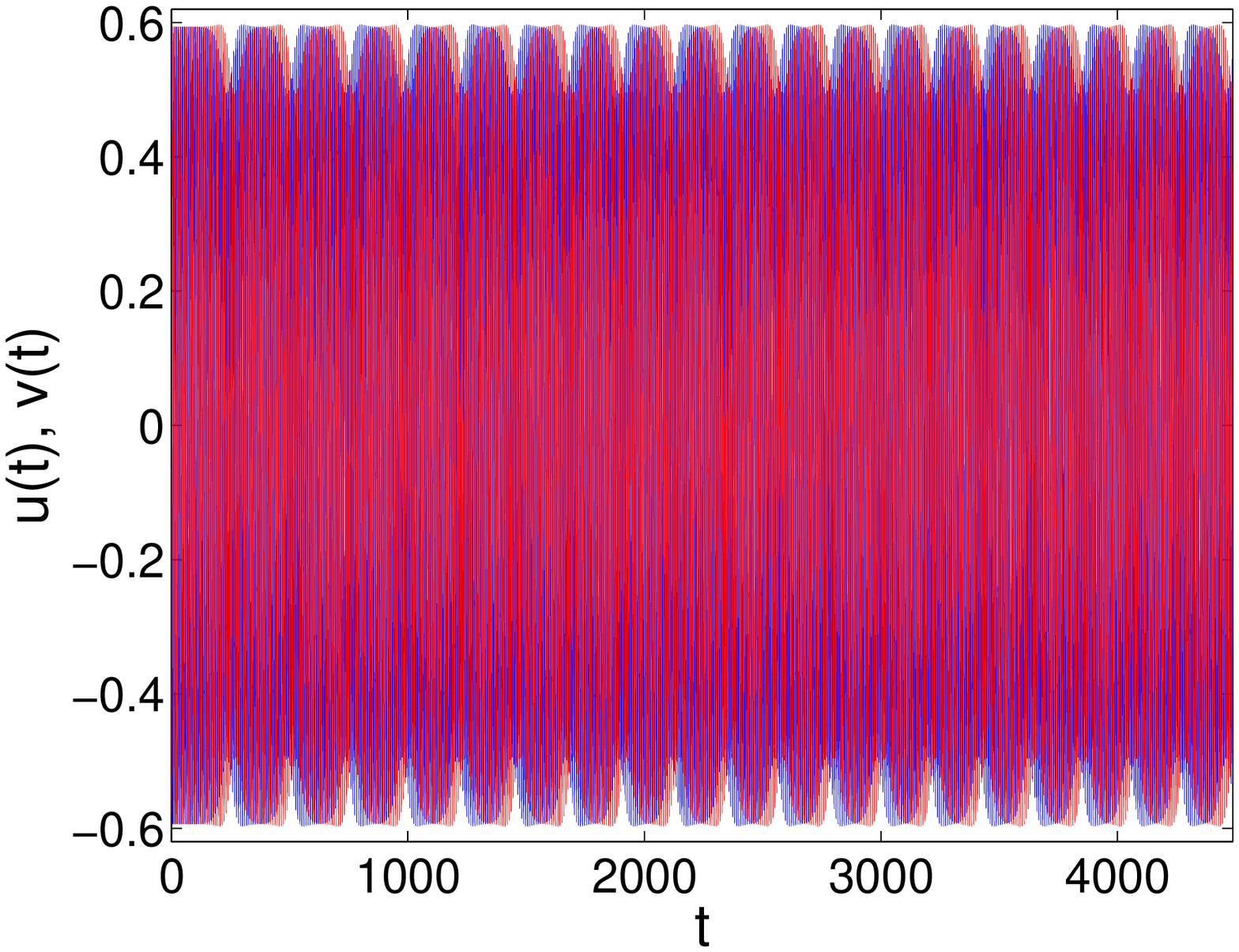} \\
\end{tabular}
\caption{Evolution of unstable solutions for the soft potential. Three
examples provide the different combination examples
where the oscillator amplitudes may grow indefinitely, while the fourth
example presents a bounded apparently quasi-periodic scenario.}
\label{fig:dyn1}
\end{center}
\end{figure}

In the soft potential, unstable solutions are mostly prone to blow-up, even in the $\delta=3\epsilon$ case where $H_2$ is conserved. This blow-up could consist of both sites tending to $\infty$ or $-\infty$ at the same time (specially in \So and \Ao solutions), or one site going to $\infty$ and the other one to $-\infty$ mostly in \Ap and \Sp solutions. \Ao solutions can exhibit both behaviors. For small values of $\gamma$, the instabilities can lead to quasi-periodic oscillations, whenever the solution at $\gamma=0$ is stable (if the solution is unstable at $\gamma=0$, it is prone to blow-up). Figure \ref{fig:dyn1} shows several examples of the dynamics of soft potentials.

\begin{figure}[t]
\begin{center}
\begin{tabular}{cc}
    $\delta=\epsilon$, $\epsilon=-1$, $\wb=1$, $\gamma=0.16$ (\So) &
    $\delta=\epsilon$, $\epsilon=-1$, $\wb=2$, $\gamma=0.46$ (\So) \\
    \includegraphics[width=6cm]{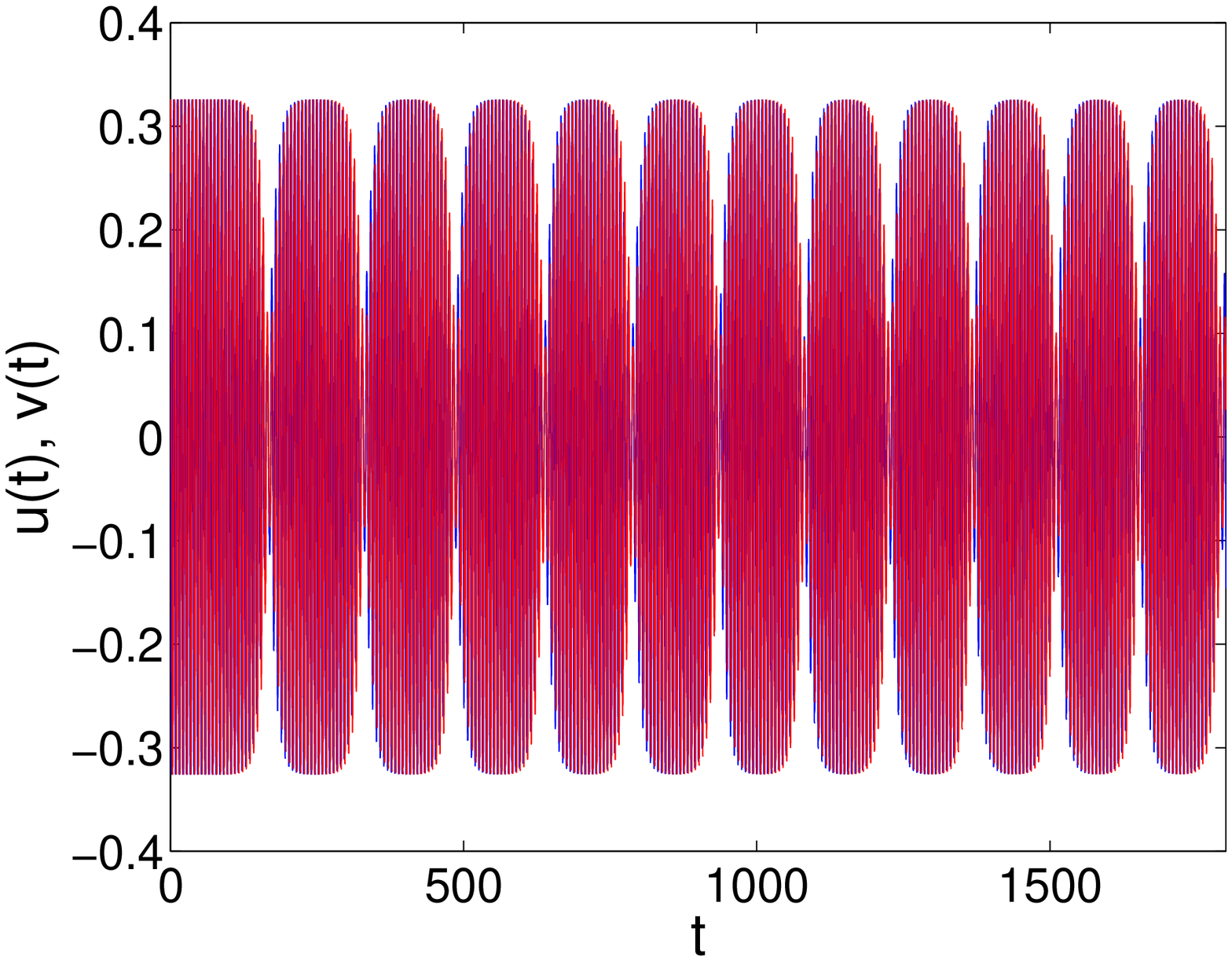} &
    \includegraphics[width=6cm]{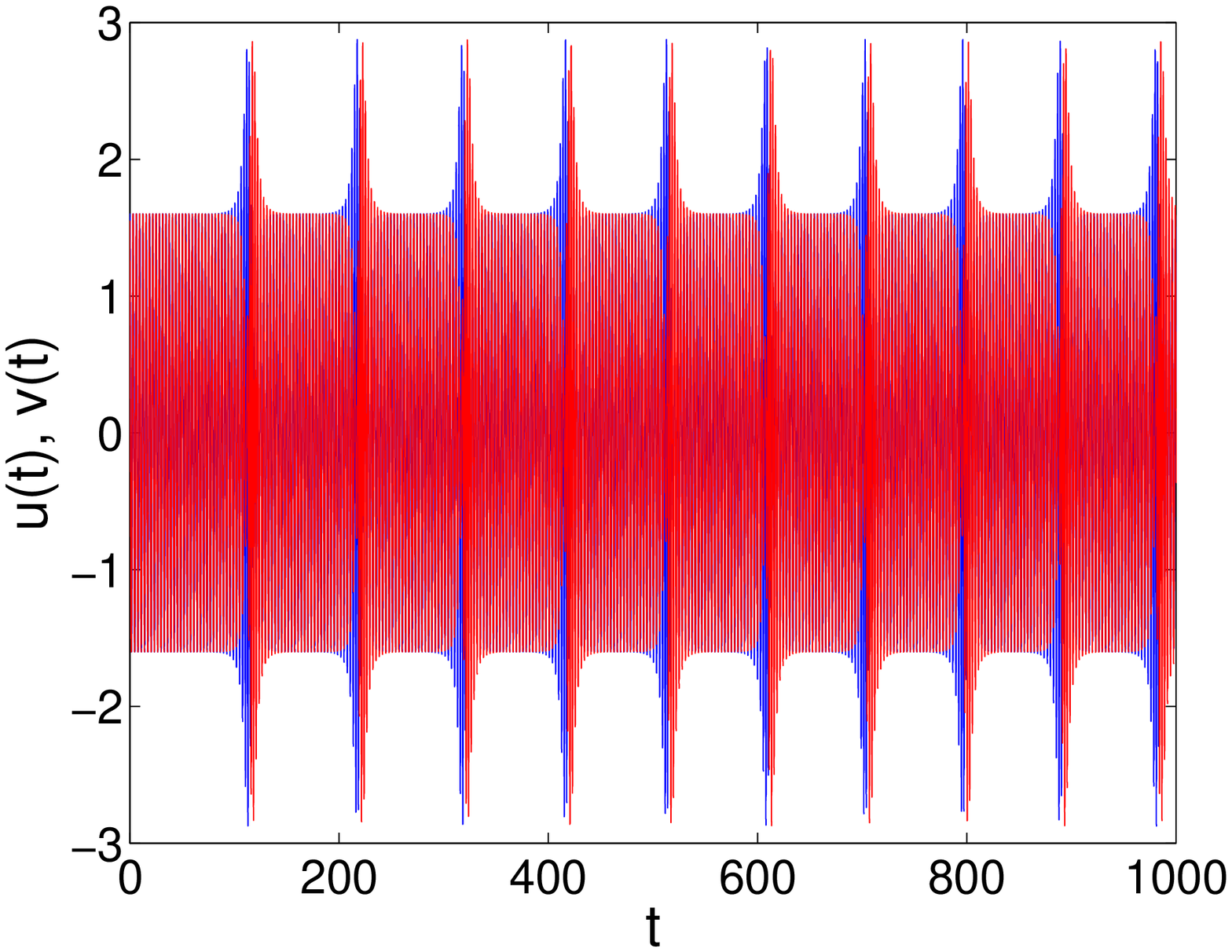} \\
    $\delta=3\epsilon$, $\epsilon=1$, $\wb=2$, $\gamma=0.2$ (\Ap) &
    $\delta=3\epsilon$, $\epsilon=1$, $\wb=2$, $\gamma=0.3$ (\Sp) \\
    \includegraphics[width=6cm]{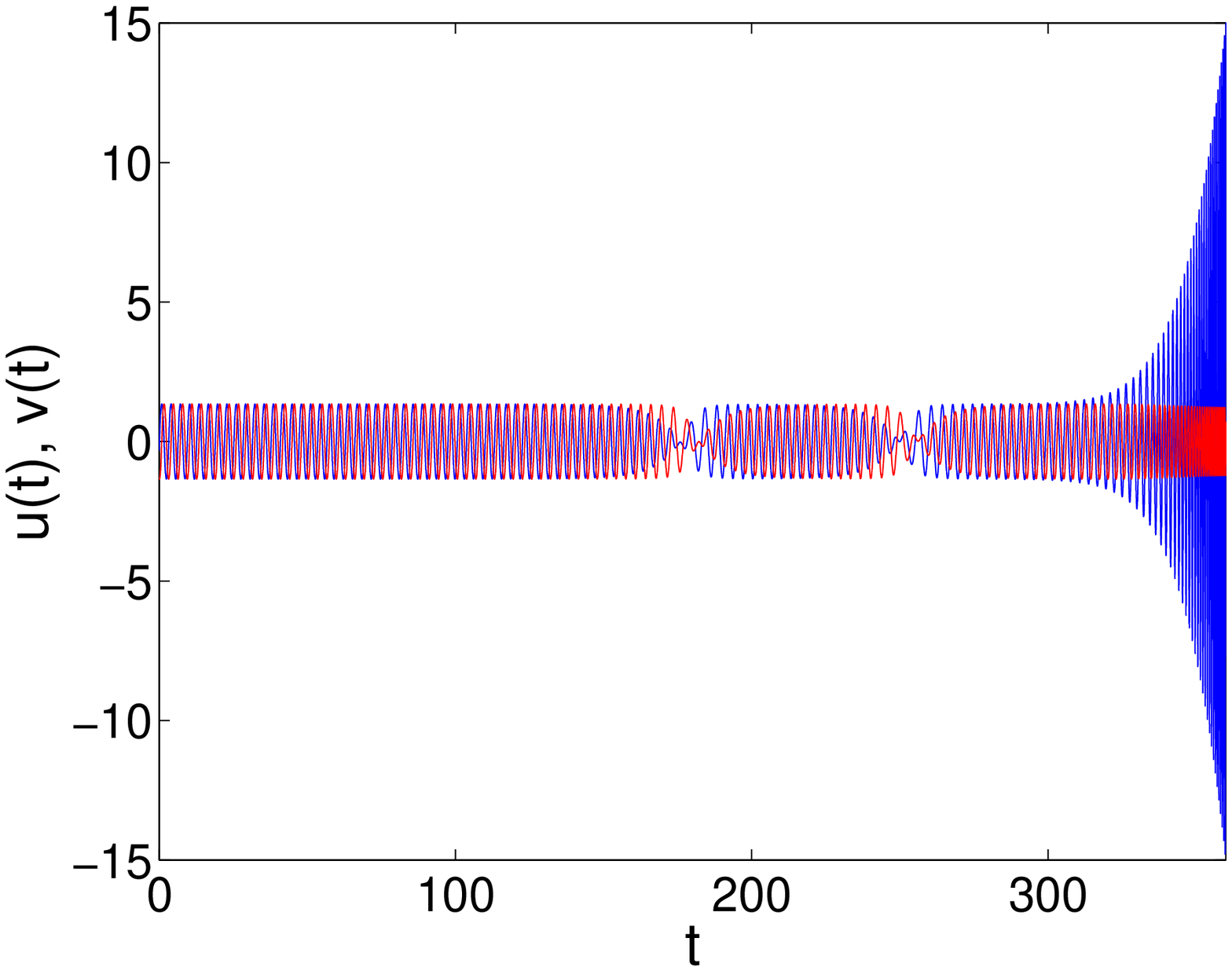} &
    \includegraphics[width=6cm]{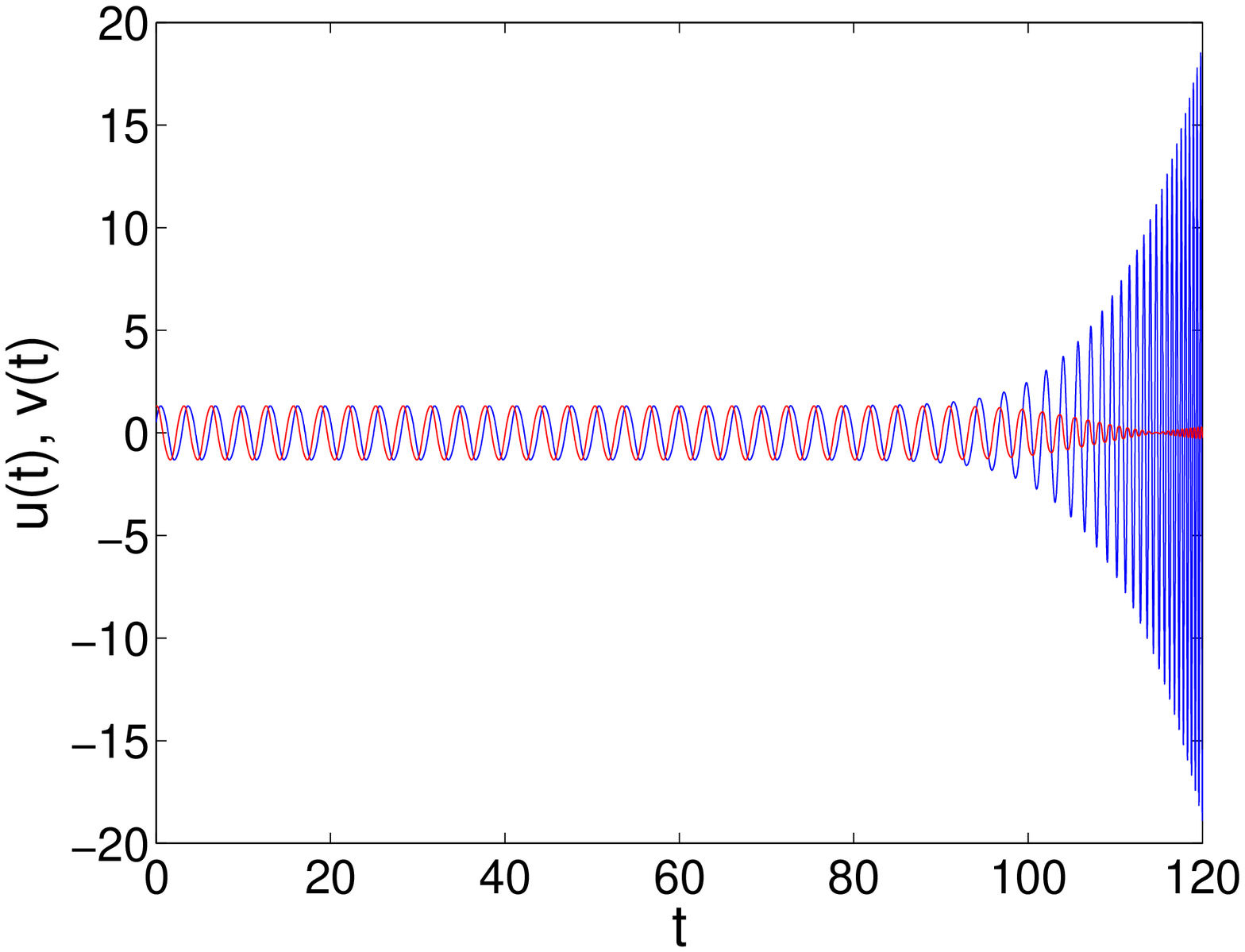} \\
\end{tabular}
\caption{Evolution of unstable solutions for the hard potential.
The top panels feature examples of quasi-periodic oscillations,
while the bottom panels illustrate indefinite growth of one of the oscillators
coupled with a decaying oscillation of the other (possibly very slowly, as in the case of the bottom left panel).}
\label{fig:dyn2}
\end{center}
\end{figure}

In the hard potential case, there are some differences between the dominant behavior when $\delta=\epsilon$ with respect to $\delta=3\epsilon$, as shown in Fig. \ref{fig:dyn2}. In the former case, where the instabilities arise from the \So solutions, we have observed quasi-periodic oscillations with amplitude peaks when $\wb>\wph$ and without these peaks if $\wb<\wph$. In the latter case, although quasi-periodic oscillations are present (mainly for small growth rates), the dominant behavior is an apparent (modulated) exponential growth on the anti-damped site, associated with a decay on the damped site. This decay is very much slower when the instability arises from the \Ap solution.

\section{Conclusion}

In the present work, we have studied various exact solutions and their stability for a generalized $\cP\cT$-symmetric coupled nonlinear oscillator system.
Complementing earlier works both at the linear level~\cite{bend_gian}
(describing a recent experiment~\cite{pt_whisper}) and at the nonlinear
level~\cite{PRA,bar_gian}, we have examined a variety of cases regarding
the relative strength of the self- and cross-interaction between our
nonlinear oscillators. In our earlier work~\cite{PRA}, only self-interactions
were considered, while in the important recent work of~\cite{bar_gian},
only a specific value of the cross interaction was considered ($\delta=3
\epsilon$), revealing remarkably the Hamiltonian nature of the model,
and then restricting consideration to its DNLS analogue. Here, we have
extended considerations to three relevant cases, namely
$\delta=\epsilon$, $\delta=3 \epsilon/2$ and $\delta=3 \epsilon$,
exploring how the existence, nonlinear bifurcation and even
dynamical trends develop as we move from weaker to stronger
cross-interaction between the nonlinear oscillators.
Importantly, the relevant pictures were developed not only
for the rotating wave approximation model of the DNLS form,
but also for the full model of the coupled oscillators. Generally,
the two cases, namely the monochromatic approximation and the
full system were very similar, except for the highly nonlinear
regime, especially in the soft nonlinearity case. Numerous important
features were identified along the way including, e.g., new families
of solutions at relative phase angles other than $0$ and $\pi$
(introduced by the cross-interaction between oscillators), as well
as solutions tractable solely in a numerical form from the four
principal families explored. Yet another feature was the
existence in the oscillator system of families of solutions not
only in the $\gamma=0$ but even in the $\gamma \neq 0$ case
in explicit form; one such pair of families appears to ``defy''
the $\cP\cT$ phase transition (in the $\delta=\epsilon$ case),
existing for all values of the gain/loss parameter $\gamma$.
Finally, the instabilities identified in the analysis were
monitored in the full dynamics of the system, revealing the
possibility of either indefinite growth or that of bounded
quasi-periodic oscillations, as the pertinent dynamical
outcome.

There are numerous questions that naturally emerge as a result of the
present work. Among the most immediate ones, it is worthwhile to
extend considerations to the case of, e.g., three oscillators
and perhaps even to that of four such, forming effectively
a two-dimensional plaquette and a building block for the
consideration of higher dimensional systems, in the spirit also
of~\cite{Guenter}. Furthermore, here only the case of cubic nonlinearities
has been explored, but it might be also of interest, as another
prototypical nonlinear system to examine the case of
quadratic nonlinearities and how their nonlinear states
are ``deformed'' in the presence of gain and loss.
At a perhaps deeper level, however, there are also some
intriguing questions that we feel are raised. For one,
an apparently $\cP\cT$-symmetric and viewed as a gain-loss
bearing system at $\delta=3 \epsilon$ is found to be Hamiltonian.
This raises the natural yet difficult question: can we discern
such a potential Hamiltonian nature and classify a system
as Hamiltonian (and not $\cP\cT$) possibly through an
appropriate (to be identified) transformation? If so,
what is the relevant criterion and how can we exclude
the presence of a yet-unknown transform that may convert
a system classified as $\cP\cT$ into one which is genuinely
Hamiltonian in a different set of variables~? Potential
progress along these veins will be reported in future work.

\vspace{5mm}

P.G.K. acknowledges support from the National Science Foundation under grants CMMI-1000337,
DMS-1312856, from FP7-People under grant IRSES-606096, from the US-AFOSR under grant FA9550-12-10332 and from the Binational (US-Israel) Science Foundation
through grant
2010239.  This work was supported in part by the U.S. Department of Energy. A.K. acknowledges financial support from Dept. of Atomic Energy, Govt. of India through a Raja Ramanna Fellowship.
P.G.K. also acknowledges useful discussions with Igor Barashenkov.

\appendix

\section{Numerical analysis of periodic orbits}\label{app:numerics}

In order to calculate periodic orbits, we make use of a Fourier space implementation of the dynamical equations and continuations in frequency or gain/loss parameter are performed via a path-following (Newton-Raphson) method. Fourier space methods are based on the fact that the solutions are $\Tb$-periodic; for a detailed explanation of these methods, the reader is referred to Refs.~\cite{AMM99,phason}.
The method has the advantage, among others, of providing an explicit, analytical form of the Jacobian. Thus, the solution for the two nodes
can be expressed in terms of a truncated Fourier series expansion:

\begin{equation}\label{eq:series}
    u(t)=\sum_{n=-n_m}^{n_m} y_n\exp(\ii n \wb t),\qquad v(t)=\sum_{n=-n_m}^{n_m} z_n\exp(\ii n \wb t) ,
\end{equation}

\noindent with $n_m$ being the maximum of the absolute value of the running index $k$ in our Galerkin truncation of the full Fourier series
solution. In the numerics, $n_m$ has been chosen as 21. After the introduction of (\ref{eq:series}), the dynamical equations (\ref{eq:dyn}) yield a set of $2\times(2n_m+1)$ nonlinear, coupled algebraic equations:

\begin{eqnarray}\label{eq:Fourier}
    F_{n,1} &\equiv& -\wb^2n^2y_n-\ii\gamma\wb n y_n+\F_n[V'(u,v)]-k z_n=0 , \\
    F_{n,2} &\equiv& -\wb^2n^2z_n+\ii\gamma\wb n z_n+\F_n[V'(v,u)]-k y_n=0 ,
\end{eqnarray}

\noindent with $V'(u_1,u_2)=u_1-\epsilon u_1^3-\delta u_1u_2^2$. Here, $\F_n$ denotes the Discrete Fourier Transform:

\begin{equation}
    \F_n[V'(u)]=\frac{1}{N}\sum_{q=-n_m}^{n_m}V'\left(\sum_{p=-n_m}^{n_m}y_p\exp\left[\ii \frac{2\pi p q}{N}\right]\right)
    \exp\left[-\ii \frac{2\pi n q}{N}\right],
\end{equation}
with $N=2n_m+1$. The procedure for $\F_n(v)$ is similar to the previous case. As $u(t)$ and $v(t)$ must be real functions, it implies that $y_{-n}=y^*_n,\ z_{-n}=z^*_n$.

In order to study the spectral stability of periodic orbits, we introduce a small perturbation $\{\xi_1,\xi_2\}$ to a given solution
$\{u_0,v_0\}$ of Eqs. (\ref{eq:dyn}) according to $u=u_{0}+\xi_1$, $v=v_0+\xi_2$. Then, the equations satisfied to first order in $\xi_n$ read:

\begin{eqnarray}\label{eq:stab}
    \ddot\xi_1 &=& (3\epsilon u_0^2+\delta v_0^2-1)\xi_1 + \gamma \dot\xi_1 + (k+2\delta u_0v_0)\xi_2 , \nonumber \\
    \ddot\xi_2 &=& (3\epsilon v_0^2+\delta u_0^2-1)\xi_2 - \gamma \dot\xi_2 + (k+2\delta u_0v_0)\xi_1 ,
\end{eqnarray}

\noindent or, in a more compact form: $\mathcal{N}(\{u(t),v(t)\})\xi=0\,$, where $\mathcal{N}(\{u(t),v(t)\})$ is the relevant linearization operator.
In order to study the spectral (linear) stability analysis of the relevant solution, a Floquet analysis can be performed if there exists $T_b\in\mathbb{R}$ so that the map $\{u(0),v(0)\}\rightarrow \{u(\Tb),v(\Tb)\}$ has a fixed point (which constitutes a periodic orbit of the original system). Then, the stability properties are given by the spectrum of the Floquet operator $\mathcal{M}$ (whose matrix representation is the monodromy) defined as:
\begin{equation}\label{eq:monodromy}
    \left(\begin{array}{c} \{\xi_{n}(\Tb)\} \\ \{\dot\xi_{n}(\Tb)\} \\ \end{array}
    \right)=\mathcal{M}\left(\begin{array}{c} \{\xi_{n}(0)\} \\ \{\dot\xi_{n}(0)\} \\ \end{array}
    \right) .
\end{equation}

The $4\times4$ monodromy eigenvalues $\Lambda=\exp(\ii\theta)$ are dubbed the {\em Floquet multipliers} and $\theta$ are denoted as {\em Floquet exponents}
(FEs). This operator is real, which implies that there is always a pair of multipliers at $1$ (corresponding to the so-called phase and growth modes) and that the eigenvalues come in pairs $\{\Lambda,\Lambda^*\}$. As a consequence, due to the ``simplicity'' of the FE structure (one pair always at $1$ and one additional pair) there cannot exist Hopf bifurcations in the dimer, as such bifurcations would imply the collision of two pairs of multipliers and the consequent formation
of a quadruplet of eigenvalues which is impossible here. Nevertheless, in the present problem, the motion of the pair of multipliers can lead to an instability through exiting (through $1$ or $-1$) on the real line leading to one multiplier (in absolute value) larger than $1$ and one smaller than $1$.

\end{document}